\documentclass[aps,prb,twocolumn,superscriptaddress,showpacs]{revtex4}

\usepackage{color}
\usepackage{times}
\usepackage{epsfig}
\usepackage{amsmath}
\usepackage{amssymb}
\usepackage{amsmath, amsthm, amssymb,  mathrsfs, epsfig}
\usepackage{dcpic, pictex, setspace, rotating}
\usepackage{faktor}
\newcommand{\be}{\begin{equation}}
\newcommand{\ee}{\end{equation}}

\usepackage{graphicx}

\begin{document}

\title{Weakly-Coupled non-Abelian Anyons
in Three Dimensions}

\author{Michael Freedman}
\affiliation{Microsoft Research, Station Q, Elings Hall, University of California, Santa Barbara, CA 93106}
\author{Matthew B. Hastings}
\affiliation{Microsoft Research, Station Q, Elings Hall, University of California, Santa Barbara, CA 93106}
\author{Chetan Nayak}
\affiliation{Microsoft Research, Station Q, Elings Hall, University of California, Santa Barbara, CA 93106}
\affiliation{Department of Physics, University of California, Santa Barbara, CA 93106}
\author{Xiao-Liang Qi}
\affiliation{Microsoft Research, Station Q, Elings Hall, University of California, Santa Barbara, CA 93106}
\affiliation{Department of Physics, Stanford University, Stanford, CA 94305, USA}

\begin{abstract}
We introduce a Hamiltonian coupling Majorana fermion degrees of freedom to a quantum dimer model.  We argue that, in three dimensions, this model has
deconfined quasiparticles supporting Majorana zero modes
obeying nontrivial statistics. We introduce two effective field theory
descriptions of this deconfined phase, in which the excitations have
Coulomb interactions. A key feature of this system is the existence
of topologically non-trivial fermionic excitations, called ``Hopfions"
because, in a suitable continuum limit of the dimer model, such excitations
correspond to the Hopf map and are related to excitations
identified in Ref. \onlinecite{Ran10}.
We identify corresponding topological invariants of the
quantum dimer model (with or without fermions) which
are present even on lattices with trivial topology.
The Hopfion energy gap depends upon the phase of the model.
We briefly comment on the possibility of a phase
with a gapped, deconfined $\mathbb{Z}_2$ gauge field,
as may arise on the stacked triangular lattice.
\end{abstract}

\maketitle


\section{Introduction}
Recently, it has been shown that three-dimensional systems of free fermions can have defects with Majorana zero-modes\cite{Teo10}.  These defects
display a ghostly remnant of braid statistics: even though the defects
are free to move in three dimensions, there are two inequivalent ways to interchange a pair of defects. This situation was analyzed further in Ref. \onlinecite{Freedman11},
where it was shown that the statistics of these particles is governed
by an extension of the permutation group which was dubbed
the `ribbon permutation group.' Motions of these defects
realize a {\it projective representation} of the ribbon permutation group
which endows them with a non-Abelian anyonic character.
We will call them 3D non-Abelian anyons, although this
is a slight abuse of the terminology.
These results were generalized to arbitrary
dimension and symmetry class.

Refs. \onlinecite{Teo10} and \onlinecite{Freedman11} both
considered systems of free Majorana fermions coupled to a
position-dependent mass term.  This mass term was treated
as a classical degree of freedom, with no quantum fluctuations,
which begs the question of what happens
when the mass term is also allowed to fluctuate.
If the mass term fluctuates about an ordered ground state,
then the defects which carry Majorana zero modes interact via a linear confining potential.  Therefore, it is natural to seek a model without long-range interaction
between defects. In such a model, 3D non-Abelian anyons would
be the weakly-coupled low-energy quasiparticles of a system -- truly
a higher-dimensional analogue of anyons in the fractional quantum Hall
effect. In this paper, we succeed partially in this goal
by constructing a microscopic model and
presenting arguments that it has deconfined
defects with Majorana zero modes. The interaction between
defects has the power-law decay characteristic
of Coulomb interactions.  The fermionic excitations of this model
are the zero modes associated with
defects, gapped modes associated with excitations of
bulk fermions, and topologically nontrivial
configurations of the mass field (the ``Hopfions") which carry fermion number.
In the ordered phase (in which the 3D non-Abelian
anyons are confined) the Hopfion will be gapped, as
in the model of Ran, Hosur, and Vishwanath\cite{Ran10} (whose
terminology we adopt), who
identified it as a gapped fermionic soliton.
We present arguments, based on two different
effective field theory descriptions,
that the Hopfion is gapless in the Coulomb phase.
In the first effective field theory, the order parameter
is scrambled by quantum fluctuations so that the order
disappears. We rotate the fermions to the local
direction of the order parameter so that, even when the order parameter disorders
and the system enters a Coulomb phase, the fermionic band
remains gapped although, we argue, there are gapless fermionic excitations
in the form of Hopfions.
In the second field theory, we gauge the order parameter, thereby
suppressing its gradient energy. We do this in an anomaly-free way
by introducing a fourth spatial dimension, of finite extent, so that
the physical space is one surface of the four-dimensional slab.
When the order parameter condenses, the hedgehogs\footnote{We will
use the terms hedgehogs, monopoles, defects, and quasiparticles
interchangeably to refer to these excitations which support Majorana
zero modes. The context will usually dictate which term we use.}
 become
magnetic monopoles, interacting via a Coulomb force
so long as there are gapless fermionic excitations -- Hopfions --
at the other surface of the four-dimensional slab.

In Section \ref{sec:dimer-topology}, we derive
new topological invariants of dimer models.
The motivation comes from the model
of free fermions coupled to dimers
studied in Section II of Ref. \onlinecite{Freedman11}.
In order to understand the statistics of quasi-particles in this model,
these invariants will be essential.
However, these results are explained in a self-contained manner,
and they should also be relevant to physicists interested in studying more traditional dimer models without fermions. While it is
well-known that these
dimer models on a torus have different topological sectors corresponding to different winding numbers, we show that there are additional
invariants present even on lattices with trivial topology,
such as a cubic lattice with open boundary conditions. Understanding
these additional invariants may be important in simulations
of dimer models, as these invariants imply that
simulations using plaquette flips will not be ergodic even in a given winding number sector.  In the discussion (Section \ref{sec:discussion}),
we raise some open problems regarding
the energy of these different topological sectors which could be addressed using quantum Monte Carlo simulations.

In Section \ref{sec:FDM}, we
add dimer dynamics to the the model
of free fermions coupled to dimers
studied in Section II of Ref. \onlinecite{Freedman11}
(where the dimers were non-dynamical). We argue
that this model inherits a Coulomb phase from the
ordinary (i.e. without fermions) quantum dimer model,
and discuss the statistics of the Majorana zero-mode-carrying
defects in this phase.
In Section \ref{sec:field-theories}, we give two different
field theories which, we believe, govern the universality class
of the Coulomb phase of the fermion dimer model.
These field theories predict that the Hopfion is gapless,
and explain how the Coulomb phase evades potential obstructions
such as anomalies. Finally, in the discussion section, we summarize
our results and discuss open problems.

\section{Topological Features of Three-Dimensional Dimer Models}
\label{sec:dimer-topology}

It is well-known that the Rokhsar-Kivelson dimer Hamiltonian $H_{RK}$ has different topological sectors on a torus, corresponding to
different winding modes of the dimers.  However, it has additional topological invariants even on a lattice with
trivial topology, so long as we consider either a finite lattice or an infinite lattice
with the boundary condition that the dimers assume a given columnar configuration at infinity.
In Section \ref{sec:Hopfion}, we present
a $Z_2$ invariant of the dimer model; we refer to a configuration in which this invariant assumes a nontrivial value as a ``Hopfion" configuration, for reasons explained later.  This invariant is present in the dimer Hamiltonian
$H_{RK}$, but its most natural physical interpretation is in the coupled fermion-dimer Hamiltonian $H$ described in Section \ref{sec:FDM}.
In Section \ref{sec:continuum}, we show that this $Z_2$ invariant
is just the parity of an integer invariant with a simple topological
interpretation in the continuum.

In this paper, we will be discussing dimer models and fermion
hopping models on lattices, primarily hypercubic lattices $\mathbb{Z}^d$.
We will take the lattice constants to be equal to $1$ to avoid
clutter in the formulas which follow. Then, in $d$ dimensions,
we will use bold-faced vectors to denote points in the lattice,
${\bf r}\in \mathbb{Z}^d$. Sometimes, we will instead use latin indices
$i, j, k, \ldots$ to denote points in the lattice, assuming some arbitrary
ordering of the points. In three dimensions, which is the main
focus of this paper, we will use $\hat{\bf x}_i$, $i=1,2,3$
or $\hat{\bf x}, \hat{\bf y}, \hat{\bf z}$
to denote basis vectors of the lattice in the directions of the three Cartesian
axes.

\subsection{Hopfion}
\label{sec:Hopfion}

\begin{figure}[tb]
\centerline{
\includegraphics[scale=0.3]{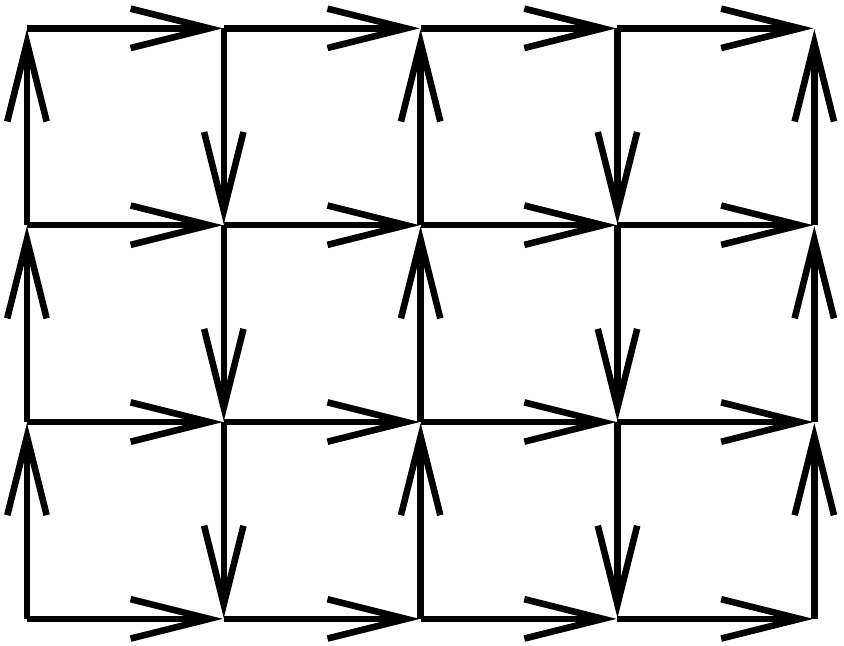}}
\caption{Illustration of signs in matrix $M$ in two-dimensions.  Arrows between sites show the signs.  An arrow pointing from
site $i$ to site $j$ implies that $M_{ij}=+i$ and $M_{ji}=-i$.  There is $\pi$-flux around each plaquette.}
\label{mfig}
\end{figure}

To define the $Z_2$ invariant, we introduce an anti-symmetric, Hermitian matrix $M$.
This matrix has matrix elements $M_{ij}$ with $M_{ij}=0$ if  $i$ and $j$ are not nearest neighbors, and $M_{ij}=\pm i$ otherwise.
The signs are chosen so that $M$ has $\pi$-flux around all plaquettes.  That is, if $i,j,k,l$ are sites around
a plaquette, then
\be
\label{piflux}
M_{ij} M_{jk} M_{kl} M_{li}=-1.
\ee
In Fig.~\ref{mfig}, we show an illustration of matrix $M$ for $d=2$ dimensions.
Such a matrix $M$ can be found for {\it any} planar lattice.  In order to find this $M$ for a planar lattice, add additional bonds to the lattice if necessary to triangulate the lattice; then, choose phases for the $M_{ij}$ to put $+\pi/2$ flux in each triangle, giving each square plaquette $\pi$ flux; finally, remove the added bonds.  On a planar lattice, the phases to do this can be chosen inductively, by choosing them on a connected sublattice which is also triangulated and increasing the size of that sublattice by adding one triangle at a time so that it remains triangulated.  Such an $M$ can also be found for some higher
dimensional lattices such as a three dimensional cubic lattice or hypercubic lattice in higher dimensions.  To find such a matrix $M$ on a hypercubic lattice, we proceed inductively: suppose we have such a matrix $M^{(d)}$ on a hypercubic lattice in $d$ dimensions.  Then, consider a $d+1$-dimensional hypercubic lattice as stacked hyperplanes, each such hyperplane containing a $d$-dimensional hypercubic lattice.  In each hyperplane, we use the matrix $M^{(d)}$, but we alternate the sign of this matrix from one hyperplane to the next.  Then, we orient the arrows connecting the hyperplanes so that they all point in the same direction; i.e., in three dimensions, we point all the vertical arrows in the up direction and in each plane we stack matrices as shown in Fig.~(\ref{mfig}) with alternating signs.

\begin{figure}[tb]
\centerline{
\includegraphics[width=3.25in]{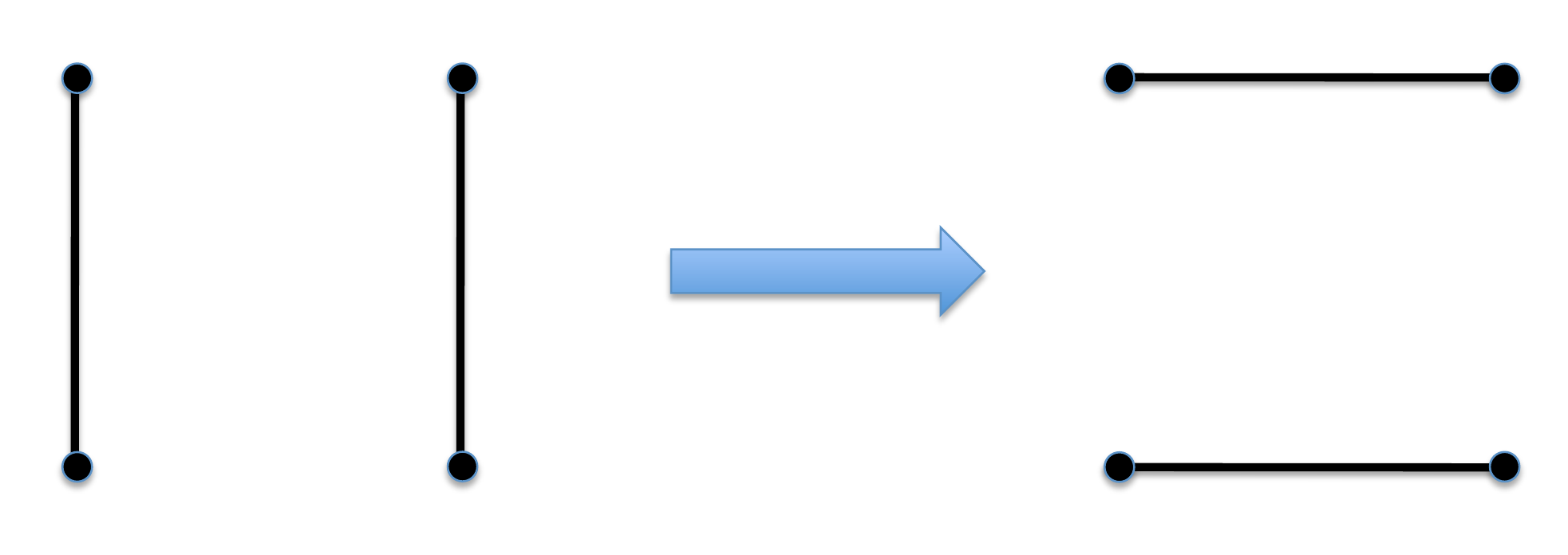}}
\caption{The basic plaquette flip move depicted above does not
change $n_H$.}
\label{fig:dimer-flip}
\end{figure}

Now consider any dimerization pattern on a lattice with no defects,
so that each site has exactly one dimer leaving it.
We specify this dimerization pattern by a set of numbers $n_{ij}$, where $i$ and $j$
label sites connected by a bond of the lattice as before and $n_{ij}=1$ if that bond is occupied by a dimer and $n_{ij}=0$ otherwise.  Then,
we define an index ${n_H}$, taking values in $\{0,1\}$:
\begin{equation}
{n_H} = [1-\text{sgn}(\text{Pf}(N))]/2
\end{equation}
where the matrix $N$ has entries given by
\be
N_{ij}=M_{ij} n_{ij}.
\ee
The matrix $n$ determines which entries of $N$ are non-zero,
and $M$ determines whether the non-zero entries are
$+i$ or $-i$. One can directly check that the sign of the Pfaffian
does not change under plaquette moves, precisely
because of the $\pi$-flux condition (\ref{piflux}) on the entries of matrix $M$.
In other words, if any plaquette containing two dimers is `flipped',
as depicted in Fig. \ref{fig:dimer-flip}, then the Pfaffian
of $N$ is unchanged. Therefore, ${n_H}$ is a $\mathbb{Z}_2$ invariant
of a dimer configuration.

\begin{figure}[tb]
\centerline{
\includegraphics[width=3.25in]{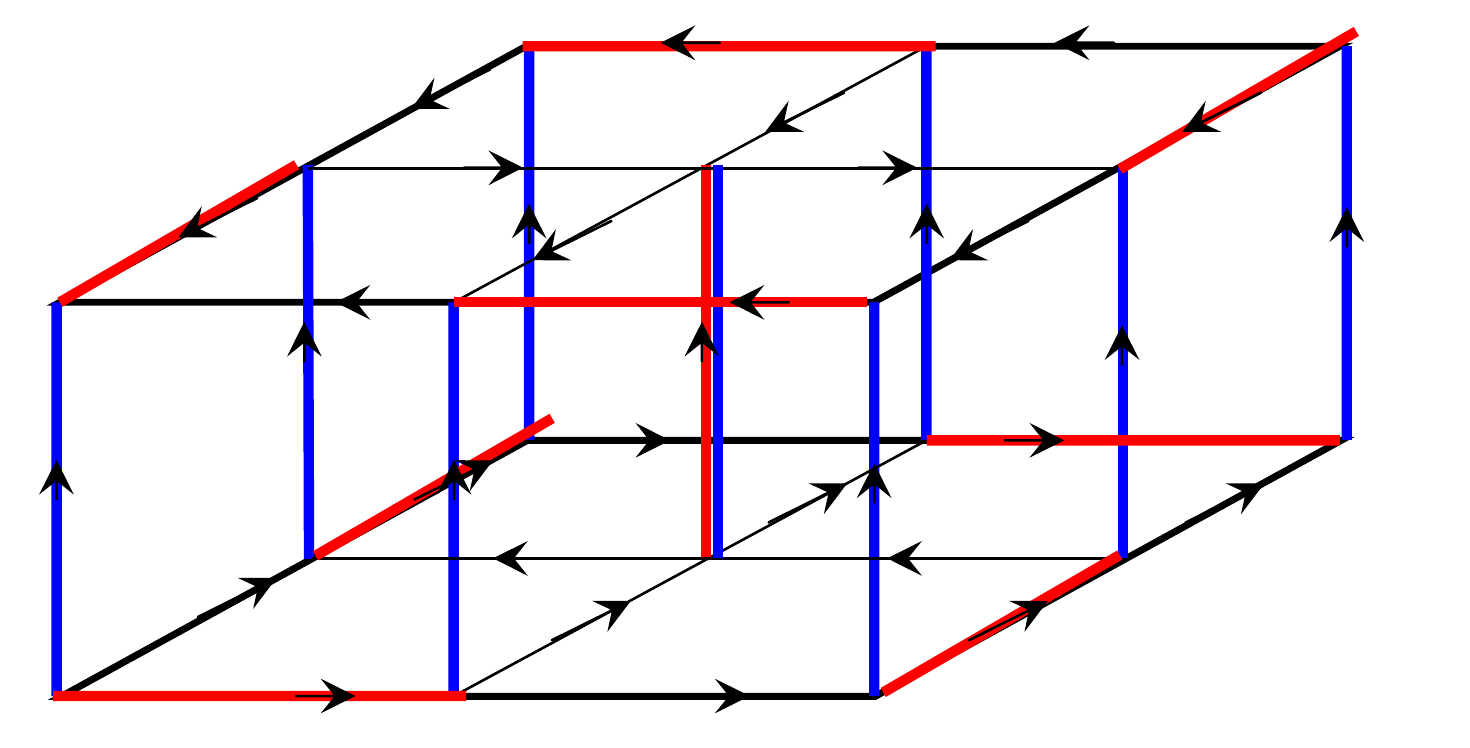}}
\caption{The red configuration of dimers is the
simplest dimer arrangement which has ${n_H}=1$.
We call such a configuration a ``Hopfion'', following
Ref. \onlinecite{Ran10}. The blue configuration
has ${n_H}=0$. The blue pattern is assumed to be repeated
outside of this 18 site section of the lattice. The arrows
are determined according to the the $\pi$-flux rule.}
\label{fig:latticeHopfion}
\end{figure}

As shown by Fortuin and Kasteleyn, the number of
dimer coverings of a planar lattice is equal to
$\text{Pf}(M)$: every single term in the Pfaffian contributes
a $+1$.  Stated in our language, Fortuin and Kasteleyn showed that $n_H=0$ for any dimer configuration on any planar lattice.
For a non-planar lattice, however,
the situation is not as simple, and we can have ${n_H}=1$.
In fact, a bilayer lattice suffices, as
the simplest configuration with ${n_H}=1$
is given in Fig.~\ref{fig:latticeHopfion}.
It has $18$ sites, arranged in a $3\times 3\times 2$ lattice.  We refer to configurations such as this as ``Hopfions", due to
their connection (explained later) with the Hopf map.
The existence of the Hopfion explains {\it why} it is hard to count dimer coverings of a three dimensional lattice: $\text{Pf}(M)$ counts the difference between the
number of configurations without a Hopfion and those with a Hopfion,
rather than their sum.

Note that while $n_H$ invariant is indeed invariant under plaquette moves,
it is {\it not} invariant under permutations
of the dimers around longer loops.  For example, in the pattern in the figure, permuting the dimers around the $8$ sites along the outside
of one of the squares in a given layer changes $h$.
Thus, one may wonder how relevant this invariant is
for the physics of the dimer model.  After all, in any experimental realization of a quantum dimer model, there will likely be some
amplitude for longer loop moves.  Also, in the original motivation for the dimer model as an approximation to the behavior of
spin-$1/2$ systems, there also is some amplitude for longer loops moves.
However, in the coupled dimer-fermion model which we will introduce
in the next section, we will
see that the invariance of ${n_H}$ is protected by superselection rules.

A crucial question is the energy of a Hopfion.  This is discussed in
Sections \ref{sec:FDM}, \ref{sec:dimer-eft} and whether
the Hopfion has a nonzero energy or not depends crucially
upon whether or not the dimers are in an ordered phase.

As a final remark, it is also possible to define $n_H$ on infinite lattices with boundary conditions that the dimers assume
a fixed columnar configuration at infinity.  This is necessary when comparing to topological results on the continuum model in the next two subsections.  We define the configuration on an infinite lattice in which all dimers are in a columnar configuration to have $n_H=0$.  Then, given any other dimer configuration which assumes the given columnar configuration at infinity, since the two configurations only differ on a finite set of sites we can compute $n_H$ by the relative sign of the Pfaffians on that finite set of sites.

\subsection{Relation between the Dimer Model and
the O(3) Non-Linear $\sigma$ model}
\label{ref:dimers-vectors}

In this subsection and the next, a {\it dimer cover}, or equivalently a {\it dimerization}, will mean one without defects.

\subsubsection{General Theory}

Let us assign a unit vector to each point ${\bf r}$ of the cubic lattice
according to the rule:
\begin{equation}
\label{eqn:dimers-vectors}
\vec{n}({\bf r}) = \sum_{i=1}^3
(-1)^{\hat{\bf x}_i \cdot{\bf r}}\, \hat{\bf x}_i \,
(n_{{\bf r},{\bf r}+\hat{\bf x}_i}-n_{{\bf r},{\bf r}-\hat{\bf x}_i})
\end{equation}
The vector points either in the direction of the dimer which touches
that site or the opposite direction, with the sign alternating from
site to site in the direction of the dimer. In this way, we can
represent a dimer configuration by a unit vector field which
points along one of the axes or, in other words, by a map
from the lattice to the octahedron
$\vec{n}_{\rm lat}~:~{\mathbb{Z}^3}~\rightarrow~${\em Octahedron}. More generally in $d$ dimensions, if the dimer lies parallel to the $i$-th direction, $1\leq i\leq d$, assign the vector $\pm \frac{\partial}{\partial x_i}$ at the two ends of the dimer according to the rule: $+$ if the $i$-th coordinate is even and $-$ if odd, giving a map $\mathbb{Z}^d\longrightarrow d$-{\em Octahedron}. Since the two assignments agree at both ends of the dimer, we extend the vector field $\pm \frac{\partial}{\partial x_i}$ to be constant on the dimer. To connect to the theory of topological defects, we would like to go further and define a canonical smooth extension $\vec{n}_{\rm cont}:\mathbb{R^d}\longrightarrow S^{d-1}$.

\begin{figure}[tb]
\centerline{
\includegraphics[width=3.7in]{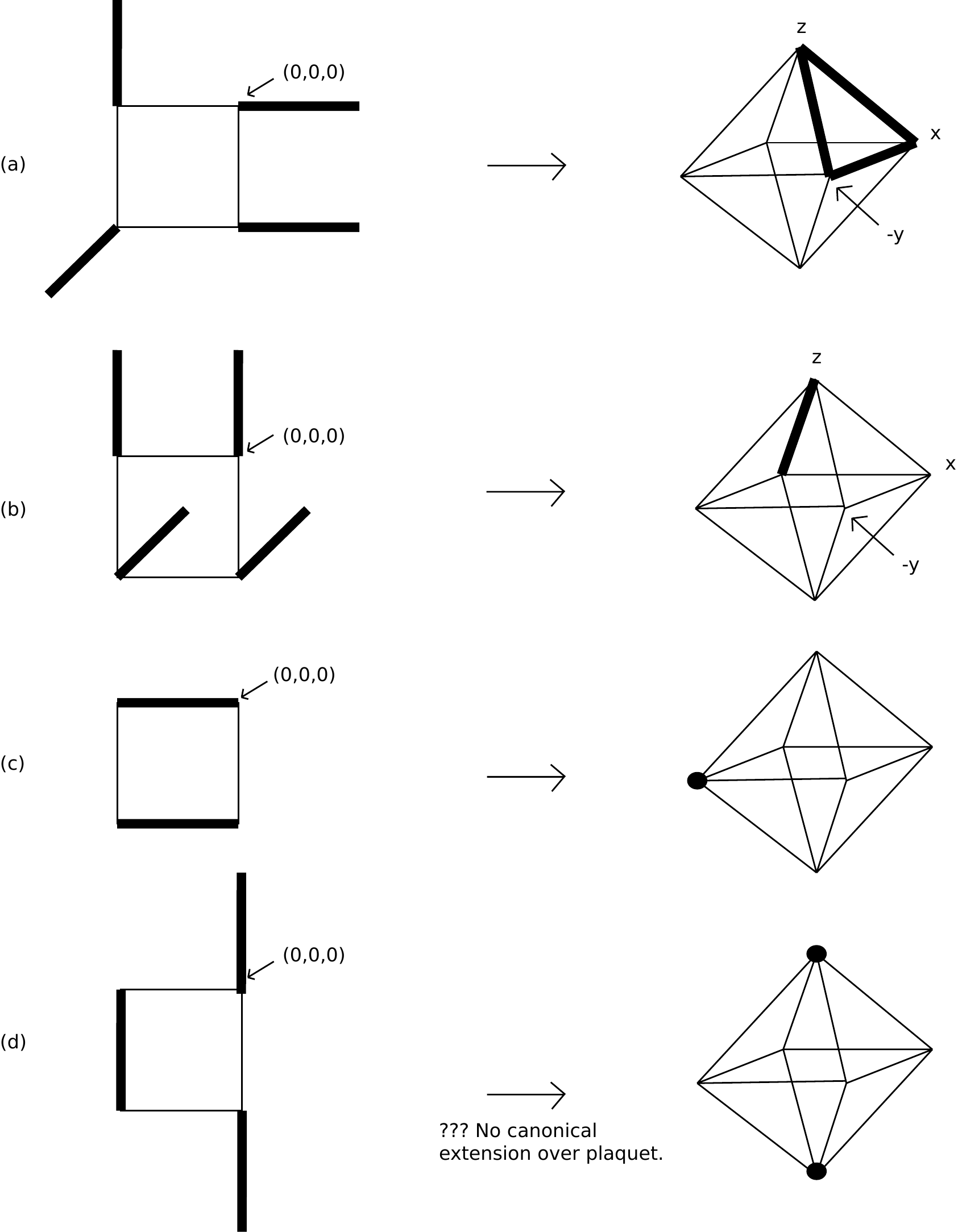}}
\caption{Examples of dimer configurations
on a plaquette and the corresponding
simplices of the octahedron.}
\label{fig:Figure5Bold}
\end{figure}

\begin{figure}[tb]
\centerline{
\includegraphics[width=3.25in]{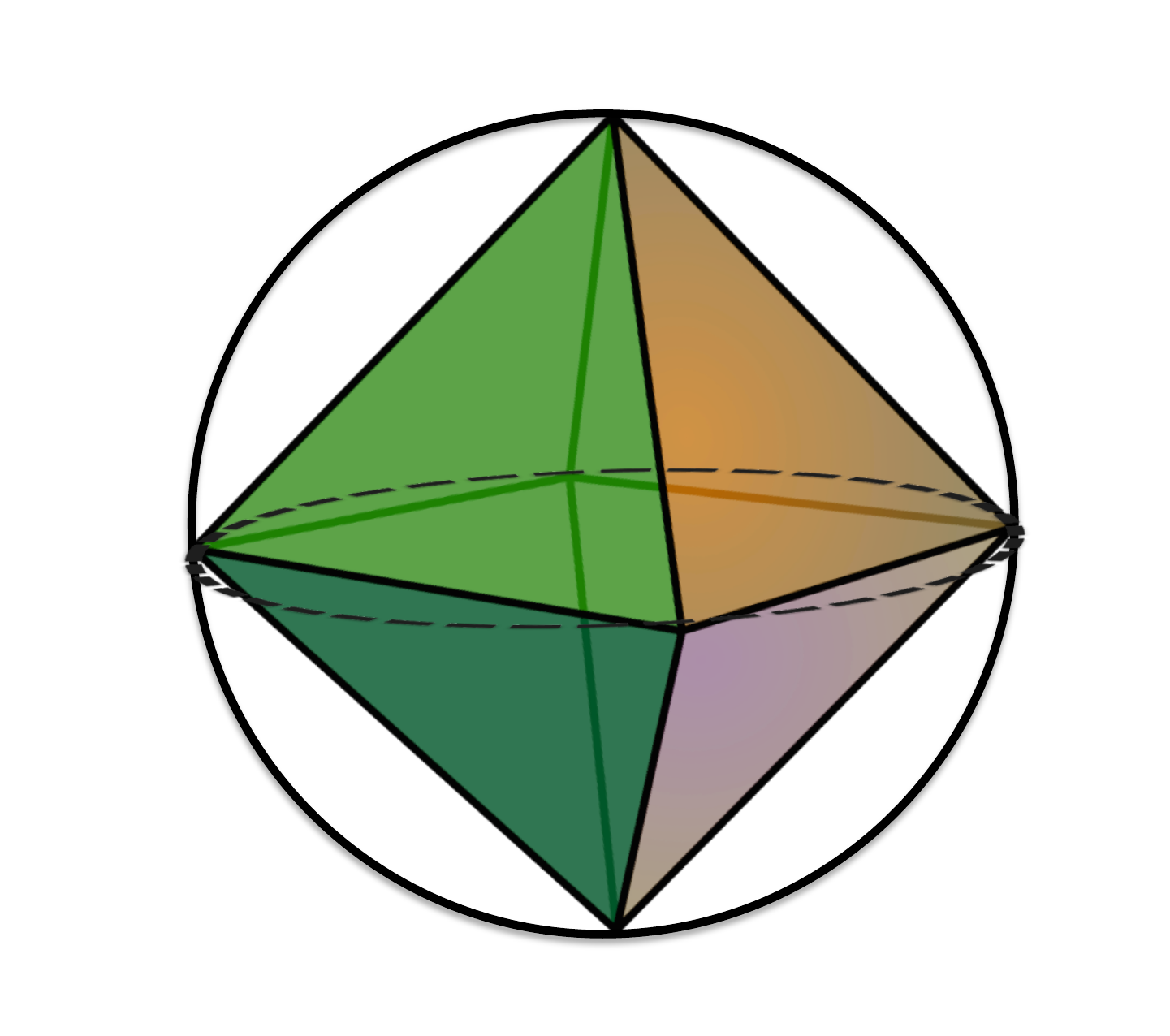}}
\caption{The octahedron is viewed as a discrete approximation
to the sphere.}
\label{fig:octahedron-sphere}
\end{figure}

The potential difficulties in constructing such an extension
are (1) that there may be multiple possible extensions locally
so that the map from $\vec{n}_{\rm lat}$ to $\vec{n}_{\rm cont}$
is a one to many map (i.e. the extension fails to be canonical) and (2) that some configurations $\vec{n}_{\rm lat}$
may necessarily have singularities in their extensions to $\vec{n}_{\rm cont}$,
i.e. points where we must have $\vec{n}_{\rm cont}=0$.
Rather than writing an obscure formula for $\vec{n}_{\rm cont}$, in the following paragraphs we discuss the issues encountered in constructing $\vec{n}_{\rm cont}$ and their resolutions. For now, we set $d=3$ and make the relevant extensions shortly. We view the octahedron as a discrete approximation for $S^2$.
Let us imagine inscribing an octahedron inside a sphere, as
shown in Figure \ref{fig:octahedron-sphere}, and then radially projecting it onto the sphere. 
Now consider any plaquette in the lattice. The four corners
of this plaquette are mapped to four (not necessarily distinct)
vertices of the octahedron.
So long as $\vec{n}_{\rm lat}$ at neighboring points does not point in antipodal
directions, call such a dimerization {\it cautious}. Moreover, we call a dimer covering {\it very cautious} if $x_i$ and $-x_i$ dimers never touch sites lying on a miniml lattice $d$-cube. A useful way to describe this is to say that the coordinate-wise $L^\infty$ distance
 of points $p$ and $q$ on the lattice
must be greater than one if $p$ and $q$ touch oppositely-directed dimers.
If the dimers are {\it very cautious}, then
the piecewise linear chain connecting the four corners of a plaquette unambiguously
bound a simplex (point, edge, or face) of the octahedron. Some examples are given in Figure \ref{fig:Figure5Bold}a,b,c.

We now assume our dimer covering $\delta$ is {\it very cautious} and, since the construction is general, we work in $d$-dimensions with $\mathbb{Z}^d$, $S^{d-1}$, the $d$-cube, and the dual $d$-octahedron, both of which are inscribed in $S^{d-1}$. We want to avoid all arbitrary choices so that our constructions will work in
parameter families of dimer coverings. The simplest construction seems to be a two step process. First, again using the $L^\infty$ distance,
``thicken'' each dimer into a rectangular solid.
Expand each of the rectangular solids in all directions
until $R^d$ is fully packed by rectangular solids. Let $f(\delta)$ be the step function (multi-valued at interfaces) which takes each rectangular solid to the vertex of the $d$-octahedron to which its core dimer has been previously assigned. Note that $\delta$ being very cautious implies the $f$-image of each fundamental $d$-cube of $\mathbb{Z}^d$ lies in a simplex of the $d$-octahedron. Second, smoothen out $f(\delta)$ by convolving it with a smooth function of very small support, $\varepsilon>0$, to produce $g$. The convolution uses the convex structure of each simplex of the $d$-octahedron.
This defines a unit vector field $\vec{n}_{\rm cont}$ on
$\mathbb{R}^3$ corresponding to any very cautious dimer configuration.

As an aside, we have checked that for $d=3$ it is sufficient merely to assume $\delta$ is cautious in order to construct a canonical extension, however the construction is somewhat different. In particular, apply the previous paragraph {\it only} to the plaquettes in $\mathbb{R}^3$ to get an extension with numerous cubical {\it holes}. The boundary of each hole has six faces, and a combinatorial argument shows the image of such a boundary under $f$ does not cover the octahedron ($6<8$) and in fact is a contractible subset $Z$ of the octahedron. Compressing the image of the boundary of a hole under $f$ radially away from $c=$ centroid $(S^2\setminus Z)$ to $\overline{c}=$ antipote$(c)$ defines a canonical extension over the hole.

The second step is a {\it refinement} trick which sends the general dimer cover to a very cautious dimer cover. We have already constructed $\vec{n}_{\rm cont}$ for very cautious dimer coverings and indeed a problem appears to arise when $\vec{n}$ points in antipodal directions
at neighboring points: there are many possible ways to
interpolate between $\vec{n}$ and $-\vec{n}$, and it is not clear which
one to choose. Such a situation occurs when the dimers are in a ``staggered'' configuration, such as depicted in
Figure \ref{fig:Figure5Bold} (d). Our solution is to imagine that there
there is a more refined lattice with $1/3$ the lattice spacing (in any dimension) of
the original lattice so that the physical lattice is a subset of the refined lattice.
For notational simplicity we temporarily revert to $3$-dimensions, although {\it refinement} clearly applies for all dimensions $d$. If the points of the refined lattice are indexed by three integers
$(m,n,p)$, then the points of the physical lattice are the points
$(3r,3s,3t)$, where $m,n,p,r,s,t \in \mathbb{Z}$.
We now make the following assignment of dimers
on the refined lattice. If the original lattice has a dimer between sites
$(3r,3s,3t)$ and $(3r+3,3s,3t)$, then the refined lattice
has a dimer between sites $(3r,3s,3t)$ and $(3r+1,3s,3t)$
and a dimer between sites $(3r+2,3s,3t)$ and $(3r+3,3s,3t)$.
The dimer on the original lattice has become two dimers
on the refined lattice. If there is no dimer between sites
$(3r,3s,3t)$ and $(3r+3,3s,3t)$ on the original lattice,
then the refined lattice has a dimer between
sites $(3r+1,3s,3t)$ and $(3r+2,3s,3t)$.
The analogous rule holds for links lying along the $\hat{\bf y}$
and $\hat{\bf z}$ directions.

\begin{figure}[tb]
\centerline{
\includegraphics[width=3.2in]{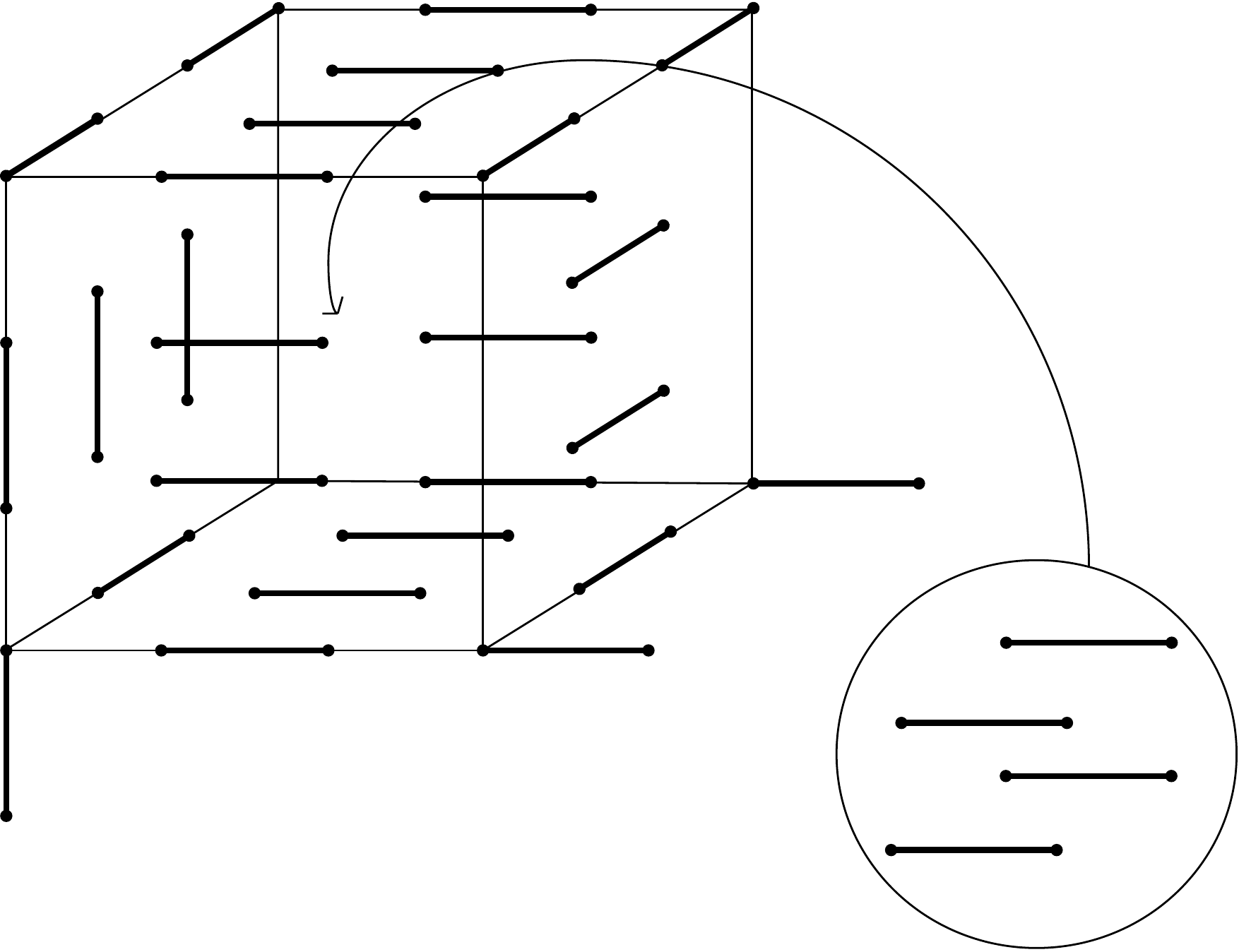}}
\caption{An example of the refinement of the dimer configuration
of a cube. A configuration which is dimerized in the $|x_1|$ direction
is inserted for the central cube.}
\label{fig:dimer-refinement}
\end{figure}

So far our refinement procedure takes a complete dimer covering into one that is incomplete. This can be rectified by dimerizing each central plaquette in any way which does {\it not} reintroduce an incautious pair, i.e. antipodal vectors on sites at dist$_{L^\infty}=1$. This is always possible since each plaquette of the unrefined lattice has at least two unfilled opposite sides which define the permissible directions for the dimers on the central plaquette in the refined lattice. For concreteness, we order the coordinates $|x|<|y|<|z|$ (or, more generally, $|x_1|<|x_2|<|x_3|<\dots<|x_d|$) and dimerize central plaquettes in the lowest possible direction. Next, central cubes (and then central $4$-cubes,$\dots$, $d$-cubes) can all be dimerized in the $|x_1|$ directions (see Figure \ref{fig:dimer-refinement}). This completes the construction of {\it dimer refinement}.

Notice the general dimer cover, after one step of refinement, becomes {\it very cautious}. Thus for general dimer coverings, the composition $\vec{n}_{\rm{cont}}\circ (\text{refinement})$ defines a mapping, call it $r_d$: \begin{align*}r_d:\{\text{dimer coverings of $\mathbb{Z}^d$}\}  \longrightarrow & \text{Maps}(\mathbb{R}^d\longrightarrow S^{d-1})\\
& :=\mathcal{M}(d).\end{align*}

\begin{figure}[b]
\centerline{
\includegraphics[width=3.2in]{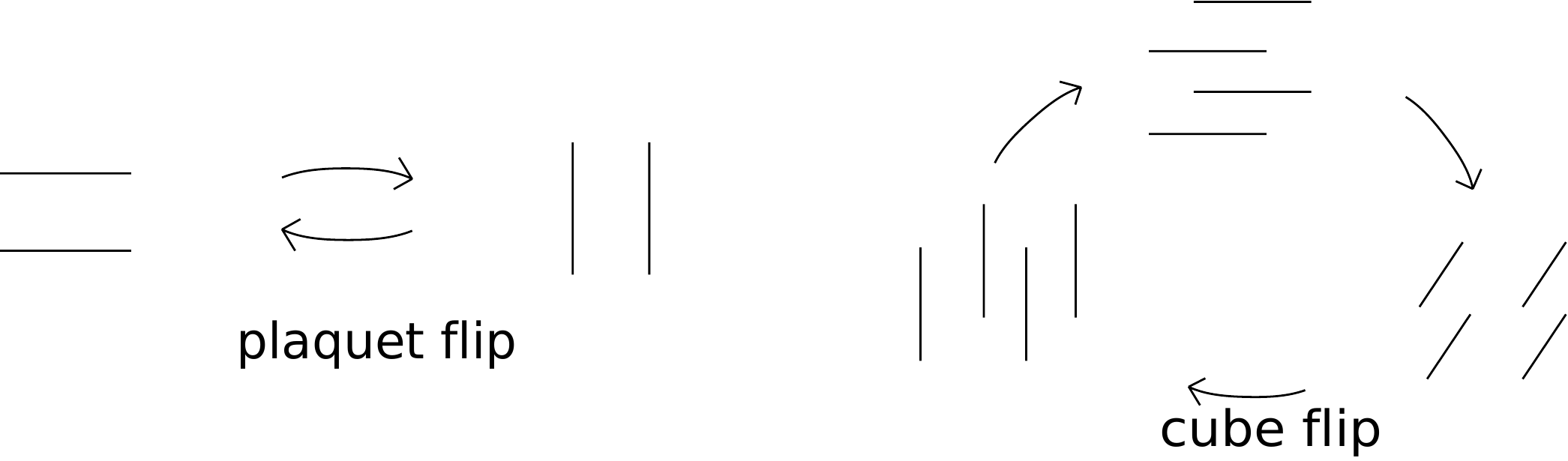}}
\caption{A plaquette flip is depicted on the left. A cyclic sequence of
cube flips is depicted on the right. Each arrow corresponds to
a single cube flip.}
\label{fig:plaquetandcubeflips}
\end{figure}

\begin{figure}[tb]
\centerline{
\includegraphics[width=3.5in]{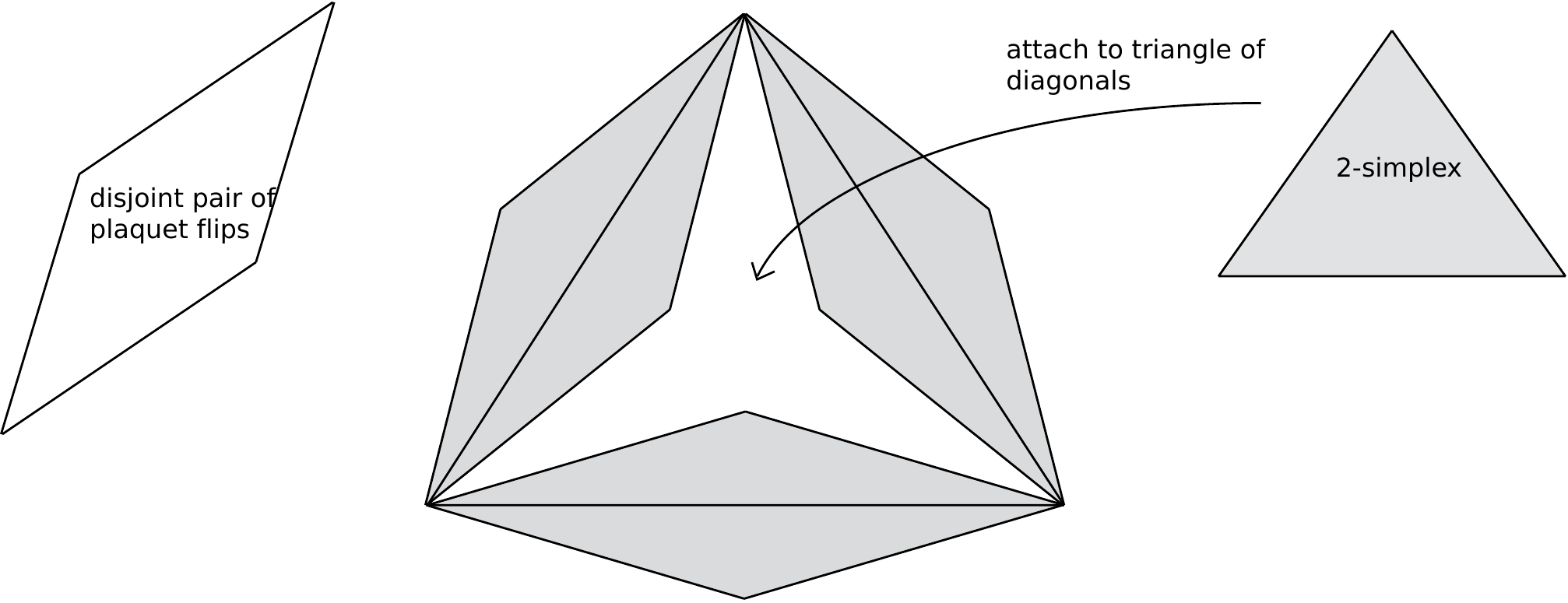}}
\caption{The $2$-simplex is associated to ``cube flip.''}
\label{fig:CubeFlipsCommute}
\end{figure}

The right hand side of the above mapping is a topological space with the $C^\infty$ compact-open topology, while the left hand side is a discrete set. To better compare the two, we would like to endow the set of dimer coverings of $\mathbb{Z}^d$
with a topology. It is customary to think of the set of dimer configurations
as a graph, with dimer configurations at the vertices and
links connecting dimer configurations which can be connected
by the plaquette flip depicted in Fig. \ref{fig:dimer-flip}.
However, we can go further and promote it to a
cell complex $\mathcal{D}_d$ by attaching $j$-cells, $j=1,2,3,\dots$,
to the discrete set of dimer configurations.
There will be primitive cells: $1$-simplices (the links of
the graph mentioned above), $2$-simplices,
$3$-simplices, $\dots$, $d-1$-simplices, corresponding to: plaquette flips, cube flips, $\dots$, $d$-cube flips, and further cells corresponding to arbitrary products of all such flips wherever these flips are realized disjointly.
A cube flip is composed of two plaquette flips, as depicted
in Fig. \ref{fig:plaquetandcubeflips}. A $2$-simplex is a
triangle which we associate to a set of three cube flips,
such as the set depicted in the diagram on the right of Fig. \ref{fig:plaquetandcubeflips}.
Note that the sides of this triangle are {\it not} attached to
three links in the graph. This is because
each arrow in the cube flip is a composition of two plaquette flips on an opposing pair of faces. Therefore, the corresponding $2$-simplex should, in order to
preserve symmetry, be attached to the {\it diagonal} of the square
corresponding to the product (i.e. disjoint occurrence of) those two plaquette flips
(i.e. to a link which is not present in the graph because it corresponds
to two plaquette flips), as shown in Figure \ref{fig:CubeFlipsCommute}.
Higher simplices are defined analogously.

As one moves across these primitive cells and their Cartesian products, starting with the map to $\mathcal{M}_d$ on the boundary, one inductively interpolates (within $\mathcal{M}_d$) the extension of $r_d$ across each cell. The result is a continuous map $$\overline{r}_d:\mathcal{D}_d\longrightarrow \mathcal{M}_d.$$

Now we restrict to dimer coverings $\mathcal{D}_d^0$ which are columnar in the $+x$-direction near infinity (i.e., $+x_1$-columnar except for finitely many dimers) and simultaneously to $\mathcal{M}_d^0$, the space of maps $\mathbb{R}^d\longrightarrow S^{d-1}$ which  takes a neighborhood of infinity to $(1,0,\dots,0)\in S^{d-1}$. This yields $$r_d^0:\mathcal{D}_d^0\longrightarrow \mathcal{M}_d^0.$$

If the O(3) non-linear $\sigma$ model is to capture the physics
of the cubic lattice quantum dimer model, then the topology of the
space of dimer coverings which are columnar in the $+x$ direction
near infinity must be the same as the topology of unit vector fields which
are equal to ${\bf \hat x}$ near infinity. A more precise way of expressing
this is that we need $r_3^0$ to be a homotopy equivalence.
More generally, one can ask whether
$r_d^0$ is a homotopy equivalence for all $d\geq 1$.

We do not know if this is true.
We can, however, prove two theorems, both of which are physically significant. \\

\noindent\textbf{Theorem 1}. {\it $r_2^0:\mathcal{D}_2^0\longrightarrow \mathcal{M}_2^0$ is a homotopy equivalence. In fact, both $D_2^0$ and $M_2^0$ are contractible.} \\

\noindent\textbf{Theorem 2}. {\it $r_d^0:\mathcal{D}_d^0\longrightarrow \mathcal{M}_d^0$ has a weak right homotopy inverse $s_d^0:\mathcal{M}_d^0\longrightarrow\mathcal{D}_d^0$, i.e. $r^0\circ s_d^0 \simeq \text{id}_{\mathcal{M}_d^0}$ for all $d\geq 1$. {\it Weak} means that technically mapping properties of the target space $\mathcal{M}_d^0$ are only tested over finite-dimensional complexes mapped into $\mathcal{M}_d^0$.} \\

\noindent\textbf{Corollary}. {\it For all $k\geq 0$ and $d\geq 1$, the map $(r_d^0)_k:\pi_k(\mathcal{D}_d^0)\longrightarrow \pi_k(\mathcal{M}_d^0)$ is onto. Thus, the dimer space has ``at least as much topology'' as the mapping space.}\\

\noindent{\it Proof}. Since $\pi_k$ is a homotopy functor, $(s_d^0\circ r_d^0)_k=\text{id}_{\pi_k(\mathcal{D}_d^0)}$, so $(r_d^0)_k$ must be an epimorphism. \qed\\

It seems possible that the $r_d^0$ are actually homotopy equivalences (which would imply $\pi_k(\mathcal{D}_d^0)\simeq \pi_k(\mathcal{M}_d^0)$ for all $k,d\geq 1$) but we could not prove this. However, a modest extension of the proof we will give for Theorem 2 shows that if we let $\hat{\mathcal{D}}_d^0$ be the direct limit of the refinement sequence $$\mathcal{D}_d^0 \xrightarrow{\text{refinement}}\mathcal{D}_d^0\xrightarrow{\text{refinement}}\mathcal{D}_d^0\xrightarrow{\text{refinement}}\dots,$$ then in fact $r_d^0$ extends to a (weak) homotopy equivalence: $$\hat{\mathcal{D}}_d^0\xrightarrow{\hat{r}_d^0}\mathcal{M}_d^0.$$

Theorem 1 will be proven in the following subsection on $d=2$ dimer coverings. For completeness, note the following trivial analog of Theorem 1 in $d=1$:
$\mathcal{D}_1^0$ and $M_1^0$ are each single points with $r_1^0$
the only possible map.

We remark that $\mathcal{M}_3^0$ is homotopy equivalent to $\text{Maps}(S^3,S^2)$ and $\pi_k(\text{Maps}(S^3,S^2))\cong \pi_{k+3}(S^2)$. The homotopy groups of $S^2$ are well studied and completely computed up to $\pi_{64}(S^2)$. This wealth of information translates directly via Theorem 2 to detect $k$-parameter families of dimer coverings of $\mathbb{Z}^3$
(columnar near $\infty$) modulo (simultaneous) plaquette flips for all $k$, such that $0<k\leq 61$. We hope there are creative ways to use this wealth of information in condensed matter.

Let us now sketch a proof of Theorem 2. For a dimer covering $\delta$ of $\mathbb{Z}^d$, the continuous map $m:\mathbb{R}^d\longrightarrow S^{d-1}$ is characterized (up to irrelevant choices from a contractible space) by the hypersurfaces - we call them {\it walls} - marking the frontiers in $\mathbb{R}^d$ between the ``colors'' $+x_1$, $-x_1$, $+x_2$, $-x_2$,$\dots$, $+x_d$, $-x_d$. These colored regions of $\mathbb{R}^d$ are the preimages of the $(d-1)$-dimensional faces of the $d$-cube dual to the vertices of the $d$-octahedron also regarded as projected to $S^{d-1}$.\\

\noindent\textbf{Definition}. Let $\mathcal{M}_d^{0,t}\subset \mathcal{M}_d^0$ be the subspace of maps corresponding to domain wall configurations so that points $p$ and $q$ with opposite colors $x_i$ and $-x_i$ (i.e. $m(p)$ and $m(q)$
lie on antipodal faces) must satisfy dist$_{L^\infty}(p,q)>t$.\\

\noindent\textbf{Lemma}. {\it For all $t$, the inclusion $\mathcal{M}_d^{0,t}\subset \mathcal{M}_d^0$ is a homotopy equivalence.}\\

\noindent{\it Proof}. Given $m:\mathbb{R}^d\longrightarrow S^{d-1}$, there is a radial expansion $t(m):\mathbb{R}^d\longrightarrow\mathbb{R}^d$, $t(m)(\rho,\vec{\theta})=(\bar{t}(m)\rho,\theta)$, for an appropriated monotonically increasing function $\bar{t}(m):\mathbb{R}^+\cup\{0\}\longrightarrow \mathbb{R}^+\cup\{0\},$ so that the compositions $m\circ t(m)\in \mathcal{M}_d^{0,t}$. Since $\bar{t}(m)$ can be chosen continuously in $m$, this defines the required (weak) homotopy inverse, $\tilde{t}:\mathcal{M}_d^0\longrightarrow\mathcal{M}_d^{0,t}$, $\tilde{t}(m)=m\circ t(m)$. \qed

The (weak) right homotopy inverse will be a composition: $s_d^0=s_d'\circ\tilde{t}$, for a map $s_d'$ which we construct next. As we define $s_d'$ the reader may notice a curious ambiguity which relates in a precise way to the cell structure on $\mathcal{D}_d^0$.

Each bond of $\mathbb{Z}^d$ has a {\it type} equal to $\pm x_i$, determined by the direction $(i)$ in which it lies and the sign: $+$ ($-$) if its smallest $i$ coordinate is even (odd). Each plaquette has a {\it type} consisting of the set containing the two distinct types of bonds in its boundary. Similarly, for all unit $k$-cubes of $\mathbb{Z}^d$, $1\leq k\leq d$, culminating with a set of $d$ colors representing the type of a $d$-cube.

Pick $t>2$ and $m'=\tilde{t}(m)\in\mathcal{M}_d^{0,t}$. Let us begin by defining an over-complete dimer covering $\bar{\delta}(m')$. For each closed unit lattice, $k$-cube $c$ of $\mathbb{Z}^d$ for $1\leq k\leq d$, we put dimers on all the bonds of $c$ precisely when the set of colors (i.e. $m'^{-1}(\text{faces of }S^{d-1})$) present in $c$ is identical to its type. Among such $k$-cubes, we call those which are maximal under inclusion {\it active}. To make $\bar{\delta}(m')$ merely complete, polarize each active $k$-cube in the direction of its centroid's color. There are two things to notice about this rule. First, the rule never produces conflicting instructions from two active $k$-cubes, $k=k_1,k_2$, for the reason that two distinct active $k$-cubes, $k=k_1,k_2$, never intersect. To see this, notice that if two $k$-cubes, $k=k_1,k_2$, intersect, the requirement that opposite colors have distance $>2$ implies that the union of the two color sets is also a consistent color set (i.e. no antipodal pairs of colors). Thus the span of the two $k$-cubes would itself have been a larger active $k$-cube, thereby contradicting maximality. The second thing to notice is that the rule is not continuous and does not always define a unique dimerization. The two are actually aspects of the same phenomenon: as a domain wall between colors sweeps across the centroid $\bar{c}$ of a $k$-cube, there will be instances where $\bar{c}$ acquires two or more colors. It is precisely the role of the primary cells of $\mathcal{D}_d^0$ to provide the target space to interpolate between the dimerizations defined by the potential polarizations of an active $k$-cube when $\bar{c}$ lies in multiple color domains. The product cells provide a similar target space for disjoint, simultaneous crossing of domain boundaries by two or more active centroids. Technically the above discussion only defines $s_d'$ on the strata of finite codimension. This is why we are only constructing a {\it weak} right homotopy inverse. This completes the description of $s_d':\mathcal{M}_d^{0,c}\longrightarrow\mathcal{D}_d^0$ for $t>2$.

To prove Theorem 2, it suffices to show that $r_d^0\circ s_d'\circ {\tilde t}=r_d^0\circ s_d^0\simeq\text{id}_{\mathcal{M}_d^0}$, i.e. that the composition is homotopic to the identity. It is sufficient to show for $m'\in\mathcal{D}_d^{0,t}$ that $r_d^0\circ s_d'(m'(\vec{r}))$ and $m'(\vec{r})$ are never antipodal, $\vec{r}\in\mathbb{R}^d$, or even oppositely colored in $S^{d-1}$, for then the two are joined by a canonical homotopy. But, consequent on our definitions of $r_d^0$ and $s_d'$, $r_d^0\circ s_d'(m'(\vec{r}))$ will have the color of $m'(\vec{r})$ for some $\vec{r}$ where dist$_{L^\infty}(\vec{r},\vec{r}')\leq 1$.

Since $t>2>1$, the two maps are never antipodal, completing the proof of Theorem 2.

In the direct limit or {\it stable} context ($\verb|^|$) the preceeding arguments localize and the opposite composition also becomes homotopic to the identity. For $p+q=d$, $p$ and $q$ non-negative integers, let $\mathcal{D}_{p,q}^{0}$ be the space of stable dimers on $\mathbb{Z}^d$ periodic in $q$-coordinates and $+x_i$-columnar near $\infty$ (unless $q=d$), $T^q$ the $q$-torus, and Maps$((T^q\times B^p,T^q\times S^{p-1});(S^{d-1},*))$ are denoted $M_{p,q}^0$.

\noindent\textbf{Theorem 3}. {\it There is a weak homotopy equivalence $\hat{r}_{p,q}:\hat{\mathcal{D}}_{p,q}^{0}\longrightarrow \mathcal{M}_{p,q}^0$.} \qed

\subsubsection{Special Knowledge When $d=2$}

Although our emphasis in this paper is on $d=3$, we have developed the general theory, and so will also explain how the Conway-Thurston\cite{Thurston90}
``height function'' allows a parallel approach for $d=2$, which can yield additional information.

Dimer coverings are, of course, the same as domino tilings.
Domino tilings may be understood as follows. For any group $G$
and set of generators $S$, we can define the Cayler graph which has a vertex
for each element of $G$ and a link between any two vertices
$g$, $g'$ if $g=g's$ for some $s\in S$ (with different colored links
corresponding to different elements in $S$). For instance,
the square lattice is the free group $F(x,y)$ on two letters $x,y$
modulo the relation $[x,y]=xyx^{-1}y^{-1}=e$ ($e$ is the identity element).
We can identify a square plaquette at the origin with the relation
$xyx^{-1}y^{-1}$ since it contains links corresponding to
$x$, $y$, $x^{-1}$, and $y^{-1}$. We can identify any other
square plaquette with a conjugate of the relation $g\,xyx^{-1}y^{-1}\,g^{-1}$:
from the origin, it goes first to vertex $g$, follows the links
corresponding to $x$, $y$, $x^{-1}$, and $y^{-1}$, and then returns
to the origin.
Now consider $F(x,y)$ with the two relations $r_1=[x^2,y]=e$, or $r_2=[y^2,x]=e$.
The quotient $G$ is a ``Conway group''.
The Cayley graph of $G$ is a three-dimensional graph.
Now consider a domino at the origin which lies in the $x$-direction.
We can identify it with ${x^2}yx^{-2}y^{-1}$ since it is twice
as long in the $x$-direction as the $y$ direction. Similarly,
a domino at the origin which lies in the $y$ direction can
be identified with the other relation ${y^2}xy^{-2}x^{-1}$.
Furthermore, any domino can be identified with a
conjugate $g\,{x^2}yx^{-2}y^{-1}\,g^{-1}$ or
$g\,{y^2}xy^{-2}x^{-1}\,g^{-1}$ of one of the relations.
In this way, any domino can be lifted into the
three-dimensional Cayley graph of $G$.
Any domino tiling lifts to a ``rough surface'' in the graph.

\begin{figure}[tb]
\centerline{
\includegraphics[width=3in]{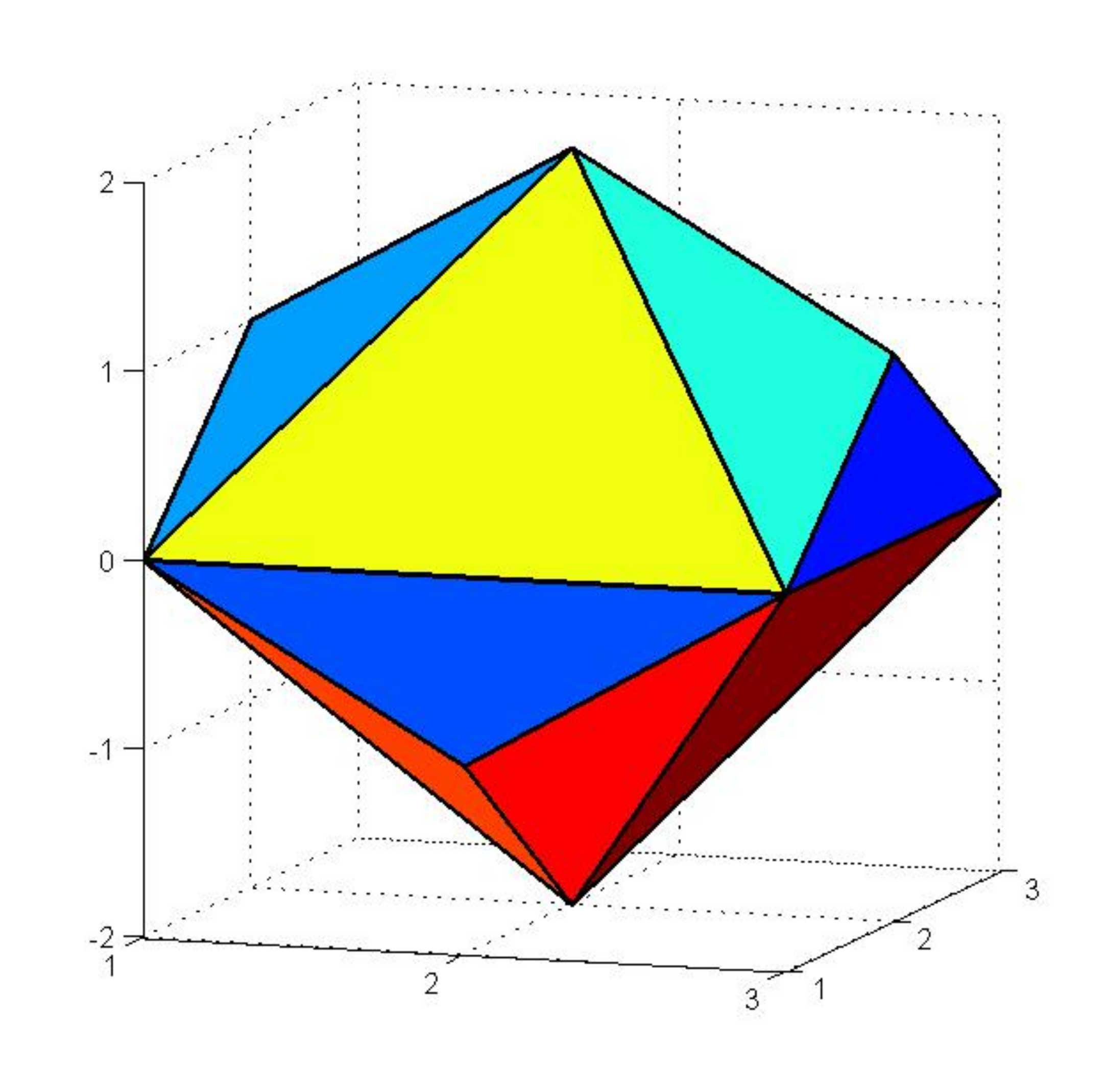}}
\caption{(Color online) A $3$-dimensional rendering of two possible height functions defining the top and bottom half of the {\it body}.}
\label{fig:3DHeightFunction}
\end{figure}

The group $G$ fits into the following (non-split) short exact sequence:
\[\begindc{0}[3]
    \obj(10,30)[a]{$1$}
    \obj(22,30)[b]{$\mathbf{Z}$}
    \obj(40,30)[c]{$G$}
    \obj(64,30)[d]{$\mathbf{Z}\oplus\mathbf{Z}$}
    \obj(79,30)[e]{$1$}
    \mor{a}{b}{}
    \mor{d}{e}{}
   \cmor((37,31)(31,33)(26,31)) \pleft(31,35){$h$}[\atleft,\dasharrow]
   \cmor((26,29)(31,27)(37,29)) \pright(51,25){}
    \cmor((57,31)(50,33)(44,31)) \pleft(51,35){$\ell_D$}[\atleft,\dasharrow]
   \cmor((44,29)(50,27)(57,29)) \pright(51,25){}
    \obj(22,25)[b1]{\begin{sideways}$\in$\end{sideways}}
    \obj(40,25)[c1]{\begin{sideways}$\cong$\end{sideways}}
    \obj(63,25)[c1]{\begin{sideways}$\cong$\end{sideways}}
    \obj(22,20)[b2]{$\scriptstyle{[x,y]}$}
    \obj(40,20)[d2]{$\scriptstyle{\{x,y | [{x^2},y],[{y^2},x]\}}$}
    \obj(63,20)[e2]{$\scriptstyle{\{x,y | [x,y]\}}$}
\enddc\]
This short exact sequence can be understood as follows.
The first group, $\mathbb{Z}$, gives the possible values of
the height function. The second group, $G$, enapsulates the domino tilings.
Each element of $G$ corresponds to a particular domino in a domino
tiling of the square lattice. The map $h$ gives the height of that domino.
The third group, $\mathbb{Z}\oplus\mathbb{Z}$ is the square lattice.
There is no natural map back from $\mathbb{Z}^2=\mathbb{Z}\oplus\mathbb{Z}$ back into $G$, however if the plane is tiled by horizontal and/or vertical dominoes in pattern $D$ (and the origin is marked), then this information defines a set-theoretic splitting $\ell_D$ (not a homomorphism) as above. The values of $\ell_D$ mod $4\mathbb{Z}$ are independent of $D$, but $\ell_D$ itself reflects the tiling. Similarly, there is a set-theoretic splitting $h$ (above) determined by the condition that $\gamma^{-1}[a,b]\gamma \longrightarrow \pm 4$, the sign depending on whether the total number of symbols $a$, $b$, $a^{-1}$, $b^{-1}$ in $\gamma$ is even or odd. The composition $h\circ\ell_D$ is the ``height'' function on $D$. It can be described as follows. Checkerboard color the $1\times 1$-squares of the plane. Starting at $e=(0,0)$, move (in any way) along the edges of the dominoes to a site $p\in\mathbb{Z}^2$. There are four possibilities as you travel along each length one lattice bond and for each one add a term $\pm 1$ according to the rules: traversing a bond with a white (black) square on your left add $+1$ ($-1$). The sum of those signs is the height at $p$, given $D$. The first observation is that a dimer flip $D\longrightarrow D'$ acts on the graph of the height function by laying on, or cutting away, a certain $3D$ ``body'' from the graph. In the case of dimers on the honeycomb, the body is a cube and these pictures are well-known. For the square lattice, the body is also convex and is pictured in Figure \ref{fig:3DHeightFunction}.

The second observation\cite{Thurston90} (see caption to Figure \ref{fig:CubeFlipsCommute}) is that with a fixed loop $\gamma$ in $\mathbb{Z}^2$ as boundary condition, interior dimer coverings (if they exist) correspond to a discrete version of the Lipschitz functions (Manhattan, or $L^1$-metric) with Lipschitz constant $\leq 1$. In the case of dimer covers fixed to $+x$-columnar near infinity this becomes: ``Lipschitz functions consntant near infinity with Lipschitz constant $\leq 1$.'' In Ref. \onlinecite{Thurston90} an algorithm for constructing the unique ``lowest'' dimer cover filling of a boundary condition is given (provided a dimer cover exists), and it is shown that every dimer cover is connected to the lowest one by a sequence of flips. In fact, the most direct (monotone) sequence of flips amounts to removing ``bodies'' until the graph is lowest. This process is canonical except for the occurrence of disjoint (hence commuting) moves where two or more bodies are simultaneously removable. In the language of this paper, there is a dimer space $\mathcal{D}_2^\gamma$ associated to filling $\gamma$ with dominoes that has vertices for each such filling, edges for flips, and $k$-cubes for $k$ simultaneously flips. The arguments of Ref. \onlinecite{Thurston90} actually show:\\

\noindent\textbf{Theorem A}. {\it For all $\gamma$ either $\mathcal{D}_2^\gamma$ is empty or contractible.} \qed\\

\begin{figure}[tb]
\centerline{
\includegraphics[width=2.5in]{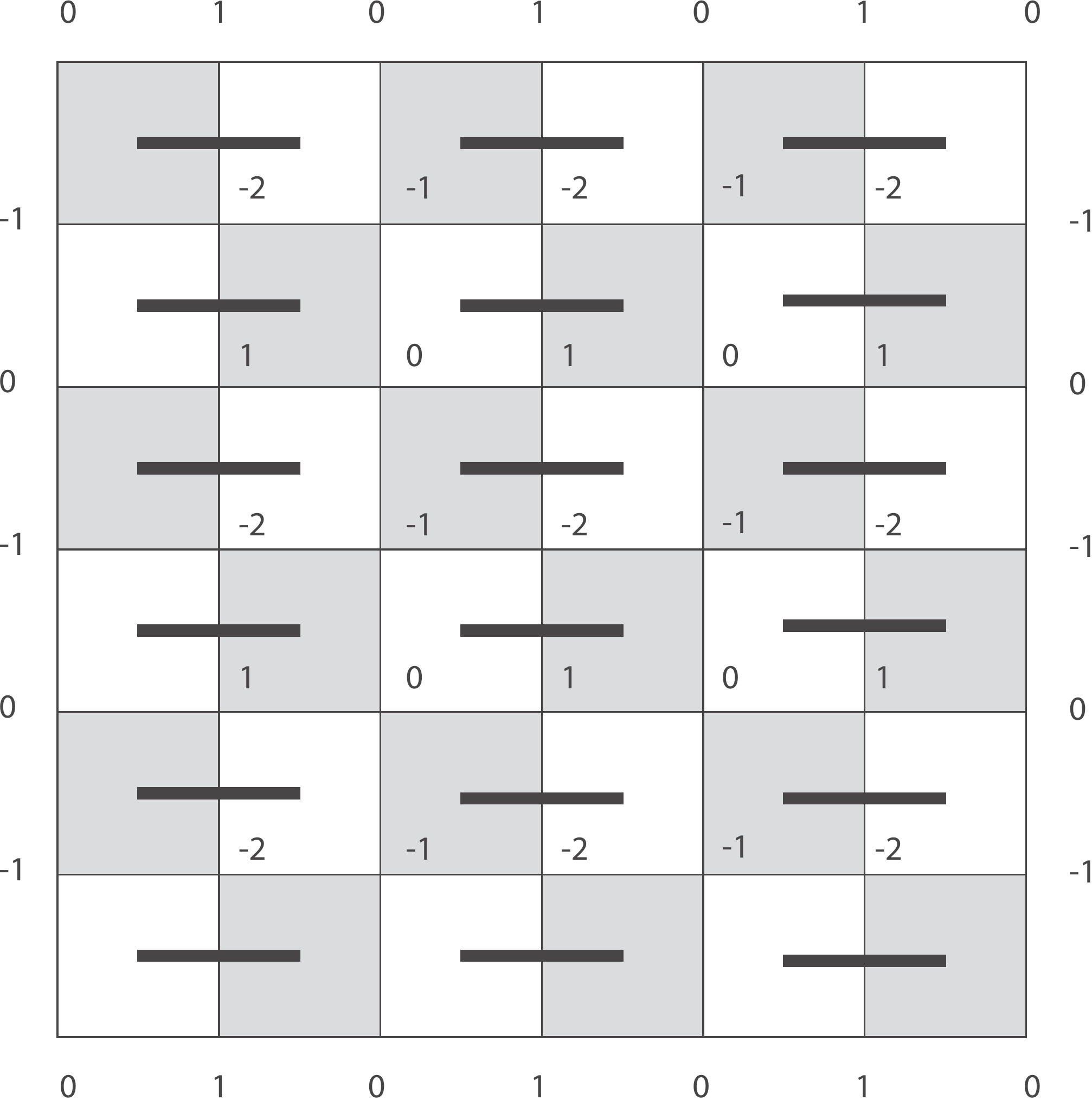}}
\caption{The height function of an $x_1$ columnar configuration of dimers.}
\label{fig:DimerizedCheckerboard}
\end{figure}

In the non-compact case, where the sharp boundary $\gamma$ is replaced with $x_1$-columnar near $\infty$, dimer covers correspond to $1$-Lipschitz functions approximately constant near infinity as seen from the height function in Figure \ref{fig:DimerizedCheckerboard}.

While there is now no ``lowest function,'' any compact family $K$ of dimer covers, each a finite alteration of $x_1$-columnar, can be canonically lowered to a lowest function constant outside a sufficiently large compact set $S\subset\mathbb{R}^2$, where $S$ depends on $K$. Thus we have:\\

\noindent\textbf{Theorem B}. {\it $\mathcal{D}_2^0 \cong *$, that is $\mathcal{D}_2^0$ is contractible.} \qed\\

Note that, in a coarse sense, the relation between the Conway-Thurston height function $h$ and our vector field $v:=r_d(\delta)$ is $v=(\sin 2\pi h/4,\cos 2\pi h/4)$.

Since the mapping space $M_2^0=\text{Maps}((S^2,*),(S^1,*))$ is also contractible, Theorem B implies Theorem 1 of the previous subsection. \qed

\subsubsection{Periodic Boundary Conditions}
\label{sec:periodic}

The height function method can be adapted to periodic boundary conditions (torus) and partially periodic boundary conditions (cylinder). Let $X$ be either a torus or cylinder. The new fact in this context is that now $h$ will be multi-valued along essential closed trajectories. $h$ is now a twisted function realizing a plaquette-flip invariant holonomy representation $\rho:\pi_1(X)\longrightarrow\mathbb{R}$. In other language $h$ is now a section of a trivial real line bundle equipped with a possibly non-trivial flat connection.

\begin{figure}[tb]
\centerline{
\includegraphics[width=3in]{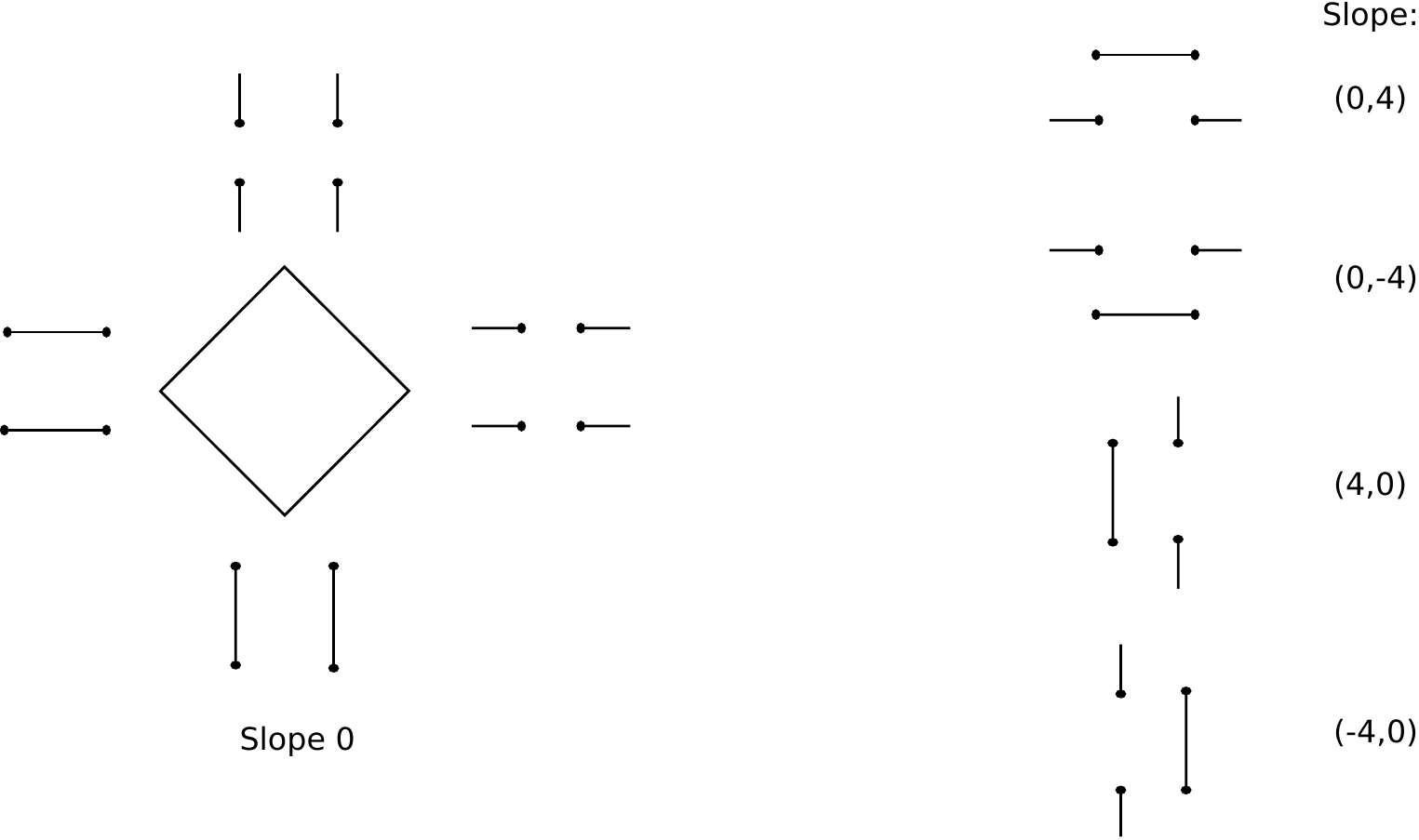}}
\caption{The octahedron is viewed as a discrete approximation
to the sphere.}
\label{fig:HarmonicRepresentations}
\end{figure}

Picking a base point vertex $x_0$ lying on the boundary of a domino, $x_0\in X$, there will be a unique lowest (relative to the connection $\mathcal{A}$) $h$ with $h(x_0)=0$. This requires checking that the lowest tiling in the neighborhood of $x_0$ is defined by the unique model shown in Figure 4.4 in Ref. \onlinecite{Thurston90}. If $X$ is non-compact, a fixed neighborhood of infinity can be specified to be $x_1$-columnar and then the notion of {\it lowest} is defined with respect to this restriction. Arguing similarly to the planar case with boundary condition $\gamma$, one may see that each realized representation $\rho$ corresponds to a connected component of the dimer space $\mathcal{D}(X)$ naturally associated to $X$. Figure \ref{fig:HarmonicRepresentations} computes $\mathcal{D}(X)$ for $X=$ $4$ points arranged as a $2\times 2$ square torus. There are eight dimer coverings organized into: one circle containing $4$ (with slope=$(0,0)$) and $4$ point components each with different non-trivial slopes.

We expect for general tori (and cylinders) $X$ the result will be analogous. Components of $\mathcal{D}(X)$ of maximal slope will be contractible while the other components will be homotopy circles (corresponding revolving the average angular direction of $\vec{n}_{\text{cont}}$ around the target circle). This will certainly be true in some coarse sense but may be true on the nose. A promising tool to study this question would be a {\it combinatorial Laplacian} $\Delta h$ of $h$. One would like to study dimer coverings by deforming to the {\it harmonic} representatives, but we leave this to later work.

\subsection{Continuum Description of Hopfions}
\label{sec:continuum}

In the previous subsection, we have argued that
dimer model configurations can be viewed as
configurations of a field whose target space is the octahedron.
If the octahedron, in turn, is approximated by the sphere,
then the dimer model can be viewed as a lattice regularization
of the O(3) non-linear $\sigma$ model.  We now make a natural relation between the invariant
$n_H$ and the so-called Hopf invariant of the O(3) model.

\begin{figure}[tb]
\centerline{
\includegraphics[height=3.25in]{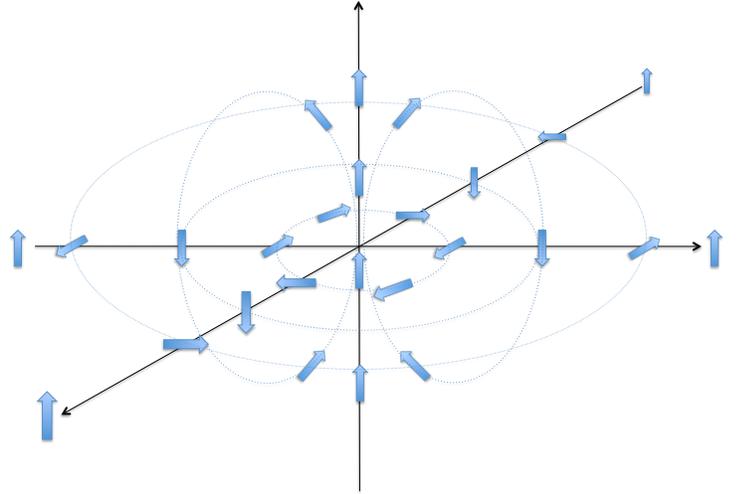}}
\caption{A unit vector field $\vec{n}$ with
${N_H}=1$.}
\label{fig:Hopf-continuum}
\end{figure}

Configurations of the O(3) non-linear $\sigma$ model are defined
by maps $\vec{n}(\vec{x})$ from three-dimensional Euclidean
space $\mathbb{R}^3$ to the sphere $S^2$. Let us consider maps
which are constant at infinity, e.g. let us consider field configurations
such that $\vec{n}(\vec{x}) \rightarrow \hat{z}$ as $|\vec{x}|\rightarrow\infty$ corresponding to the
boundary condition on dimers that they assume a fixed columnar configuration at infinity.
Then we can add a point at $\infty$ so that we have a map
from $S^3$ to $S^2$. The homotopy classes of such maps
are in one-to-one correspondence with the
integers, ${\pi_3}({S^2})=\mathbb{Z}$. The integer $N_{\rm Hopf}$
which indexes these homotopy classes is the linking number between
the pre-images of any two points on $S^2$ (the pre-images
are closed curves) which may be computed from the integral formula:
\begin{equation}
\label{NHdef}
{N_H} = -\frac{1}{4\pi} \int {d^3}x \,\epsilon_{\mu\nu\lambda}\,
a_\mu \,\vec{n} \cdot \partial_\nu \vec{n} \times \partial_\lambda \vec{n}
\end{equation}
where $a_\mu = \frac{i}{2}(\partial_\mu z^\dagger \, z - z^\dagger \partial_\nu z)$
if we write $\vec{n}$ in the form $\vec{n} = z^\dagger \vec{\sigma} z$.
Alternatively, we can define $a_\mu$ by
$$
\partial_\nu a_\mu - \partial_\nu a_\mu =
\frac{1}{4\pi} \vec{n} \cdot \partial_\nu \vec{n} \times \partial_\lambda \vec{n};
$$
the ambiguity in defining $a_\mu$ from this relation does not affect the integral.
A configuration of a unit vector field $\vec{n}$ with
${N_H}=1$ is depicted in Figure \ref{fig:Hopf-continuum}.

The resemblance between the configurations
in Figs. \ref{fig:latticeHopfion} and \ref{fig:Hopf-continuum},
and the fact that the configuration of Fig.~\ref{fig:latticeHopfion} maps to a vector field with $n_H=1$,
are not accidental:

\noindent\textbf{Theorem 4}.
\begin{equation}
\label{Hopfionisparity}
{n_H} \equiv {N_H} \text{mod}\, 2,
\end{equation}

\noindent\textbf{Proof}.
We have shown that different dimer configurations with different values
of $N_H$ cannot be connected by dimer moves consisting of plaquette flips.
We have not shown that $N_H$ is the {\it only} obstruction to connecting
two different dimer configuration. However, theorem 3, for $(p,q)=(d,0)$, says precisely this when ``dimer refinement" is added to ``plaquet flip" as an allowable move. In particular, $\pi_0\left(\hat{\mathcal{D}}_{3,0}^0\right)\simeq\mathbb{Z}$ is generated by the dimerization $\delta$ in Fig. \ref{fig:latticeHopfion}. For a general space $\pi_0$ is merely a set, but $\pi_0\left(\hat{\mathcal{D}}_{3,0}^0\right)$ is an abelian group. The ``sum" of two configurations is represented by taking two balls containing the nontrivial, {\it i.e.}, non-$x_1$-columnar part of each configuration respectively and shifting the two balls relative to each other by some large even vector $V=(2x_0,2y_0,2z_0)$ so that they become disjoint, and finally completing the dimer covering outside the two balls by $x_1$-columnar dimers. This definition mimics the abelian operation on $\pi_3(S^2)$: the requirement that $V$ is even preserves the parity rules used to go from dimerization to maps.

The map $n_H$ factors as follows:
\[\begindc{0}[3]
    \obj(80,26)[f]{(T1)}
    \obj(15,30)[a]{$n_H:$}
    \obj(22,30)[b]{$\mathcal{D}_3^0$}
    \obj(22,20)[c]{$\hat{\mathcal{D}}_{3,0}^0$}
    \obj(55,30)[d]{$\mathbb{Z}_2$}
    \obj(37,22)[e]{$\hat{n}_H$}
    \obj(22,25)[b1]{\begin{sideways}$\hookleftarrow$\end{sideways}}
    \mor{b}{d}{}
    \mor{c}{d}{}
\enddc\]

We have checked this by (laboriously) producing the sequence of plaquet flips carrying $\delta$ to its refinement $\delta'$. But the vector fields $n_{\rm cont}$ and $n'_{\rm cont}$ associated to any dimerization and its refinement are (canonically) homotopic, as they always lie within $90^\circ$ after scaling ($\times 1/3$) the refined lattice back to the original size. Thus $N_H$ also factors

\[\begindc{0}[3]
    \obj(80,26)[f]{(T2)}
    \obj(15,30)[a]{$N_H:$}
    \obj(22,30)[b]{$\mathcal{D}_3^0$}
    \obj(22,20)[c]{$\hat{\mathcal{D}}_{3,0}^0$}
    \obj(55,30)[d]{$\mathbb{Z}$}
    \obj(37,22)[e]{$\hat{N}_H$}
    \obj(22,25)[b1]{\begin{sideways}$\hookleftarrow$\end{sideways}}
    \mor{b}{d}{}
    \mor{c}{d}{}
\enddc\]

and by Theorem 3, $\hat{N}_H$ induces an isomorphism on $\pi_0$. So we have a diagram

\[\begindc{0}[3]
    \obj(80,26)[f]{(T3)}
    \obj(22,30)[a]{$\pi_0\left(\mathcal{D}_3^0\right)$}
    \obj(22,15)[b]{$\pi_0\left(\hat{\mathcal{D}}_{3,0}^0\right)$}
    \obj(55,30)[c]{$\mathbb{Z}$}
    \obj(55,15)[d]{$\mathbb{Z}_2$}
    \obj(40,27)[e]{$\hat{N}_H\simeq$}
    \obj(37,33)[g]{${N}_H$}
    \obj(37,12)[h]{$\hat{n}_H$}
    \obj(45,22)[j]{${n}_H$}
    \obj(18,24)[k]{${\rm inc}_0$}
    \mor{a}{b}{}
    \mor{b}{c}{}
    \mor{a}{c}{}
    \mor{a}{d}{}
    \mor{b}{d}{}
    \mor{c}{d}{}
\enddc\]

in which three triangles are known to commute: lower-left (T1), upper-left (T2), and lower-right (a consequence of $\hat{N}_H$ being an isomorphism and $\hat{n}_H$ an epimorphism). By abstract nonsense, the forth, upper-right triangle must also commute, which was to be proved.\qed

%
%

\subsection{Effective Field Theory for the 3D Dimer
Model on the Cubic Lattice}
\label{sec:dimer-eft}

Let us now suppose that the dimers are governed by the Rokhsar-Kivelson
Hamiltonian,
\begin{multline}
H_{RK}^f=V\sum_{\square} \Bigl( \Bigl| || \Bigl\rangle \Bigr\langle || \Bigr|
+ \Bigl| = \Bigl\rangle \Bigr\langle = \Bigr| \Bigr) \\ - t\sum_{\square} \Bigl( \Bigl| || \Bigl\rangle \Bigr\langle = \Bigr|
+ h.c. \Bigr).
\end{multline}
Then, it is known that for $V>t$ the system is in a staggered phase
with no flippable plaquettes. At the RK point, $t=V$, the ground state
can be found exactly: it is an equal amplitude superposition of
all possible dimer configurations. For ${V_c}<V<t$, the system is in a Coulomb
phase with power-law dimer-dimer correlations \cite{Huse03}.
For $V<V_c$, the system spontaneously breaks translational symmetry and
orders, depending on the value of $V$, in either a columnar,
plaquette, or cube phase.

We now reinterpret these results in light of our conclusion
that a configuration of the 3D cubic lattice dimer model can be represented by
a unit vector field $\vec{n}$, at least at a course-grained level.
We begin by writing the action for a continuum field theory description
of the Rokhsar-Kivelson on the cubic lattice in the form $S[\vec{n}]$.
In the low-energy, long-wavelength
limit, we can perform a gradient expansion for the action for this field:
\begin{equation}
S = \int {d^3}x\, d\tau \left[ \frac{1}{2g}(\partial_\mu \vec{n})^2
+ V(\vec{n})\right]
\label{eqn:NLSM}
\end{equation}
Here, the potential $V(\vec{n})$ is non-zero because
the dimers do not have O($3$) symmetry. Different possible
forms for the potential $V(\vec{n})$ will favor different possible
ordered states.
For instance, if we take
\begin{equation}
V(\vec{n}) = a(n_x^4 + n_y^4 +n_z^4) + b(n_x^4 + n_y^4 +n_z^4)^2
+ c n_x^2 n_y^2 n_z^2
\end{equation}
then there will be regimes of the couplings $a, b, c$ in which
each of the three phases columnar, plaquette, and cube occurs.
$V(\vec{n})$ is clearly highly-relevant when $\vec{n}$ orders since
it determines which ordered phase the system settles into.
However, it may be irrelevant at the critical point to the Coulomb phase
and it is certainly irrelevant in the Coulomb phase itself.

It will prove to be very useful to use
a different form for the action
which is equivalent to (\ref{eqn:NLSM}) in the ordered phase.
We will use the CP$^1$ representation for the order parameter:
\begin{multline}
S = \int {d^3}x\, d\tau \left[ \frac{1}{2g}\left|(i\partial_\mu + a_\mu){z_a}\right|^2 +
\lambda(z_a^* z_a -1) \right.\\
\left. + \, V(z^\dagger \vec{\sigma}z)\right]
\label{eqn:cubic-dimer-CP1}
\end{multline}
In this equation, $\vec{n}=z^\dagger \vec{\sigma}z$. The coupling
constant $g$ controls the strength of fluctuations. For $g$ large,
there are large fluctuations in the direction of
$\vec{n}=z^\dagger \vec{\sigma}z$; for $g$ small,
fluctuations are small.
The field $\lambda$ is a Lagrange multiplier
which enforces the constraint $z_a^* z_a = 1$.
The gauge field $a_\mu$ has no kinetic term; it eliminates
the phase degree of freedom of $z$, which does not
enter into $\vec{n}$. Hedgehogs in $\vec{n}$ are magnetic
monopoles of $a_\mu$:
\begin{equation}
\frac{1}{2\pi} \int_S m_\lambda \epsilon_{\mu\nu\lambda}
\epsilon_{ijk}
{n_i} \partial_\mu {n_j} \partial_\nu {n_k} =
\frac{1}{2\pi} \int_S m_\lambda \epsilon_{\mu\nu\lambda}
f_{\mu\nu}
\end{equation}
where $f_{\mu\nu} = \partial_\mu a_\nu - \partial_\nu a_\mu$,
$a_\mu = \frac{i}{2}(\partial_\mu z^\dagger \, z
- z^\dagger \partial_\nu z)$, and $m_\lambda$ is the unit normal
to the surface enclosing the monopole.

In the ordered phase, $\left\langle z_a \right\rangle \neq 0$.
The gauge field, $a_\mu$, is massive by the Anderson-Higgs
mechanism. Consequently, monopoles/hedgehogs are confined.
In the disordered phase, $\left\langle z_a \right\rangle = 0$.
The field $z_a$ can be integrated out, thereby generating a
Maxwell term for the gauge field $a_\mu$.
Thus, the disordered phase of this model is a Coulomb
phase with a gapless photon.

This is a slight refinement of the description of the
U(1) gauge field description of the cubic lattice
quantum dimer model\cite{Huse03} which defines
the magnetic field through:
\begin{equation}
{B_i}({\bf r}) = (-1)^{{\bf Q}\cdot{\bf r}}\,\left(n_{{\bf r},{\bf r}+{\bf \hat i}} -
\frac{1}{6}\right)
\end{equation}
where ${\bf Q}=(1/b,1/b,1/b)$, $b$ is the lattice constant,
and $i=x,y,z$. The Lagrangian for this field is, in the continuum limit,
\begin{equation}
{\cal L} = (\partial_t {\bf a})^2 - r (\nabla\times {\bf a})^2
- \kappa (\nabla\times\nabla\times {\bf a})^2
\label{eqn:cubic-dimer-lagrangian}
\end{equation}
where ${\bf B} \equiv \nabla\times {\bf a}$. At the Rokhsar-Kivelson
point, $t=V$, $r=0$. Otherwise, we can neglect the $\kappa$
term in the infrared limit. The Lagrangian (\ref{eqn:cubic-dimer-lagrangian})
is the low-energy limit of (\ref{eqn:cubic-dimer-CP1}) in $a_0=0$
gauge. The action (\ref{eqn:cubic-dimer-CP1})
also specifies the low-lying massive (in the Coulomb phase)
degrees of freedom and gives a picture for the gapped phases
in terms of the condensation of these degrees of freedom.
This description is also reminiscent (albeit in a different dimension)
of the 2+1-D non-compact CP$^1$ model of Motrunich and Vishwanath
\cite{Motrunich04}, which describes the paramagnetic phase of an O(3) $\sigma$-model
with hedgehogs suppressed. (See also Ref. \onlinecite{Motrunich05}.)

A Hopfion configuration can be written in terms of
the CP$^1$ representation as a configuration of
$z$ which is topologically-equivalent to:
\begin{equation}
\label{eqn:Hopfion-z}
z = \frac{1}{r^2 + \lambda^2}
\begin{pmatrix} 2\lambda r \sin\theta e^{i\phi} \cr
2\lambda r \cos\theta + i (r^2-\lambda^2) \end{pmatrix}
\end{equation}
In the ordered phase, there is a gradient energy $\int {d^3}x\,(\partial\vec{n})^2$
for such a configuration. Alternatively, we can describe this
energy as the Meissner energy $\int {d^3}x\,a_\mu a^\mu$
which results when $z$ condenses, with
$a_\mu = \frac{i}{2}(\partial_\mu z^\dagger \, z
- z^\dagger \partial_\nu z)$.
This energy scales as $b$ if we rescale a Hopfion
$\lambda \rightarrow b\lambda$; therefore, a Hopfion minimizes its energy
by becoming as small as possible. The minimal-energy Hopfion will be only
a few plaquettes in size -- essentially a fluctuating version
of the classical configuration depicted in Figure \ref{fig:latticeHopfion} --
and its energy will be determined by lattice-scale dynamics.

In the Coulomb phase, on the other hand, there is no
stiffness for the field $\vec{n}$. Instead, there is simply
the Maxwell energy
$$
U = \int {d^3}x\, (E^2 + B^2),
$$
which energy scales as $U \rightarrow U/b$ under a dilatation
$\lambda \rightarrow b\lambda$ of the Hopfion
(\ref{eqn:Hopfion-z}). Therefore, an infinite-size
Hopfion will have vanishing energy in the Coulomb phase of
the $3D$ cubic lattice dimer model.

\section{Fermion Dimer Model}
\label{sec:FDM}

In this section we define the fermion dimer model.  We define this model in
Sections \ref{sec:Model-Def1}, \ref{sec:Model-Def2} in a general fashion,
such that the definition can be applied to any arbitrary lattice for which
we can find a matrix $M$ obeying the $\pi$-flux condition of Eq.~(\ref{piflux}).
We then specialize to two dimensional and three dimensional examples in later subsections, showing how to realize deconfined Majorana
excitations in three dimensions.

\subsection{Free Fermions on the Cubic Lattice with $\pi$ flux}
\label{sec:free-fermions}

In Ref. \onlinecite{Freedman11}, we introduced a model of
fermions hopping on a hypercubic lattice with $\pi$ flux through
each plaquette. Specializing to the case of three dimensions,
there is a single Majorana fermion operator
$\gamma_{\bf r}$ at each site ${\bf r}\in \mathbb{Z}^3$ of the lattice.
We will have a unit cell with 8 sites, so this can be
viewed as a model of spin-$1/2$ electrons with a
two-fold orbital degeneracy per unit cell (and charge non-conserving
terms, such as would be present in a superconductor).
The Hamiltonian takes the following form
if the hopping has uniform magnitude
(with signs determined by the $\pi$-flux rule):
\begin{equation}
H = \sum_{{\bf r}\in\mathbb{Z}^3} \sum_{i=1}^3
i\, t_{{\bf r},{\bf r}+\hat{\bf x}_i} \gamma_{\bf r}
\gamma_{{\bf r}+\hat{\bf x}_i}
\end{equation}
where $t_{{\bf r},{\bf r}+\hat{\bf x}_1}= {t_x}
(-1)^{{\bf r}\cdot(\hat{\bf x}_2 + \hat{\bf x}_3)}$,
$t_{{\bf r},{\bf r}+\hat{\bf x}_2}= {t_y}(-1)^{{\bf r}\cdot(\hat{\bf x}_3)}$, and
$t_{{\bf r},{\bf r}+\hat{\bf x}_3}={t_z}$. Here, $\gamma_{\bf r}$ is a real
fermionic operator, $\gamma_{\bf r}=\gamma_{\bf r}^\dagger$.
This Hamiltonian is translationally-invariant
since the flux, which is gauge-invariant, is the same through each plaquette.
However, it is convenient to take an 8-site unit cell on the cubic lattice
so that the lattice constant is now $2$. Then, we can group the $8$
Majorana fermion operators in each unit cell, $\gamma_{{\bf r}+n\hat{\bf x}_i}$,
with ${\bf r}\in(2\mathbb{Z})^3$, $n=0,1$, and $i=1,2,3$, into a single
$8$-component spinor:
\begin{equation}
\chi_{\alpha\beta\gamma}({\bf r}) \equiv  \gamma_{{\bf r}+\alpha\hat{\bf x}_1+
\beta\hat{\bf x}_2+\gamma\hat{\bf x}_3}
\label{eqn:chi-def}
\end{equation}
with $\alpha,\beta,\gamma = 0,1$.
In the continuum limit, the Hamiltonian takes the form
\begin{equation}
H = \chi \,\alpha_j i\partial_j \chi
\label{eqn:chi-Ham}
\end{equation}
where
\begin{eqnarray}
\label{eqn:alpha-def}
\alpha_1 &=& \sigma_x \otimes \sigma_z \otimes \sigma_z \, , \cr
\alpha_2 &=& I \otimes \sigma_x \otimes \sigma_z \, , \cr
\alpha_3 &=& I \otimes I \otimes \sigma_x \, .
\end{eqnarray}
Here, we have specialized to the case $t_x = t_y = t_z$,
but taking anisotropic hoppings will simply make
the velocity anistropic in the continuum limit, so long as
the hopping strengths remain non-zero.
In Eq. \ref{eqn:chi-def}, the first Pauli matrix acts on the index $\alpha$ in
Eq. \ref{eqn:chi-def}, the second Pauli matrix acts on $\beta$,
and the third on $\gamma$.

If we define
\begin{eqnarray}
\lambda^1_{\mu\nu} &\equiv&
\frac{1}{\sqrt{2}}\left(\chi_{0\mu\nu} + \chi_{1\mu\nu}\right)\cr
\lambda^2_{\mu\nu} &\equiv&
\frac{1}{\sqrt{2}}\,(\sigma_x)_{\mu\alpha}\left(\chi_{0\alpha\nu} - \chi_{1\alpha\nu}\right)
\label{eqn:lambda-def}
\end{eqnarray}
then Eq. \ref{eqn:chi-Ham} can be re-written in the form:
\begin{equation}
H = \lambda^1 \,{\tilde \alpha}_j i\partial_j \lambda^1 +
\lambda^2 \,{\tilde \alpha}_j i\partial_j \lambda^2
\label{eqn:chi-Ham-Majorana}
\end{equation}
where ${\tilde \alpha}_1 =  \sigma_z \otimes \sigma_z$,
${\tilde \alpha}_2 = \sigma_x \otimes \sigma_z$,
${\tilde \alpha}_3 = I \otimes \sigma_x$.
From Eqs. \ref{eqn:chi-def}, \ref{eqn:lambda-def} and the reality of $\gamma_{\bf r}$,
we see that $\lambda^i = (\lambda^i)^*$. Therefore,
(\ref{eqn:chi-Ham-Majorana}) is the Hamiltonian
of two 4-component Majorana spinors.
In Eq. \ref{eqn:chi-Ham}, they have been
been put together into the 8-component field $\chi$
defined in Eq. \ref{eqn:chi-def}.

Equivalently, $\lambda^{1,2}$ form a single 4-component Dirac spinor:
\begin{equation}
\label{eqn:psi-def}
\psi \equiv \frac{1}{\sqrt{2}}( \lambda^1 + i \lambda^2 )
\end{equation}
with Dirac Hamiltonian
\begin{equation}
H = \psi^\dagger \,{\tilde \alpha}_j i\partial_j \psi
\label{eqn:chi-Ham-Dirac}
\end{equation}
The reason that we have introduced the 8-component field $\chi$
defined in Eq. \ref{eqn:chi-def} and the Hamiltonian (\ref{eqn:chi-Ham})
is that we will soon be considering mass terms (and, eventually,
interactions with other fields) which are not invariant under
the U(1) symmetry respected by Eq. \ref{eqn:chi-Ham-Dirac}.

Now suppose that the hopping strengths do not have uniform magnitude.
If they are staggered in the $z$-direction, so that
$t_{{\bf r},{\bf r}+\hat{\bf x}_3}=1+ (-1)^{{\bf r}\cdot\hat{\bf x}_3} \,m$
while $t_{{\bf r},{\bf r}+\hat{\bf x}_3}=1$ and
$t_{{\bf r},{\bf r}+\hat{\bf x}_3}=1$ are unchanged,
then a mass gap opens. The Hamiltonian can be written in the form:
\begin{equation}
\label{eqn:beta-mass-terms}
H = \chi \,\alpha_j i\partial_j \chi + m \chi \beta_3 \chi
\end{equation}
where $\beta_3 = iI\otimes I\otimes\sigma_y$.
If, instead, we had staggered the hopping strengths
in the $x$-direction or $y$-direction,
we would have added a mass term with
$\beta_1=i\sigma_y \otimes\sigma_z \otimes\sigma_z$
or $\beta_2=iI \otimes\sigma_y \otimes\sigma_z$, respectively.

Suppose now that we have staggered the hopping strength
in the $z$-direction by an amount $m$, as above.
We decrease the hopping strength in the $z$-direction until
we reach ${t_z}=m$, at which point
the hopping strengths in the $z$-direction
alternate between $2m$ and $0$. This does not close the gap.
We can now reduce ${t_x}, {t_y}$, without closing the gap,
until we reach ${t_x}={t_y}=0$. Then, the system breaks up
into an array of $2$-site systems, with a $2$-level system
$i \gamma_{\bf r} \gamma_{{\bf r}+\hat{\bf x}_3} = \pm 1$ for
${\bf r}\cdot\hat{\bf x}_3 \in 2\mathbb{Z}$ on each
such $2$-site systems (or for ${\bf r}\cdot\hat{\bf x}_3 \in 2\mathbb{Z}+1$
if we had taken ${t_z}=-m$).
We could similarly stagger the hopping strength in the
$x$-direction and tune to the point ${t_x}=m, {t_y}={t_z}=0$
or stagger the hopping strength in the $y$-direction
and tune to ${t_y}=m, {t_x}={t_z}=0$.
In these extreme limits, the links with non-zero hopping strength
form a dimer configuration. The goal of this paper is to analyze
what happens when these dimers become dynamical.

\subsection{Non-Dynamical Dimers}
\label{sec:Model-Def1}

The Hilbert space of the system is the tensor product of a Majorana fermion Hilbert space and a dimer Hilbert space.
The Majorana fermion Hilbert space has one Majorana mode per site of the lattice.  The dimer Hilbert space has one two-state
system on each bond.  On each bond of the system, a dimer can be present or absent.  Thus, the Hilbert space includes states in which a given site has multiple dimers or zero dimers; however, we will add a penalty term to the Hamiltonian to penalize these terms.
States with distinct dimerization patterns are orthogonal.
The Hamiltonian we study is
\be
H=H_{\rm penalty}+H_{ff}+H_{\rm RK}^f,
\ee
where $H_{\rm penalty}$ is a term penalizing any state
which does not have one dimer per site.  We choose
\be
H_{\rm penalty}=U_0 \sum_i W_i,
\ee
where $U_0>>1$; $i$ ranges over the sites of the lattice; and
\be
W_i=\Bigl(\sum_{<ij>} \hat n_{ij}-1 \Bigr)^2,
\ee
Here, the sum over $j$ ranges over sites which neighbor site $i$,
and $\hat n_{ij}$ is $+1$ if there is a dimer on the bond
connecting site $i$ to site $j$, and $0$ otherwise.
Thus, $W_i$ is diagonal in the dimer basis described
above and equals $0$ if there is one dimer touching site $i$ and is positive
otherwise.

The term $H_{ff}$ couples the fermions to the dimers.
In the previous section, we defined a matrix $M$ obeying the $\pi$-flux
condition of Eq.~(\ref{piflux}).
We use this matrix to define
\be
H_{ff}=\sum_{ij} M_{ij} \hat n_{ij} \gamma_i \gamma_j,
\ee
where $\gamma_i$ is the Majorana fermion operator on site $i$.

Before defining the term $H_{\rm RK}^f$,
we pause to describe the physics of the Hamiltonian
\begin{equation}
H_0 \equiv H_{\rm penalty}+H_{ff} .
\label{eqn:H_0}
\end{equation}
This Hamiltonian is diagonal in the dimer basis, so eigenstates of $H_0$ can be chosen to be eigenstates of the dimer
number operators $\hat n_{ij}$.  For any given dimerization pattern (i.e., for any choice of whether each bond has a dimer or not),
the eigenstates of $H_0$ are
\be
|\psi_d \rangle \otimes |{\psi_f}(d) \rangle,
\ee
where $\psi_d$ is a vector in the dimer Hilbert space
with the given dimerization pattern $d$, and ${\psi_f}(d)$ is
an eigenvector of the free fermi Hamiltonian
\be
H_{ff}(d)\equiv \sum_{ij} M_{ij} n_{ij}(d)\, \gamma_i \gamma_j.
\ee
where the numbers $n_{ij}(d)$ are equal to $0$ or $1$ and are eigenvalues of the corresponding
operator $\hat n_{ij}$ so that
\be
\hat n_{ij} |\psi_d \rangle = n_{ij}(d) |\psi_d \rangle.
\ee

For sufficiently large $U_0$, the ground states of $H_0$ have dimerization patterns with no defects, so that each
site has exactly one dimer touching it.  To see this, note that for any dimerization pattern $d$, the smallest eigenvalue of
$H_{ff}(d)$ is greater than or equal to $-\sum_{ij} n_{ij}(d)$,
i.e. $-1$ times the number of dimers in the given
dimerization pattern.
If the pattern has no defects, the smallest eigenvalue is exactly equal to $-1$ times the number of defects, and
the corresponding eigenvector is:
\begin{equation}
|\psi^0_f(d)\rangle
\end{equation}
which is defined by the condition
\begin{equation}
\label{eqn:psi_f^0-def}
M_{ij} n_{ij}(d)\, \gamma_{i} \gamma_{j}|\psi^0_f(d)\rangle =
- |\psi^0_f(d)\rangle
\end{equation}
for all $i,j$ with $n_{ij}=1$.
In a dimerization pattern with defects
at which more than one dimer touches a site,
the smallest eigenvalue of $H_{ff}(d)$ is strictly greater
than $-1$ times the number of dimers.
Thus, while a dimerization pattern with defects can reduce the lowest
eigenvalue of $H_{ff}(d)$ below $-N/2$, for sufficiently large $U_0$ the
penalty term for such patterns is larger than the reduction in fermionic energy and so
for sufficiently large $U_0$,
the ground states of $H_0$ indeed have no defects.

\begin{figure}[tb]
\centerline{
\includegraphics[scale=0.3]{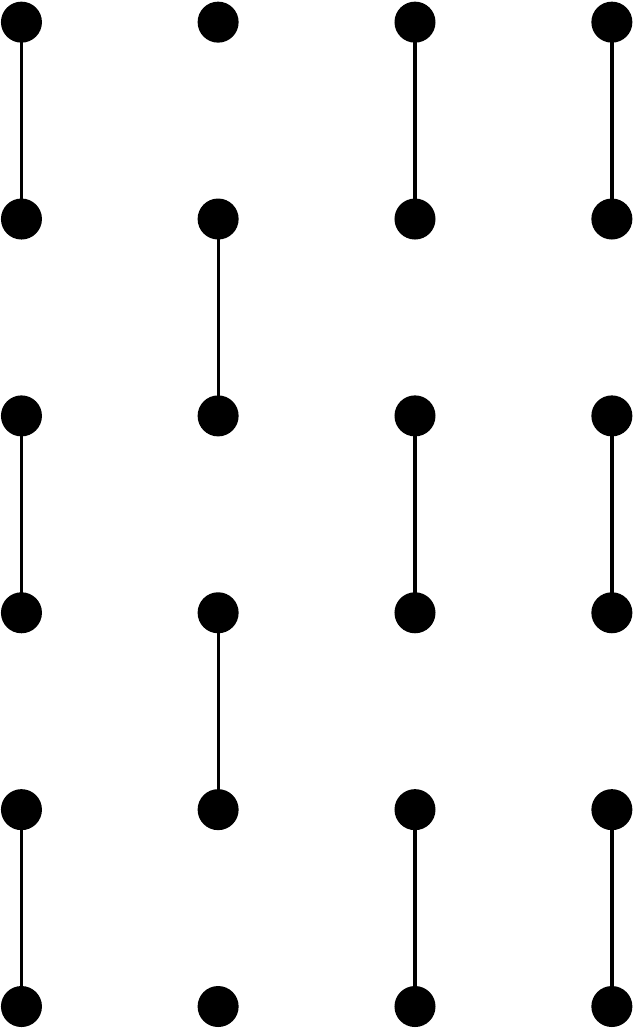}}
\caption{Dimerization pattern with two defects with zero modes on each defect.}
\label{figdefect}
\end{figure}

The Hamiltonian $H_0$ has a highly degenerate ground state subspace,
in one-to-one correspondence with the dimer configurations of the lattice.
For large $U_0$, the low-energy excitations of $H_0$ (\ref{eqn:H_0})
above any of these ground states are generated by acting with
Majorana fermion operators.
\begin{multline}
\left| {\psi_d}\right\rangle\otimes \left| {\psi_f}(d); {i_1},{i_2},\ldots,{i_p}\right\rangle
\equiv \\ \gamma_{i_1}\,\gamma_{i_2}\, \ldots \, \gamma_{i_n}
\left| {\psi_d}\right\rangle\otimes \left| {\psi^0_f}(d)\right\rangle
\end{multline}
where $n_{{i_k}{i_l}}=0$ for all $1 \leq k,l \leq p$. Such a state
has energy $2p - \sum_{ij}n_{ij}(d)$. Therefore, $H_0$ has
an energy gap equal to $2$ above the ground state subspace.

\subsection{Dimer Dynamics}
\label{sec:Model-Def2}

The Hamiltonian $H_0$ partly fulfills our goal, since monomers,
which support Majorana zero modes,
are deconfined, as shown in Fig.~\ref{figdefect}.
However, due to the enormous ground state degeneracy, this situation with deconfined defects is really a result of fine-tuning the
Hamiltonian.  Adding local interactions between dimers can lift the ground state degeneracy and produce interaction between
defects.  We now add such interactions between dimers by adding to
the Hamiltonian a term $H_{RK}^f$.
Our goal is to construct $H_{RK}^f$ such that, for $d=3$, the
system has a unique ground state (or possibly an $O(1)$ or polynomial ground state degeneracy, depending on the topology of the lattice), with
deconfined defects, and
also such that the system is stable to adding additional weak interactions.  The resulting Hamiltonian will have gapless
gauge modes; in the discussion we consider the question of whether it is possible to construct a gapped system.

\begin{figure}[tb]
\centerline{
\includegraphics[scale=0.3]{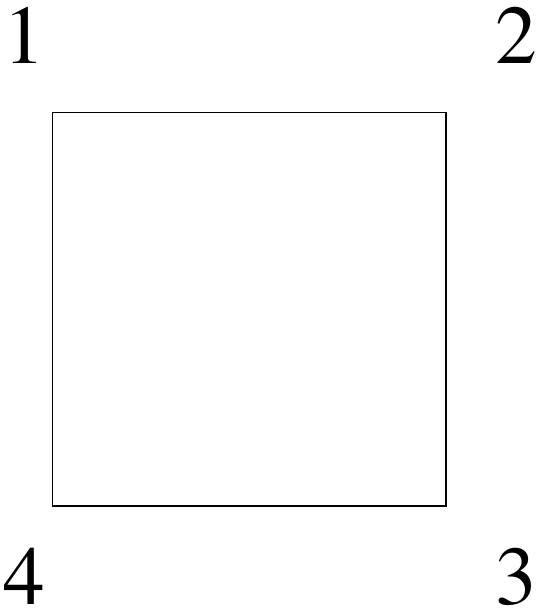}}
\caption{Labelling of points in a plaquette.}
\label{plaqlabel}
\end{figure}

The term $H_{RK}^f$ that we add is strongly reminiscent of the RK
dimer model \cite{Rokhsar88}.
It is equal to
\begin{multline}
H_{RK}^f=V\sum_{\square} \Bigl( \Bigl| || \Bigl\rangle \Bigr\langle || \Bigr|
+ \Bigl| = \Bigl\rangle \Bigr\langle = \Bigr| \Bigr) \\ - t\sum_{\square} \Bigl( \Bigl| || \Bigl\rangle \Bigr\langle = \Bigr| \otimes
{\rm Swap}_{13}+ h.c. \Bigr).
\end{multline}
We include the ``$f$" in the superscript of $H_{RK}^f$ to denote that this Hamiltonian couples the fermions and dimers, while
in later discussions we use $H_{RK}$ to denote the RK dimer model Hamiltonian.
In the above equation, the sum ranges over all plaquettes of the lattice.  The first term is the usual RK
potential energy term.  For $V>0$, it penalizes configurations with parallel dimers on a given plaquette.
The second term is very similar to the RK kinetic term, with one twist.  It is the product of two different operators
on the plaquette, one acting on the dimer Hilbert space and the other acting on the fermion Hilbert space.  The first term, which acts on the
dimer Hilbert space, does a plaquette move of the dimers.  The term ${\rm Swap}_{13}$
acting on the fermion Hilbert space interchanges the fermions on
sites $1,3$ on opposite corners of the plaquette
(we label the sites around the plaquette in clockwise order $1,2,3,4$ starting at the top left as shown in Fig.~\ref{plaqlabel}) so that
\begin{eqnarray}
\label{interchange}
\gamma_1 & \rightarrow & \pm \gamma_3, \\ \nonumber
\gamma_3 & \rightarrow & \mp \gamma_1,
\end{eqnarray}
with the signs chosen so that if the configuration has no defects, and if the plaquette move changes the dimer configuration from $d$ to $d'$,
then if the fermions are in the ground state of the Hamiltonian $H_{ff}(d)$ before
the move, then they are in the ground state of $H_{ff}(d')$ after the move.
To explicitly specify the signs,
if the initial configuration has $n_{12}=n_{34}=1$, and $n_{14}=n_{23}=0$, then the plaquette move turns this into a configuration
with $n_{14}=n_{23}=1$ and $n_{12}=n_{34}=0$, while the swap operator applies
\begin{eqnarray}
\gamma_1 & \rightarrow & (M_{34}/M_{14})\gamma_3=-M_{34}M_{14}\gamma_3, \\ \nonumber
\gamma_3 & \rightarrow & (M_{12}/M_{23})\gamma_1=-M_{12}M_{23}\gamma_1.
\end{eqnarray}
Due to the $\pi$-flux rule (\ref{piflux}), we indeed have
$\gamma_1 \rightarrow \pm \gamma_3$ and $\gamma_3 \rightarrow \mp \gamma_1$.

Note that $H_{RK}^f$ requires there to be $\pi$-flux
in each plaquette in the matrix $M$ because the interchange
of two Majorana fermions requires one of them acquire a minus sign as in
Eq.~(\ref{interchange}) in order to preserve fermion parity, respecting the superselection rules.

Note also that the kinetic term in $H_{RK}^f$ does not change the total number of defects in the dimerization pattern.
Further, in the sector with no defects, the terms $H_{RK}^f$ and $H_{ff}$ commute.
Thus, to determine the ground state of the Hamiltonian $H=H_0+H_{RK}^f$ in the no defect sector, we can directly use previous results
on the quantum dimer model.
If $\psi_0=\sum_d A(d) |\psi_d \rangle$ is the ground state of the RK
Hamiltonian with given $t,U_0$, where
$\psi_d$ is a state with given dimerization pattern and $A(d)$
are complex amplitudes, then
\be
\sum_d A(d) \left|\psi_d \right\rangle \otimes \left| \psi_f^0(d)\right\rangle,
\ee
is the ground state of $H$, where $\psi_f^0(d)$ is the ground state of $H_{ff}(d)$.
That is, the fermions ``go along for the ride" as the dimers move.

Consider the RK point, $t=V$, at which the ground
state of the fermion dimer model may be found exactly:
\be
|0\rangle = {\cal N}{\sum_d}  |{\psi_d}\rangle \otimes |\psi_f^0(d)\rangle.
\label{eqn:FDM-gs}
\ee
Here, ${\cal N}$ is a normalization constant.
This state has $H_{FDM}$ eigenvalue $-N/2$.
The ground state(s) is/are an equal amplitude superposition of
all dimer configurations which can be obtained from a fixed
one by acting with plaquette moves. In three dimensions,
there will be different degenerate states corresponding to different Hopf numbers,
according to the discussion of Section \ref{sec:dimer-topology}.
We view these as Hopfion excitations ``above the ground state'',
albeit with zero energy. In addition, if the lattice has periodic
boundary conditions, then there will be different
winding number sectors which are not mixed by plaquette moves
and, therefore, there will be different degenerate ground states.

Now consider the excited states
\be
|{i_1}, {i_2}, \ldots, {i_m}>
= {\cal N}\sum_{d'}  |\psi_{d'}\rangle \otimes |\psi_f^0(d)\rangle.
\label{eqn:FDM-monomers}
\ee
where the dimerizations $d'$ all have $m$ monomers
at fixed sites ${i_1}, {i_2}, \ldots, {i_m}$. $H_{RK}^f$ does
not move the monomers, and $|{i_1}, {i_2}, \ldots, {i_m}>$
is an eigenstate of $H_{FDM}$ with energy
$-N/2 + m U_0$. This energy does not depend on
the distance between the monomers.
Thus, the fermion dimer model has deconfined monomers
at the RK point (as the ordinary dimer model does).
Moreover, the states $\gamma_{i_k} \ldots \gamma_{i_p} |{i_1}, {i_2}, \ldots, {i_m}>$
also have identical energy $-N/2 + m U_0$. Thus, the deconfined
monomers support Majorana zero modes.

We do not know exact fermionic excited states of the fermion dimer model,
but we can consider the ansatz:
\begin{eqnarray}
|{a_i}\rangle &=& {\sum_i}{a_i} {\gamma_i}{\sum_d}  |{\psi_d}\rangle \otimes |\psi_f^0(d)\rangle\cr
&=& {\sum_i}{a_i} {\sum_d}  |{\psi_d}\rangle \otimes |\psi_f^0(d);i\rangle.
\end{eqnarray}
Such a state is not an eigenstate of $H_0 + H_{RK}^f$,
but it is an eigenstate of $H_0$ with eigenvalue $-N/2 + 2$.
From the definition of  $H_{RK}^f$, we see that
the commutator $[H_{RK}^f, \gamma_i] $ is a fermionic operator which
acts only on the plaquettes neighboring site $i$. Thus,
\begin{equation}
\langle{a_i}| H_{FDM} |{a_i}\rangle \geq -N/2 + 2 - c \, t
\label{eqn:fermon-excited}
\end{equation}
where $c$ is a constant which is roughly equal to the probability
in the state $|0\rangle$ that an arbitrary plaquette has two parallel dimers.
This implies that for $t\ll 1$ (but $V=t$), there is still a gapped band of
fermionic excitations even when the dimers resonate, albeit
a gap which is reduced compared to $H_0$.
Of course, it is possible that a completely different fermionic state
has lower energy than the ansatz (\ref{eqn:fermon-excited}).
However, this is unlikely since the equal-time fermion Green function
is short-ranged: $\left\langle 0 | \gamma_i \gamma_j | 0\right\rangle = 0$ unless
$i$ and $j$ are on the same plaquette. Furthermore, the existence
of zero modes at defects implies that there is a fermionic gap in
the bulk. Thus, it is likely that the fermions remain gapped at
the RK point of the Hamiltonian $H_0 + H_{RK}^f$.
Indeed, this same argument says that the fermions remain gapped even away
from the RK point, so long as $U_0$ is sufficiently large and $t$ is sufficiently
small; this is important for the next two subsections where we discusses
the dimer dynamics in liquid phases near the RK point.

On the other hand, there is a gapless Hopfion at the RK point:
it is the equal amplitude superposition over all dimer configurations
with $N_H=1$:
\be
|H \rangle = {\cal N}\sum_{d \ni {N_H}(d)=1}  |{\psi_d}\rangle \otimes |\psi_f^0(d)\rangle .
\ee
In the fermion dimer model, the $n_H = N_H \text{mod} 2$
is simply the fermionic parity of the ground state of $H_{ff}(d)$.
To see this, let us define the fermion parity as
\begin{equation}
(-1)^{N_F} = \prod_{i\in E} M_{i,i+\hat{x}} {\gamma_i} \gamma_{i+\hat{x}}
\end{equation}
Here $E$ is a sublattice of the original lattice which contains half
of the sites; $\hat{x}$ is one of the basis vectors of the lattice;
and $E, \hat{x}$ are chosen so that
if $i\in E$ then $i+\hat{x}\not\in E$ is not. Then, in
the ground state of $H_{ff}(d)$ defined by Eq.
\ref{eqn:psi_f^0-def},
\begin{equation}
(-1)^{N_F} = \text{Pf}(N)
\end{equation}
Therefore, at least at the RK point, the Hopfion is a gapless
fermionic excitation. However, this gapless fermionic
excitation does not affect the Majorana zero modes
any more than the gapless bosonic mode which is present at the
RK point does. In Section \ref{sec:braiding}, we will discuss whether this
survives beyond the RK point.

Finally, we note that since the fermion parity is necessarily a conserved quantity,
if the $Z_2$ invariant $\text{Pf}(N)$ of the dimer configuration
does not match the fermion parity, then the system
cannot be in the ground state of $H_{ff}(d)$. Instead, an energetic price
must be paid since the fermions must be in an excited state for
the given dimerization pattern. Thus,
although it is possible to add terms to the ordinary dimer model
which mix different $Z_2$ sectors -- namely, terms
which move dimers around loops longer than a single plaquette --
the $Z_2$ invariant is protected by superselection rules
in the fermion dimer model.

\subsection{Two-dimensional Lattices}

At $t=V$, the ground state of $H_{RK}$ is given, in any dimension, by the equal amplitude superposition for all dimer configurations. As we have
argued above, at this point, there are deconfined monomers which
support Majorana zero modes while the fermions remain gapped.
We now consider what happens when we move away from the RK point.
In two dimensions, for $t>V$ or $t<V$, the dimers order\cite{Fradkin03}.  In both cases, this leads to a confinement of the defects.  Consider
for example the configuration shown in Fig.~\ref{figdefect}, which displays columnar order as occurs for large negative $V$; there
is a linear confining potential between defects.  Further, it has been argued that even the point $t=V$ is very finely tuned\cite{Fradkin03}, and is
unstable to adding other local perturbations to the Hamiltonian.

However, we can instead consider the model on a triangular lattice, where there is a liquid phase with deconfined monomers\cite{Moessner01}
for a range of coupling near the so-called ``RK point" $t=V$.
In this case, we expect that the fermion dimer model will show deconfined Majorana zero modes near this point. In this regime, we expect that the system is described
by a topological quantum field theory (TQFT).  This theory should be a subtheory of $Z_2 \otimes Ising$, where the $Z_2$ discrete gauge theory contains particles $1,e,m,em$ (identity, electric, magnetic, and product of electric and magnetic), and Ising contains the particles $1,\sigma,\psi$.  The present triangular lattice theory should be a subtheory of this, as it contains the magnetic particle $m$, corresponding to $Z_2$ vortices in the dimer configuration.  It also contains the $\psi$ particles (an excited state of the fermions).  It also contains the particle $e\sigma$ (a defect carries $Z_2$ charge and a Majorana zero mode).  So, the complete set of particles of the triangular lattice model should be $1,m,e\sigma,em\sigma,\psi,m\psi$.  There is one subtlety in this set of particles.  In the ordinary $Z_2$ discrete gauge theory of a dimer model, the particle $e$ corresponds to a defect site.  Given a defect site, there are two possible topological sectors: draw a line from the defect site to infinity, and consider the sectors with either an even or odd number of dimers crossing this line.  The particle $e$ corresponds to the equal superposition of those two sectors, and $em$ corresponds to the superposition of those two sectors with opposite signs.  However, in the fermion dimer model, the sectors with even or odd dimer number correspond to different Hopfion number (recall that for a planar lattice without holes, $n_H=1$ for all configurations, but the presence of the defect allows $n_H$ to assume either sign).  While we cannot take a superposition of two configurations with different fermion parity, the presence of the $\sigma$ particle on the defect allows us to fix this problem, so that the particle $e\sigma$ corresponds to the equal amplitude superposition of the even and odd dimer sectors with the sign of the $\sigma$ on the defect changed in the odd sector.  Similarly, $em\sigma$ is the opposite amplitude superposition of those two sectors, again with the sign of the $\sigma$ changed in the odd sector.

\subsection{3D Cubic Lattice}
\label{sec:3D-Cubic}

In the three dimensional cubic lattice, the situation is more interesting.
It is believed \cite{Huse03} that $H_{RK}$ on a cubic lattice has a
stable liquid phase for $V$ slightly less than $t$.
This phase has a gapless photon-like mode and is, therefore,
called the Coulomb phase. Defects have a power-law interaction
via this photon-like $U(1)$ gauge field; this power-law interaction is attractive between defects on opposite sublattices, but decays as a power of distance between the defects, so that a single defect is
deconfined: it can be taken arbitrarily far away from other
defects while still having a bounded energy.

In the fermion dimer model, $H_{FDM} = H_0 + H_{RK}^f$,
the situation is expected to be very similar. The gapped fermionic
band is expected to remain gapped in the Coulomb phase, though
the gap will be decreased by a term proportional to $t$,
just as at the RK point. Similarly,
monomers supporting Majorana zero modes are expected to be deconfined.
The main difference between the Coulomb phase and the RK point
is that the interaction energy between two monomers at distance $x$ is expected to
be
\begin{equation}
V_{RK} = \int \frac{d\omega}{2\pi}\frac{{d^2}q}{(2\pi)^2}\,(\text{const.})\,
e^{iq\cdot x} \sim \delta(x)
\end{equation}
at the RK point (note that the $\langle {a_0}(q,0)\, {a_0}(-q,0) \rangle$ correlation
function, which determines the interaction between static charges,
is independent of $q$ at zero frequency at the RK point, which is anisoptropic
between space and time) and
\begin{equation}
V_{C} = \int \frac{{d^3}q}{(2\pi)^3}\,
\frac{e^{iq\cdot x}}{q^2} \sim \frac{1}{x}
\end{equation}
in the Coulomb phase.

The Hopfion is a gapless excitation at the RK point.
We do not have a direct calculation of the Hopfion
energy in the Coulomb phase. However, we
present effective field theories in Section \ref{sec:field-theories}
which describe such a phase, and they indicate that
the Hopfion has vanishing gap throughout the Coulomb phase.

We now describe the statistics of the zero modes.  The result of the study in
Refs. \onlinecite{Teo10,Freedman11} was that in the models
considered there, there are two inequivalent ways to interchange a pair of
monomers.  Under the interchange, one of the Majorana
modes changes sign, so that the interchange of defects on sites $i,j$ leads to either $\gamma_i \rightarrow \gamma_j,\gamma_j \rightarrow -\gamma_i$ or $\gamma_i \rightarrow -\gamma_j, \gamma_j \rightarrow \gamma_i$.  Further, it was found that it was possible to produce
the transformation $\gamma_i \rightarrow -\gamma_i, \gamma_j \rightarrow -\gamma_j$, which does not exchange the defects but
changes the sign of both of them, without moving the defects but by simply evolving the mass term under an appropriate trajectory.
We will find very similar behavior in the fermion dimer model.

\begin{figure}[tb]
\centerline{
\includegraphics[scale=0.3]{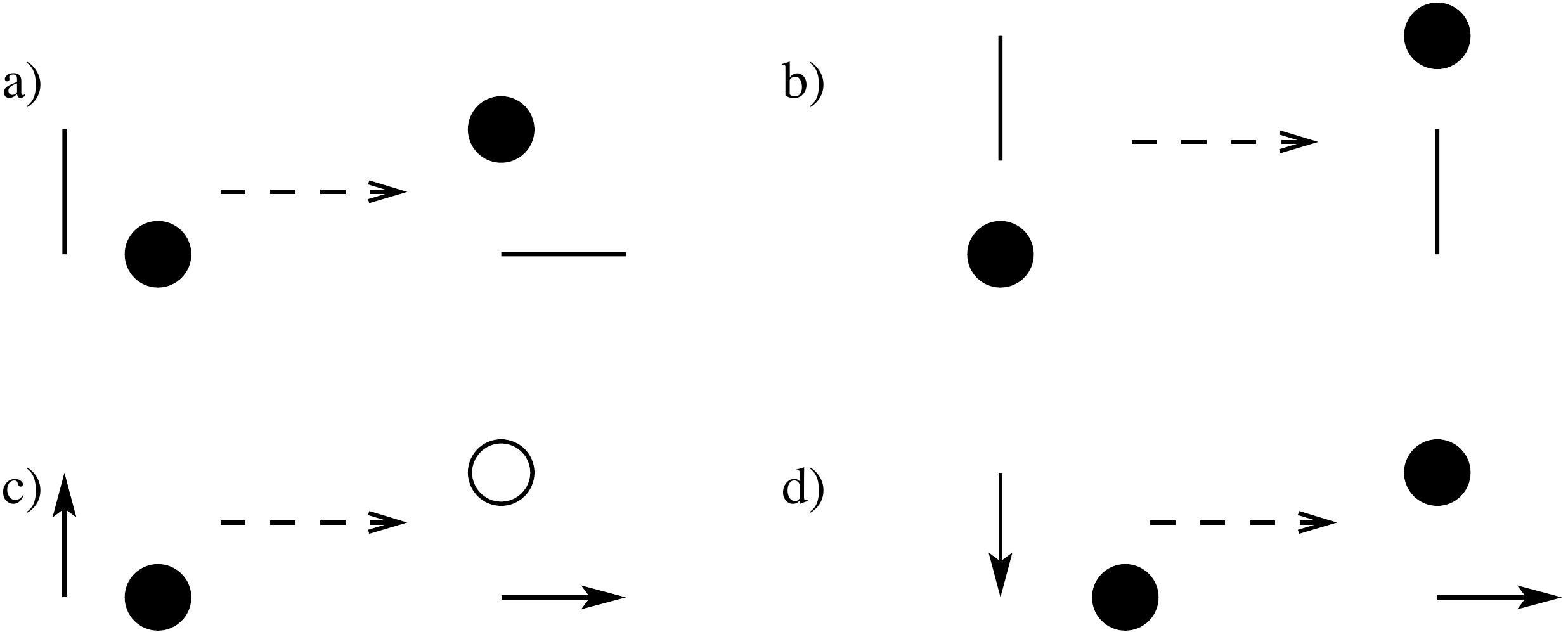}}
\caption{a),b) Illustration of possible processes moving a defect and a dimer.  The dashed arrow is used to indicate
``before" and ``after" configurations.  In (a), the defect moves along a diagonal, while in (b) it moves along a straight line.
c),d) How to compute the signs in the defect motion.  The arrows on the dimers indicate the arrows as in Fig.~\ref{mfig}, while
the dashed arrow again indicates ``before" and ``after".  Solid circles are used to indicate the defect site, while an open circle
indicates that the defect site changes sign.}
\label{defectmove}
\end{figure}

To describe the statistics, we need to modify the Hamiltonian $H$ so that the defects become mobile.  We can add terms
to the Hamiltonian allowing processes such as those shown in Fig.~\ref{defectmove}(a,b), moving the defect and the dimer (we show only some of
the possibilities in the figure, but there are others).

The process described in the figure is a product of two terms.  First, there is a term changing the dimer number from $1$ to $0$
on one bond and from $0$ to $1$ on another bond.  Second, there is a term interchanging the Majorana mode on the defect site and
on one end of the dimer.  Due to the superselection rules, this term must change the sign of one of the Majorana modes.  For example,
using the site labelling of Fig.~\ref{plaqlabel} to describe different sites, in Fig.~\ref{defectmove}(a) the defect moves from site $1$ to site $3$ so we interchange $\gamma_1 \rightarrow \pm \gamma_3,\gamma_3 \rightarrow \mp \gamma_1$.  We choose the sign in the interchange so that if fermions are in their ground state for the dimer configuration before the move,
then they remain in their ground state after the move.
In Fig.~\ref{defectmove}(c,d), we show two different cases leading to different signs; in one case the defect
changes sign and in the other it does not.  If the arrow convention is chosen
following Fig.~\ref{mfig}, then \ref{defectmove}(c) and \ref{defectmove}(d) correspond to different choices of the plaquette on which
the defect is moving, as the arrow direction depends on the plaquette.  The sign also depends upon the dimer configuration within the given plaquette.  Consider the process of Fig.~\ref{defectmove}(c); if the defect had moved on the same plaquette, but the initial configuration had instead a dimer connecting the top two sites of the plaquette (rather than connecting the left two sites as shown), then the final configuration would instead have a dimer connecting the right two sites of the plaquette, and also the sign {\it would be different}: the defect would move without a sign change.  Note that this dependence of sign upon dimer configuration only occurs when a defect moves diagonally within a plaquette, but not when it moves in a straight line as in Fig.~(\ref{defectmove})(b), where the sign is fixed.

We can also add terms to the Hamiltonian which favor the defect lying on certain sites.  These are terms which reduce the penalty for having zero or more than
one dimer touching a site.
For example, we can modify $H_{penalty}$ which was previously defined to be $U_0 \sum_i W_i$, so that it is instead
$\sum_i U_0(i) W_i$, allowing a site dependent penalty $U_0$.  By reducing $U_0(i)$ for certain sites $i$, we favor having the
defect on those sites.

By time-dependent control of these two terms, allowing defect motion and favoring certain sites for the defect, we can interchange two defects.
That is, given defects at sites $i,j$, we can perform a sequence of
transformations so that the defects end at sites $j,i$.  Let us imagine that the terms in the Hamiltonian are adjusted such that
the path that the defect follows is almost completely deterministic: for a given defect, we pick a sequence of sites $i_0,i_1,i_2,...$
We then require that the defect start at site $i$.  We then add a term to the Hamiltonian allowing the defect to move between sites $i_0,i_1$ (which are assumed to lie
a distance $2$ apart so that this motion is possible) but not to any other sites, and gradually adjust the position-dependent $U_0(i)$
so that the defect moves to sites $i_1$ after a given period of time with probability close to $1$.  We then repeat the process moving from
$i_1$ to $i_2$, and so on.  We say that this is ``almost completely" deterministic because the defect may resonate back and forth between
sites $i_0$ and $i_1$ for some time, and so on, but otherwise the sequence of sites visited is completely determined.  We also assume that in this process we adjust the terms allowing the defect motion to only allow one of the two possible choices for a defect to move between two signs which are diagonally opposite on a plaquette (i.e., we allow either the process of Fig.~\ref{defectmove}(c) where the dimer moves from the left side of the plaquette to the bottom or the process where the dimer moves from the top side to the right side, but not both processes).

We find that the sign a given defect
acquires during its motion is completely determined by the path it follows, i.e., the particular sequence of lattice sites
it moves through, as well as the choice of dimer configurations in a plaquette when it moves between two sites diagonally opposite on a plaquette.  As a result, there are, in principle,
four distinct possibilities when we interchange defects: each defect can either
acquire a minus sign or not.

In two of the cases, when either both defects acquire a
minus sign or neither defect acquires
a minus sign, a Hopfion is emitted.  To understand this
possibility, note that we can define the $Z_2$ invariant
above even in a system with defects.  We simply
compute the Pfaffian of the submatrix of the matrix $N$ containing only the sites with no defects.  If the initial configuration
has a given sign of the Pfaffian (the initial state may be a quantum superposition of different defect configurations with the given sign), then the superselection rules show that the sign of the Pfaffian changes after interchange.  One
way to prove this is to define the $Z_2$ invariant for a system with defects in a slightly different way as follows:
take the matrix $N$ defined previously and then define a matrix $N'$ which has the same matrix elements as $N$ except if sites $i,j$
are the two defect sites then $N'_{ij}=\pm i$.  Then, the matrix $N'$ has non-vanishing Pfaffian and its sign is an invariant under plaquette flips.  We choose
some arbitrary choice of the sign of $N_{ij}$ for the two defect sites initially, calling one defect site $i$ and the other site $j$ and setting $N'_{ij}=+i$.
Then, as the defects move, we define a sequence of different matrices $N'$, one such matrix for each defect position, in the natural way: for each defect configuration with defects at sites $k,l$,
we pick the sign in the definition of $N'_{kl}$ so that if the defect originally at site $i$ moved to site $k$ and changed sign a total of $n_1$ times along the process (recall that each time a defect moves, it may or may not change sign, depending on which sites it moves between) and the defect
originally at site $j$ moved to site $l$ and changed sign a total of $n_2$ times along the process, then $N'_{kl}=+i (-1)^{n_1+n_2}$.  Then, the
sign of the Pfaffian of $N'$ does not change along this process.  Thus, after interchanging the two defects, if neither one of them
acquires a negative sign, then the sign of the Pfaffian of the submatrix of $N'$ containing the signs without defects must change
sign.  That is, a Hopfion is emitted. In Section \ref{sec:braiding},
we will discuss its physical implications.

\section{Effective Field Theories of the Coulomb Phase
of the Fermion Dimer Model}
\label{sec:field-theories}

\subsection{CP$^1$ Model}

In this section, we give an effective field theory description
of the fermion dimer model.  We take the CP$^1$ description
of the ordinary cubic lattice dimer model, discussed in section \ref{sec:dimer-eft},
as our starting point:
\begin{multline}
S = \int {d^3}x\, d\tau \left[ \frac{1}{2g}\left|(i\partial_\mu + a_\mu){z_a}\right|^2 +
\lambda(z_a^* z_a -1) \right.\\
\left. + \, V(z^\dagger \vec{\sigma}z)\right]
\end{multline}
We couple this Lagrangian to fermions
on the cubic lattice with $\pi$ flux through each plaquette
As discussed in Section \ref{sec:free-fermions},
the low-energy theory of such fermions is
a single 8-component Majorana fermion:
\begin{equation}
S = \int {d^3}x\, d\tau \left[ \chi^T \partial_\tau \chi
+\chi^T \alpha_j i\partial_j \chi \right]
\end{equation}
where the $\alpha_i$s are defined in Eq. \ref{eqn:alpha-def}.

As discussed in Section \ref{sec:free-fermions}, the effect of
staggering the hopping strengths is to add mass terms to the Hamiltonian,
as in Eq.~(\ref{eqn:beta-mass-terms}).
In the fermion dimer model, the dimers effect an extreme
form of staggering on the fermion hoppings. Thus,
the natural coupling of the fermions to
the dimers is:
\begin{multline}
S = \int {d^3}x\, d\tau \left[  \chi^T \partial_\tau \chi
+\chi^T \alpha_j i\partial_j \chi +
m \, z^\dagger \sigma_k z \,\chi \beta_k \chi \right.\\
\left. + \,\frac{1}{2g}\left|(i\partial_\mu + a_\mu){z_a}\right|^2 +
\lambda(z_a^* z_a -1) \right]
\end{multline}
where
\begin{eqnarray}
\beta_1 &=& i\sigma_y \otimes\sigma_z \otimes\sigma_z \, , \cr
\beta_2 &=& iI \otimes\sigma_y \otimes\sigma_z \, ,\cr
\beta_3 &=& iI\otimes I\otimes\sigma_y \, .
\end{eqnarray}
If the dimers order in, for instance, the columnar phase
with the dimers aligned along the $z$-direction, then
the fermion action takes the form
\begin{eqnarray}
\label{eqn:columnar-fdm-action}
S &=& \int {d^3}x\, d\tau \left[  \chi^T \partial_\tau \chi
+\chi^T \alpha_j i\partial_j \chi +
m \, \,\chi \beta_3 \chi \right]\cr
&=& \int {d^3}x\, d\tau \left[ \psi^\dagger \partial_\tau \chi
+\psi^\dagger {\tilde \alpha}_j i\partial_j \psi +
m \,\psi^\dagger {\tilde \beta}_3 \psi \right]\cr
&=& \int {d^3}x\, d\tau \left[ {\overline \psi} \gamma_\mu i\partial_\mu \psi
+ m \,{\overline \psi} \psi \right]
\end{eqnarray}
where $\psi$ is defined in Eqs. \ref{eqn:lambda-def} and
\ref{eqn:psi-def}; the ${\tilde \alpha}_j$s are defined
after Eq. \ref{eqn:chi-Ham-Majorana};
${\tilde \beta}_3 \equiv I \otimes \sigma_y$;
$\gamma_0 \equiv {\tilde \beta}_3$, $\gamma_i \equiv {\tilde \beta}_3
{\tilde \alpha}_i$, ${\overline \psi} \equiv \psi^\dagger \gamma_0$.
In the columnar phase, the action for the dimer degrees of
freedom is most simply written in the form
\begin{equation}
S_{\rm dimer} = \int  {d^3}x\, d\tau  \left[ \frac{1}{2g}(\partial_\mu \vec{n})^2
+ V(\vec{n})\right]
\end{equation}
The interaction between the fermions and the dimers is:
\begin{equation}
S_{\rm int} = \int  {d^3}x\, d\tau \, m \,\psi^\dagger {\tilde \beta}_k \psi
(n_k - \delta_{k3})
\end{equation}
If the fermions are integrated out, the terms in
$S_{\rm dimer}$ are renormalized and
a topological term is generated, as shown by
Ran, Hosur, and Vishwanath\cite{Ran10}.
This topological term takes the form:
\begin{multline}
\label{eqn:S-topo}
S_{\rm topo} =\\ \frac{1}{1920\pi} \int {d^4}x\, d\tau\,
\epsilon^{\alpha\beta\gamma\mu\nu}
\text{Tr}\left( \gamma_5 V \partial_\alpha V
\partial_\beta V\partial_\gamma V\partial_\mu V\partial_\nu V\right)
\end{multline}
where $V$ takes values in SU(3)/SO(3)
and is an extension of $\vec{n}$ to a $4+1$-dimensional
manifold whose boundary is $3+1$-dimensional
spacetime; on this boundary, $V$ takes values in
$S^2 \subset$ SU(3)/SO(3). There isn't a unique extension
of $\vec{n}$ to $V$, but the resulting value for
$S_{\rm topo}$ (which must be an integral multiple of $\pi$)
is unique up to $2\pi$. Therefore,
$e^{iS_{\rm topo}}$ is either $+1$ or $-1$, depending on
whether two Hopfions are exchanged or not.

In order to discuss the Coulomb phase in which
$\vec{n}=z^\dagger \vec{\sigma}z$ is disordered, it is useful
to rotate the fermions to the local direction of the
dimer order parameter, which is well-defined so long
as hedgehogs don't proliferate. We rotate
\begin{equation}
\chi \rightarrow {\cal R} \chi
\end{equation}
where ${\cal R}$ is defined by
\begin{equation}
{\cal R}^\dagger \left(z^\dagger \sigma_k z\right) \,\beta_k\, {\cal R} = \beta_3
\end{equation}
We can write ${\cal R}$ explicitly in the form:
\begin{equation}
\label{eqn:R-explicit}
{\cal R} = {u_1} \left(I \otimes I \otimes I\right) + i {v_1} M_3
- i {u_2} M_2 + i {v_2} M_1
\end{equation}
where ${z_j}={u_j}+i{v_j}$, and
\begin{eqnarray}
M_1 &=& I \otimes \sigma_y \otimes \sigma_x \, , \cr
M_2 &=& \sigma_y \otimes \sigma_z \otimes \sigma_x \, ,\cr
M_3 &=& \sigma_y \otimes \sigma_x \otimes \sigma_z \, .
\end{eqnarray}
Note that ${\cal R}$ is a real matrix since the $M_i$s are
purely imaginary; therefore, ${\cal R}^\dagger = {\cal R}^T$.
Then, in terms of the rotated fermions, the action takes the form:
\begin{multline}
\label{eqn:rotated-action}
S = \int {d^3}x\, d\tau \left[ \chi^T \partial_\tau \chi
+\chi^T \alpha_j i\partial_j \chi +
m \, \chi^T \beta_3 \chi \right.\\
\chi^T {\cal R}^T \partial_\tau {\cal R} \chi
+\chi^T {\cal R}^T \alpha_j i\partial_j {\cal R} \chi\\
\left. + \,\frac{1}{2g}\left|(i\partial_\mu + a_\mu){z_a}\right|^2 +
\lambda(z_a^* z_a -1) \right]
\end{multline}
or, re-writing this in terms of the Dirac fermion $\psi$,
as in Eq. \ref{eqn:columnar-fdm-action}:
\begin{multline}
\label{eqn:rotated-action-psi}
S = \int {d^3}x\, d\tau \left[ \psi^\dagger \partial_\tau \psi
+\psi^\dagger {\tilde \alpha}_j i\partial_j \psi +
m \, \psi^\dagger \beta_3 \psi \right.\\
\psi^\dagger {\tilde{\cal R}}^\dagger \partial_\tau  {\tilde{\cal R}} \psi
+\psi^\dagger {\tilde{\cal R}}^\dagger {\tilde \alpha_j} i\partial_j  {\tilde{\cal R}} \psi\\
\left. + \,\frac{1}{2g}\left|(i\partial_\mu + a_\mu){z_a}\right|^2 +
\lambda(z_a^* z_a -1) \right]
\end{multline}
where ${\tilde{\cal R}}=H^T {\cal R} H$ and $H =
\frac{1}{\sqrt{2}}  \begin{pmatrix} I & \sigma_x \cr
I & -\sigma_x \end{pmatrix} \otimes I$.

We now consider the Coulomb phase, in which the dimers
are disordered. In this phase $z_a$ is gapped, so we can integrate
it out. This generates a Maxwell term for the gauge field and also
a minimal coupling between the gauge field and the fermions:
\begin{equation}
\label{eqn:QED}
S = \int {d^3}x\, d\tau \left[ \,{\overline \psi} \gamma_\mu (i\partial_\mu
+ a_\mu) \psi
+ \,m{\overline \psi} \psi + \frac{\kappa}{2} f_{\mu\nu}^2\right]
\end{equation}
$\psi$ and $a_\mu$ are minimally-coupled because
the rotated $\psi$ is linear in $z_a$. Since $z_a$ is charged
under the gauge field $a_\mu$, the rotated $\psi$ is as well.
Therefore, the effective field theory for the liquid phase of the
fermion dimer model is simply 3+1-D QED, but the fermions
and the gauge fields may have different velocities (which will,
of course, be different from the speed of light) and the fermion
mass will be different from the bare electron mass.

We note that we could have instead arrived at
the action (\ref{eqn:QED}) through a slave fermion construction.
Suppose we define a bosonic `holon' $z_a$
and `spinon' $f_{\alpha\beta\gamma}$ through
the definition:
\begin{eqnarray}
\chi &=& \frac{1}{2}\left[({z_1} + {z_1^*}) \left(I \otimes I \otimes I\right) +
({z_1} - {z_1^*}) M_3 \right.\nonumber\\
& &- \left. i ({z_2} + {z_2^*}) M_2 + ({z_2} - {z_2^*}) M_1 \right]f
\end{eqnarray}
This representation is redundant, as evinced by its invariance
under
\begin{eqnarray}
z_a &\rightarrow& e^{-i\theta} z_a ,\cr
f &\rightarrow& \left[\cos\theta\, I\otimes I\otimes I + i \sin\theta \,\sigma_y \otimes\sigma_x\otimes I \right]\,f.
\end{eqnarray}
This redundancy can be manifested, at low-energies, through the
emergence of a U(1) gauge field $a_\mu$ which couples to both
$z_a$ and $f_{\alpha\beta\gamma}$. When $z_a$ is gapped,
it can be integrated out, giving Eq. \ref{eqn:QED} with $\psi$
replaced by the Dirac fermion constructed from $f$ along
the lines of Eqs. \ref{eqn:lambda-def} and \ref{eqn:psi-def}.

Note that this field theory describes the low-energy regime
below the hedgehog (or, in QED language, monopole) gaps
(assuming that the fermion gap $m$ is smaller than the hedgehog/monopole
gap), where we were able to safely integrate out the field $z$.
In order to discuss the physics of hedgehogs,
we need to retain the field $z$ since a hedgehog is configuration
such as
\begin{equation}
z = \begin{pmatrix} \cos\frac{\theta}{2} e^{-i\phi/2} \cr
\sin\frac{\theta}{2} e^{i\phi/2} \end{pmatrix}
\end{equation}
where $r,\theta,\phi$ are polar coordinates.
However, we will, instead, consider a different but related
effective field theory in the next subsection,
in which hedgehogs are explicitly
retained, although the relation to the fermion dimer model
is less direct.

Hopfions, however, are non-singular configurations
of the field $z$. Therefore, they are part of our theory,
and are manifested as configurations
of the gauge field $a_\mu$, given by
\begin{equation}
a_\mu = \frac{i}{2}(\partial_\mu z^\dagger \, z - z^\dagger \partial_\nu z)
\end{equation}
where $z$ is a configuration which is topologically-equivalent
to a Hopfion, such as:
\begin{equation}
\label{eqn:Hopfion-z2}
z = \frac{1}{r^2 + \lambda^2}
\begin{pmatrix} 2\lambda r \sin\theta e^{i\phi} \cr
2\lambda r \cos\theta + i (r^2-\lambda^2) \end{pmatrix}
\end{equation}
As noted in Section \ref{sec:dimer-eft}, such a configuration
has vanishing energy as $\lambda\rightarrow \infty$.
However, this gapless excitation is actually a fermion,
as a result of the topological term (\ref{eqn:S-topo})
which results when the gapped fermions are integrated out.
Therefore, the action (\ref{eqn:QED}) is misleading; the
system actually has a gapless fermion -- the Hopfion.
In Section \ref{sec:braiding}, we will discuss the effect of
a gapless Hopfion. First, however, we will discuss an effective
theory in which hedgehogs and Hopfions are retained explicitly.

\subsection{$4+1$-Dimensional Model}\label{sec:4d}

In this subsection we will propose an alternative effective theory description of the Coulomb phase, which has a completely different starting point but leads to results
consistent with the one proposed in the last subsection. This approach is motivated by the t'Hooft-Polyakov approach to magnetic monopoles\cite{tHooft1974,polyakov1974}. We start from the action of Majorana fermions with a mass term:
\begin{eqnarray}
S &=& \int {d^3}x d\tau \left[\chi^T \partial_\tau \chi
+\chi^T \alpha_j i\partial_j \chi +
m n_k \chi \beta_k \chi\right.\nonumber\\
& &\left.+\frac1{2g}\partial_\mu n_k\partial^\mu n_k\right]
\end{eqnarray}
where $n_k$ is the order parameter $\vec{n}$ written in components. It is natural to ask whether one can modify the model by introducing an ${\rm SU(2)}$ gauge field, which is broken to $U(1)$ by the mass terms. If this can be done, the linearly divergent monopole energy will be screened and the interaction between monopoles assumes a Coulomb form, similar to the t'Hooft-Polyakov monopole. Such an approach turns out to be impossible due to an anomaly, as noted in Ref. \onlinecite{Freedman11}, but the problem can be cured by introducing a $(4+1)-d$ theory, as we now discuss.

To introduce an ${\rm SU(2)}$ gauge field it is convenient to rewrite the action in
terms of Weyl fermions:
\begin{eqnarray}
S&=&\int d^3xd\tau\left[c^\dagger\left(\partial_\tau + \sigma^i(-i\partial_i)\right)c\right.\nonumber\\
& &\left.+\left(m n_kc^\dagger\left(\sigma_y\otimes\tau_y\tau_k\right){c^\dagger}^T+h.c.\right)+\frac1{2g}\partial_\mu n_k\partial^\mu n_k\right]\nonumber\\
\end{eqnarray}
in which $c$ is a two-component spinor in space-time, and also carries an isospin $1/2$ representation of an internal ${\rm SU(2)}$ symmetry. In components, one can write $c_{\sigma s}$ with $\sigma=1,2$ spin indices and $s=1,2$ isospin indices. $\sigma^i$ and $\tau_i$ stands for Pauli matrices in spin and isospin indices, respectively. Physically, one can consider this action as a spin singlet superconductor formed by Weyl fermions. Due to fermionic statistics a spin singlet pair must be triplet in isospin, which is why $n_k$ carries a vector representation of the isospin symmetry. From this expression, one can see that the kinetic energy terms are ${\rm SU(2)}$ invariant, and the mass term breaks SU(2) symmetry to U(1). In this theory the monopole has linearly divergent energy. 
To avoid such a divergent energy one can gauge the theory by introducing an SU(2) gauge field $a_\mu$:
\begin{eqnarray}
S' &=&\int d^3xd\tau\left[c^\dagger\left(\partial_\tau-ia_0 + \sigma^i(-i\partial_i-a_i)\right)c\right.\nonumber\\
& &+\left(m n_kc^\dagger\left(\sigma_y\otimes\tau_y\tau_k\right){c^\dagger}^T+h.c.\right)\nonumber\\
& &+\frac1{2g}\left.D_\mu n_kD^\mu n_k\right]+\frac1{4q^2}{\rm Tr}\left[f_{\mu\nu}f^{\mu\nu}\right]
\label{S3plus1}
\end{eqnarray}
In such an action, the order parameter $n_k$ carries gauge charge of the gauge field $a_\mu$. When the order parameter $n_k$ condenses, it induces a mass term for both the fermions and, by the Higgs mechanism, for the gauge field as well.
The monopoles become U(1) magnetic monopoles of the residual U(1) gauge field, so the interaction between them is a Coulomb interaction.

However, the above reasoning fails when we take into account the axial anomaly of the Weyl fermion. 
If we introduce an ultraviolet regularization of the theory above, the regularization has to break the SU(2) symmetry. More explicitly, if we consider a lattice regularization of the theory, according to the the Nielson-Ninomiya theorem\cite{Nielsen80} there must be doubling partners $\tilde{c}$ of the Weyl fermions $c$ with opposite chirality. The UV regularization must break chiral symmetry so that the Weyl fermions and their doubling partners are coupled and there is only one SU(2) symmetry transforming both of them. To make the doubling partners gapped, there must be a mass term $\tilde{n}_k$ for $\tilde{c}$ which breaks SU(2) symmetry. Thus the effective action of the order parameter $n_k$ in general contains a coupling term $n_k\tilde{n}_k$
other than the ordinary kinetic energy and potential energy terms. With $\vec{n}$ a constant field determined by the cut-off, the monopole of $\vec{n}s$ will reobtain a linear energy cost, so that we conclude that in a properly regularized version of theory (\ref{S3plus1}), the monopoles remain confined.


Interestingly, a resolution to this problem can be found by introducing a $(4+1)$-dimensional theory. It is well-known that $(3+1)-d$ Weyl fermions can be regularized as ``domain wall fermions" living on the boundary of a $(4+1)-d$ lattice Dirac theory\cite{Kaplan92}. More generically, it was demonstrated in Ref. \cite{Qi08} that any $(4+1)$-d gapped fermion theory with a nontrivial second Chern number in the geometrical gauge field defined in momentum space has such domain wall fermions on its boundary, and the number and chirality of the Weyl fermions on the boundary is determined by the second Chern number. Such a $(4+1)$-d state is a topological insulator which is the parent state for $(3+1)$ and $(2+1)$ dimensional time-reversal invariant topological insulators.\cite{Qi08}

Here we consider such a lattice regularization by using the lattice Dirac model. In real space
\begin{multline}
S_{4+1}=\int_0^\beta d\tau\left[\sum_ic_i^\dagger\partial_\tau c_i+(M+4B)\sum_ic_i^\dagger\Gamma^0c_i\right.\\
\left.-\sum_{i,\hat{\alpha}=1,2,3,4}\left(c_i^\dagger\frac{i\Gamma^\alpha+B\Gamma^0}{2}c_{i+\hat{\alpha}}+h.c.\right)\right]
\end{multline}
with $\Gamma^{0,1,..,4}$ Hermitian Dirac $\Gamma$ matrices ${\Gamma^a}^\dagger=\Gamma^a,~\left\{\Gamma^a,\Gamma^b\right\}=\delta^{ab}$. In the continuum limit this action describes a massive Dirac fermion with mass $M$. For the mass range $-2<M/B<0$ the Dirac model has Chern number $C_2=1$ in the Brillouin zone, and on the boundary there is one Weyl fermion. To obtain the Weyl fermion SU(2) doublet in Eq. (\ref{S3plus1}) we consider an SU(2) doublet of $c_i$ in $(4+1)$-d, coupled with an SU(2) lattice gauge field and a Higgs field:
\begin{multline}
S'_{4+1}=\int_0^\beta d\tau\left[\sum_ic_i^\dagger\left(\partial_\tau-ia_0 \right)c_i+(M+4B)\sum_ic_i^\dagger\Gamma^0c_i\right.\\
\left.-\sum_{i,\hat{\alpha}=1,2,3,4}\left(c_i^\dagger\frac{i\Gamma^\alpha+B\Gamma^0}{2}e^{ia_{i\hat{\alpha}}}c_{i+\hat{\alpha}}+h.c.\right)\right.\\
+\left.\sum_i\left(m n_kc_i^\dagger\left(\mathcal{T}\otimes \tau_y\tau_k\right){c_i^\dagger}^T+h.c.\right)\right]+S_{\rm M}[a_\mu]+S_{\rm \sigma}[{\bf n}]\label{LatticeAction}
\end{multline}
with $S_{\rm M}[a_\mu]$ the Maxwell term for $a_\mu$ and $S_{\rm sigma}[{\bf n}]$ the lattice version of the sigma-model action $\frac1{2g}D_\mu n_k D^\mu n_k$. $\mathcal{T}$ is the time-reversal matrix similar to $\sigma_y$ for the two-component fermion. In the same way as in the $(3+1)$-d theory, the order parameter $n_k$ is a Higgs field
carrying a triplet representation of SU(2).

\begin{figure}[tb]
\label{fig4d}
\centerline{
\includegraphics[scale=0.3]{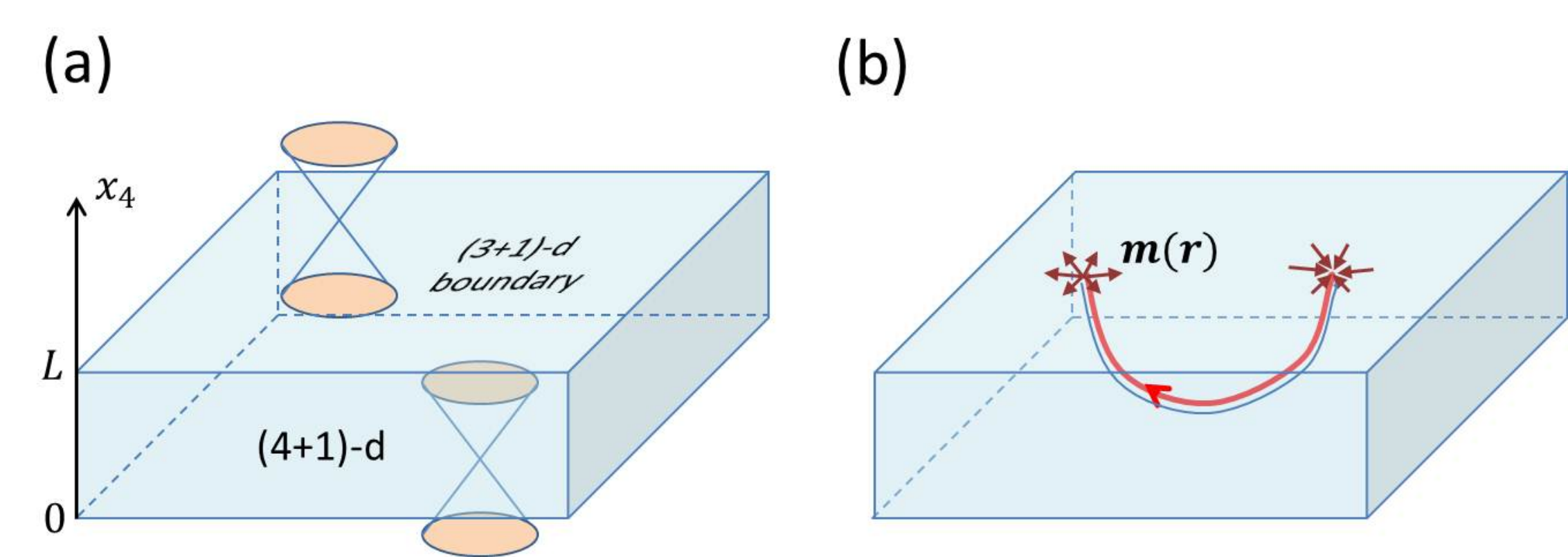}}
\caption{(a) Illustration of the $(4+1)$-d theory, which consists of lattice Dirac fermions coupled with a SU(2) gauge field described by the action (\ref{LatticeAction}). For $-2<M/B<0$ the bulk is gapped and topologically nontrivial, with one doublet of Weyl fermions on each surface. (b) When the Higgs field $\vec{n}$ is in the condensed phase, the surface Weyl fermions become gapped Majorana fermions, and the monopoles on the 3d surface are end-points of one-dimensional monopole lines in the 4d bulk.  }
\end{figure}

Consider this theory defined on a finite thickness slab ${\rm R^3\times I}$ with finite thickness along the 4-th dimension and infinite along the other three dimensions. If $n_k=0$, a doublet of massless Weyl fermions live on each of the two boundary ${\rm R^3}$ with opposite chirality, as is illustrated in Fig. \ref{fig4d} (a). We can consider the Weyl fermions on the two boundaries as the $c$ fermions in Eq. (\ref{S3plus1}) and their doubling partners $\tilde{c}$. The mass term $n_k$ breaks SU(2) symmetry and transforms the Weyl fermions into massive Majorana fermions. Since we consider the top surface as the physical system and the bottom surface as the doubling partners, we always consider the $\vec{n}$ configurations with fixed boundary condition $\vec{n}=\vec{n}_0$ at the bottom surface, so that the UV cut-off is fixed and has no topological configuration.

The advantage of the $(4+1)$-d theory is that it describes the anomaly of the boundary theory explicitly. In $(3+1)$-d, the point-like monopole of the $\vec{n}$ field can be defined since $\pi_2(S^2)=\mathbb{Z}$. In $(4+1)$-d the same homotopy group leads to one-dimensional topological defects. Thus the monopole on the boundary 3d space is always the end of 1d ``monopole lines" in the bulk, as shown in Fig. \ref{fig4d} (b). This picture immediately leads to the conclusion we obtained earlier that the monopoles are still linearly confined after the theory is gauged. Indeed, the linear confinement is due to the existence of the UV cut-off on the other boundary, since the boundary condition on the bottom surface does not allow the monopole lines to penetrate in the other surface. (Otherwise the lowest energy monopole lines should be perpendicular penetrating both surfaces when the monopoles are far away.)

\begin{figure}[tb]
\centerline{
\includegraphics[scale=0.3]{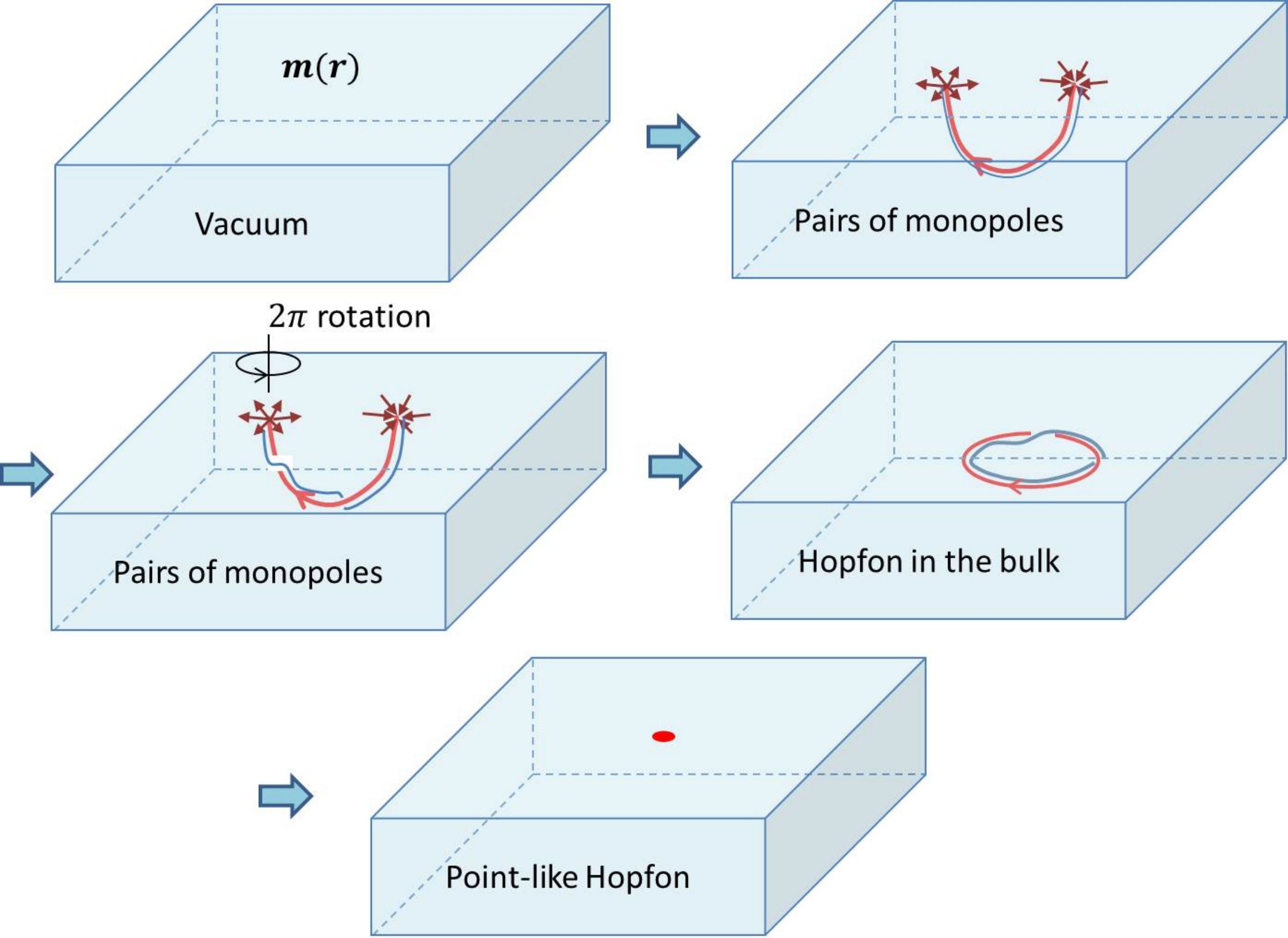}}
\caption{Illustration of the creation of a Hopfion in the bulk by creating a monopole-anti-monopole pair, rotate the monopole by $2\pi$ and annihilate the pair. The Hopfion is a point-like defect of the order parameter $\vec{n}$ in the bulk 4d space. }
\label{Hopf4d}
\end{figure}

The existence of non-Abelian statistics can also be understood explicitly in the $(4+1)$ dimensional theory. As is discussed in Ref.\onlinecite{Freedman11} and earlier in this draft, the essential fact enabling the possibility of non-Abelian statistics in $(3+1)$-d is the existence of two topologically distinct exchange process of two monopoles, which are different by a ``braidless operation", {\it i.e.} spinning one monopole by $2\pi$ while keep the other monopole fixed. If we create a monopole-anti-monopole pair and do the braidless operation, and then annihilate the pair, the configuration left is not the vacuum but the Hopf map. In the $(4+1)$-d theory, after such a process the $\vec{n}$ configuration on the top surface is a nontrivial map with winding number $1$ in $\pi_3(S^2)=\mathbb{Z}$. In the bulk, such a process will create a point-like defect in the 4d space, just as $\pi_2(S^2)=\mathbb{Z}$ leads to stable point-like defects in 3d space. Thus the ``Hopfion" we discussed earlier is a point-like defect in the $(4+1)$-d theory, which can be some distance away from our physical 3d space on the boundary. This argument is illustrated in Fig. \ref{Hopf4d}. In relation to the chiraly anomaly, it can be shown that the braidless operation corresponds to a configuration of the SU(2) gauge field with nontrivial instanton number $\frac1{32\pi^2}\int d^3xd\tau \epsilon^{\mu\nu\sigma\tau}{\rm tr}\left[f_{\mu\nu}f_{\sigma\tau}\right]=1$, which thus pumps one fermion from the boundary into the bulk. After the monopole-anti-monopole pair is annihilated, the fermion pumped by this process stays with the Hopfion in the bulk.

\begin{figure}[tb]
\centerline{
\includegraphics[scale=0.3]{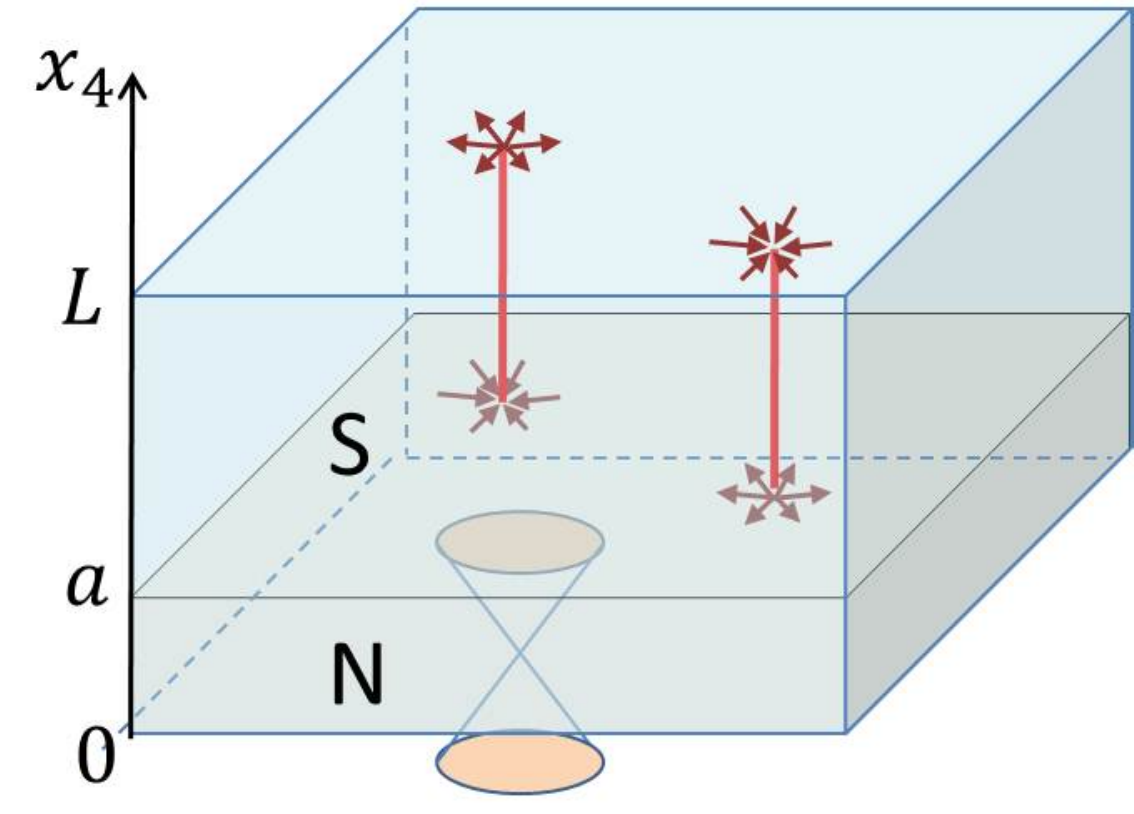}}
\caption{Illustration of the configuration with the order parameter in the Higgs phase in the S (``superconducting") region and vanishes in the N (``normal") region. Each perpendicular monopole line traps a single Majorana zero mode confined at the top endpoint, and two monopole lines are coupled by Coulomb interaction. The bottom surface state is a SU(2) doublet of gapless Weyl fermion.}
\label{fig4d2}
\end{figure}

Now we address the question of whether it is possible to have a phase with deconfined monopoles in this theory. In the $(4+1)$-d theory, the confinement of monopoles is due to the monopole line connecting two monopoles. This is the same as confinement between U(1) magnetic monopoles in a $(3+1)$-d superconductor. To make the monopoles deconfined we need to suppress the Higgs phase. In last subsection, we quantum disordered the order parameter field $z$ for this purpose. In the $(4+1)$-d theory, the deconfined phase can be introduced in a different, somewhat more explicit way by making use of the extra dimension. To this end, consider a modified version of the $(4+1)$-d theory as shown in Fig. \ref{fig4d2}, with the $(4+1)$-d slab geometry the same as that discussed above but the order parameter $\vec{n}=0$ in a region $0<x_4<a$ including the bottom surface and its neighborhood.  The region with $\vec{n}=0$ is a normal topological insulator phase (denoted by N) preserving SU(2) symmetry, and the rest of the system (denoted by S) which is a ``color superconductor" of the SU(2) isospin. The bottom surface state has a width of $\xi\simeq 1/|M|$ in the $x_4$ direction. If $a\gg 1/|M|$ the bottom surface can be considered as decoupled from the symmetry breaking order parameter field $\vec{n}$ and thus remains massless Weyl fermions. In such a geometry, a monopole on the top surface can
be attached to a perpendicular monopole line along the $x_4$ direction which terminates at the domain between the N and S regions, as shown in Fig. \ref{fig4d2}. The interaction between two such monopole lines perpendicular to the $(3+1)$-d boundary is the $(3+1)$-d Coulomb interaction, just as for ordinary t'Hooft-Polyakov monopoles. In the N region, the SU(2) gauge invariance is recovered and the SU(2) magnetic flux can spread in space. It is important to note that a Majorana zero mode is trapped at the end point of the monopole line at the top surface, but no Majorana zero mode exists at the bottom end point of the monopole line, since the fermion degree of freedom is completely gapped at the N/S boundary. Now we consider a rotation of the monopole on the top surface by $2\pi$. If the bottom end point of the monopole line is not rotated, a Hopfion will be created in the S region which is a point-like topological defect costing some energy. If the N region is in the Coulomb phase of the
SU(2) gauge field, the Hopfion will prefer to move down to the N region where it will smear in the whole space to save energy. As is discussed above, the Hopfion corresponds to a soliton configuration of the SU(2) gauge field with nontrivial second Chern number, which thus carries one fermion due to the nontrivial response of the $(4+1)$-d topological insulator. Because the bottom surface is now gapless, if we rotate the monopole slowly the Hopfion will finally propagate to the bottom surface and disappear, so that the energy cost from the gauge field strength and order parameter can be completely eliminated, and one fermion is pumped from the top surface to the gapless bottom surface. Thus, in this configuration we have realized deconfined monopoles with a single Majorana zero mode and Coulomb interaction. One can see that gapless Weyl fermions on the bottom surface are necessary for realizing this phase while still preserving the non-Abelian statistics of monopoles, without which it would be impossible to pump a fermion away from the top surface. For a slab with finite thickness $L$ there is a nonzero coupling between the top and bottom surface states. Consequently, in principle the gapless bottom surface states will mediate a coupling between two Majorana zero modes on the top surface which decays as a power law rather than exponentially in distance. However, such a coupling is suppressed by an exponential factor of $e^{-LM}$ which can be made arbitrarily small for $L\gg \xi=1/M$. Compared with the $(3+1)$-d theory discussed earlier, we see that the Weyl fermions on the bottom surface in the $(4+1)$-d theory correspond to Hopfions in the $(3+1)$-d field theory, or the lattice fermionic dimer model. The $(4+1)$-d and $(3+1)$-d theories lead to consistent results for a deconfined phase with gapless fermionic Hopfions. Moreover, the $(4+1)$-d theory provides an explicit way to describe the fermionic Hopfions and a controlled way to suppress the coupling between Majorana zero modes in the monopoles and the gapless Hopfions.

It should be noticed that the discussion above can be generalized to more generic systems with $N$ Weyl fermions coupled to an U(N) gauge field. The $(3+1)$ dimensional action (\ref{S3plus1}) can be generalized to
\begin{eqnarray}
S' &=&\int d^3xd\tau\left[c^\dagger\left(\partial_\tau-ia_0 + \sigma^i(-i\partial_i-a_i)\right)c\right.\nonumber\\
& &+\left(m c^\dagger_a\left(\sigma_y\otimes n^{ab}\right){c^\dagger}_b+h.c.\right)\nonumber\\
& &+\frac1{2g}\left.{\rm Tr}\left(D_\mu n^\dagger D^\mu n\right)\right]+\frac1{4q^2}{\rm Tr}\left[f_{\mu\nu}f^{\mu\nu}\right]
\label{actionSUN}
\end{eqnarray}
with $c$ a N-dimensional spinor carrying fundamental representation of U(N), $n_{ab}$ an $N\times N$ mass matrix satisfying $n=n^T,~n^\dagger n=1$. This mass term gives the Majorana fermion a mass $m$ and breaks SU(N) to the subgroup of SO(N). The order parameter manifold is thus $SU(N)/SO(N)$. This model can be considered as a generic model for Majorana fermions, while the SU(2) model discussed above corresponds to the special case of $N=2$. In SU(2) case we have the Hopf map coming from $\pi_3(S^2)=\mathbb{Z}$. In the case of SU(N) $(N>2)$ we have $\pi_3(SU(N)/SO(N))=\mathbb{Z}_2$ so that only a $\mathbb{Z}_2$ nontrivial generalization of the Hopf map can be defined. We also have $\pi_2(SU(N)/SO(N))=\mathbb{Z}_2$ so that monopoles still exist but there is no distinct monopole and anti-monopole. For such a model, the discussion above all holds since we can still define the $(4+1)$-dimensional model, which still have monopole lines and point-like Hopfons. By the same configuration as is discussed in Fig. \ref{fig4d2}, a phase with weakly-interacting monopoles can be found.

\section{Monopole Statistics}
\label{sec:braiding}

We now discuss what happens when monopoles are exchanged
and twisted. (We refer to any process which changes
the sign of a single Majorana zero mode operator and emits
a Hopfion as a ``twist" since it produces the same effect on the
defect as the $2\pi$ rotation in Ref.~\onlinecite{Freedman11}.)
As discussed in the context of the dimer model, when two defects
are interchanged, we have four possible choices:
either, neither, or both Majorana zero modes could change sign.
In the ordered phase (or any other phase in which
a Hopfion costs non-zero energy), one of the zero modes
must change sign (if the process is done adiabatically);
which zero mode changes sign is determined by the path.
If a single monopole is twisted, then a Hopfion must
be emitted, as we discussed in Section \ref{sec:3D-Cubic} (twisting
a single monopole changes the sign of its Majorana zero mode
so a Hopfion is created to preserve fermion parity).
In the ordered phase, the emitted Hopfion must be
virtual if the twisting is done slowly enough. Therefore, the virtual Hopfion must
be absorbed by another monopole, which also twists, as a result.
This is a process that changes the sign of both defects.

In controlled processes, we hold the monopoles fixed and
determine which monopoles twist.
However, suppose that we allow quantum fluctuations of the defects
in the fermion dimer model. (The Coulomb phase will generically
have such fluctuations.) That is, suppose that we move defects
with a position dependent $U_0(i)$, but we have non-vanishing terms
in the Hamiltonian allowing the defect to move locally. Then, an
interesting process is possible in which the
defect moves around a short loop, changing its sign and emitting a Hopfion.
This Hopfion can then be absorbed by the other defect,
changing its sign. That is, there is an amplitude for a process that spontaneously
changes the sign of both defects.
When a Hopfion is transmitted from one defect to the other,
it induces an effective interaction term $i \gamma_i \gamma_j$ in the Hamiltonian,
where $\gamma_i,\gamma_j$ are the operators corresponding to the zero modes on defect $i$ and $j$ and where the coefficient of this effective term depends upon the spacing between the defects. We expect that if the defects
are far separated, then this process will be suppressed exponentially in the spacing between the defects.

Now consider the Coulomb phase. The first question is whether the
Majorana zero modes survive at all in the presence of gapless Hopfions.
At the RK point of the fermion dimer model, the zero modes are completely
unaffected by the Hopfion, as discussed in Section \ref{sec:Model-Def2}.
However, this is a special feature of the fermion dimer model.
At a generic point in the Coulomb phase, we expect a coupling between
the Majorana zero modes and Hopfions. However, this coupling need
not destroy the Majorana zero modes. Consider the following
toy model. Suppose we have a system of non-interacting Majorana fermions, arranged on a $d$-dimensional lattice with $\pi$-flux hoppings.  Add an additional defect site which couples only to a single site, which we call site $0$, in the lattice, so that the full Hamiltonian is $H=H_0+i t \gamma_{defect} \gamma_0$, where $H_0$ is the $\pi$-flux Hamiltonian, $t$ is the tunneling strength between the defect site and site $0$, and $\gamma_{defect}$ and $\gamma_0$ are Majorana operators on the respective sites.  This system has an exact zero energy state (exact because we added only one defect).  The amplitude of this state on site $x$ is proportional to $G_{0x}$, where the Green's function $G$ is $G=H_0^{-1}$.  For $d=2$, $G_{0x}$ is proportional to $1/|x|$, so that the norm square is propotional to $1/|x|^2$ and the integrated norm squared diverges logarithmically; this implies that in $d=2$ the state is delocalized.  However, for $d=3$, $G_{0x} \propto 1/|x|^2$, so the integrated norm squared converges and we have an exactly zero energy state localized near the defect.  So, for $d>2$, there is a localized state with an infinite lifetime, despite the coupling to a gapless continuum.  Based on this toy model calculation, it seems likely that similarly the defect mode can have an infinite lifetime, despite coupling to gapless Hopfions, for $d=3$, although some caution is required since this answer is clearly sensitive to the density of states of the gapless continuum near zero energy, and the answer could change for different dispersion relations.

We can analyze the effect of the coupling of a zero mode to a continuum more
generally by considering an effective field theory of an isolated monopole
at the origin with its associated zero mode. Let us suppose that the
low-energy effective action for the Hopfion can be written as a
relativistic Dirac fermion as in the toy model above. Then the effective action for the Hopfion
and the Majorana zero mode will take the form:
\begin{multline}
\label{eqn:zero-mode-Hopfion}
S = \int d\tau {d^d}x \,\overline{\psi} \gamma^\mu i\partial_\mu \psi +
\int d\tau \gamma \partial_\tau \gamma\\
+ v\int d\tau \gamma (\overline{u} \psi(0) + c.c.)
\end{multline}
where $\psi$ is the Hopfion annihilation operator, $\gamma$ is the Majorana
zero mode, $v$ is the magnitude of the matrix element between
the zero mode and the Hopfion, and $u$ is a constant unit spinor
which is the matrix element between the zero mode and
the Hopfion divided by $v$. Then, by power counting, $v$ has scaling dimension
$1-\frac{d}{2}$. In three dimensions, the coupling between the zero
mode and the bulk fermion is irrelevant and, therefore, unimportant in the infrared.
The behavior seen at the RK point of the fermion dimer model is the
fixed point behavior of the phase. This can also be seen by an explicit
calculation of the zero mode Green function at lowest order
in $v$:
\begin{equation}
\left\langle \gamma(i\epsilon_n) \gamma(-i\epsilon_n) \right\rangle =
\frac{Z}{i\epsilon_n}
\end{equation}
where $Z^{-1} = 1 + O({v^2})$. There is still a pole at zero energy.
The zero mode is spread out over a few lattice sites due to its interaction
with the bulk, so that the probability for it to remain at the origin is
reduced by a factor of $Z$. However, it has not disappeared into
the bulk continuum, as would have happened if there had been
a constant Hopfion density-of-states at zero energy, in which
case the zero mode Green function would have had a form
$\sim (i\epsilon_n + i\Gamma \text{sgn}({\epsilon_n}))^{-1}$
for some constant $\Gamma$. In the field theories of the previous
section, the Hopfion has a relativistic spectrum and hence has vanishing density of states at zero energy; it is clearly not a Dirac particle,
since it has additional degrees of freedom (a Hopfion can stretch and
distort in various ways, as well as changing its overall size), but this actually helps us since the low energy Hopfions are necessarily large, so they will have smaller matrix elements to couple to a point-like defect.
Thus, we expect that the zero mode survives in the full theory.

In Eq. \ref{eqn:zero-mode-Hopfion}, we have assumed
that Hopfion number is conserved, as it is in the fermion dimer
model. However, if dimer moves involving more than one plaquete
are included in the Hamiltonian, then the Hopfion number will
not be conserved; it will only be conserved modulo 2, as is required
by fermion parity. Thus, the generic situation is, in fact, that
the Hopfion is a Majorana fermion.

In the Coulomb phase, the amplitude for a virtual Hopfion
to be exchanged between two monopoles
will fall off polynomially with the distance between the defects.
Therefore, the splitting between the two states of a pair of
monopoles will fall off as an inverse power of the distance between
them, rather than exponentially, as we would have in a true
topological phase.  For $d=3$, using the non-interacting toy model above,
the splitting is inversely proportional to the square of the distance.

The gaplessness of the Hopfion has another important
consequence. A single monopole can twist
and emit a zero energy Hopfion.
If $\eta$ is the creation operator for a zero-energy Hopfion (which has
infinite size since the Hopfion energy is inversely proportional to
its linear size) and $\gamma_i$ is the Majorana fermion operator
for the $i^{\rm th}$ quasiparticle, then twisting the $i^{\rm th}$ quasiparticle
implements the transformation
\begin{equation}
\label{eqn:twists-anti-commute}
R_i = \eta \,\gamma_i
\end{equation}
Meanwhile, exchanging quasiparticles $i$ and $j$ without twisting
either one (which would be necessary to heal the order parameter
in the confining phase of the hedgehogs) implements the transformation
\begin{equation}
S_{ij} = \frac{1}{2}\eta ( \gamma_i + \gamma_j )
\end{equation}
Thus, the combination of an exchange and a twist is
\begin{equation}
T_i = R_i S_{ij} = \frac{1}{2} ( 1+ \gamma_i  \gamma_j ) =
e^{\pi \gamma_i  \gamma_j /4}
\end{equation}
which is the representative of a generator of the ribbon permutation group.
Thus, the set of low-energy processes which are possible
in the Coulomb phase is larger than simply the ribbon permutation
group. Hopfion emission processes are possible; the
ribbon permutation group is the subset of low-energy
processes in which real Hopfion creation or annihilation does
not occur.

\section{Discussion}
\label{sec:discussion}

When a quasiparticle in a $2D$ topological phase
is rotated by $2\pi$, or {\em twisted}, the state of the system changes by a phase
which depends only on the particle type. If there are several particles
of the same species, then the phase acquired by the system doesn't
depend on which particle is rotated. This follows from locality: an operation
performed on a single particle cannot affect distant particles and must,
therefore, be just a phase. However,
in a system of $3D$ non-Abelian anyons -- or, to be more precise,
particles which obey {\it projective ribbon permutation statistics} --
a twist has a highly non-trivial effect. Rotating a
particle {\em anti-commutes} with rotating a different one.
In the model introduced by Teo and Kane\cite{Teo10}
and in the general framework introduced in Ref. \onlinecite{Freedman11},
there are confining long-ranged interactions $V(r) \propto r$ between
the hedgehogs which are $3D$ non-Abelian anyons. Thus, it is plausible that
an operation performed on a single particle can affect distant particles.
Indeed, the order parameter configuration which interpolates between
hedgehogs can only be healed after an exchange if one of the hedgehogs
is twisted. Consequently, exchanges
belong to the ribbon permutation group\cite{Freedman11}, rather than the ordinary
permutation group. However, this alone does not explain why
twists anti-commute. It must, furthermore, be the case --
as it is in the model introduced by Teo and Kane\cite{Teo10} --
that the ribbon permutation group is represented {\it projectively}, rather than
linearly. The multiplication rule of the ribbon permutation group is only
represented up to a sign (but hardly an innocuous sign!)
on the Hilbert space of the theory. This is presumably related to
the fact that the overall phase associated with any exchange
is not a topological quantity -- since, as a result of the long-ranged
force between hedgehogs, it depends on non-universal details of the
exchange. Therefore, it need not linearly represent the
ribbon permutation group. Thus, the fact that hedgehogs are
not quite point-like quasiparticles is doubly important: (1) since there
is an order parameter interpolating between them, their exchanges
belong to the ribbon permutation group; and (2) since there is
a long-ranged force between hedgehogs, the ribbon permutation group
is projectively represented.

Thus, it is a little difficult to imagine how $3D$ non-Abelian anyons
could be the weakly-coupled point-like quasiparticles of a system.
Nevertheless, in this paper, we have given
two complementary field theory descriptions of
a Coulomb phase of $3D$ non-Abelian anyons. In this phase,
twists of different quasiparticles do not commute because twisting
a quasiparticle causes a Hopfion, a gapless fermionic
excitation of the system, to be emitted. In fact,
two twists anti-commute, as may be seen from Eq.
\ref{eqn:twists-anti-commute}.
In this way, quasiparticles communicate their twists to distant particles
through the emission and absorption of gapless Hopfions.
The very gaplessness of the Hopfions which facilitates
the non-trivial statistics of the monopoles also reduces
the protection of the quantum information which is contained
in their zero modes. The splitting between zero modes falls
off only as a power law, rather than an exponential. It does
so as a result of Hopfion phase space restrictions, but this
is no worse than a Fermi liquid, which is also stable due to phase
space restrictions, rather than a gap.

In a recent preprint, McGreevy and Swingle\cite{McGreevy11}
argue that a magnetic monopole cannot have a single
Majorana fermion zero mode in a spontaneously-broken
SU(2) gauge theory. Our phase does not conflict with their
conclusions because they demand that the fermions be
fully-gapped in the bulk while our model necessarily has a
gapless fermionic Hopfion.

We have also constructed an explicit microscopic model
which has both confining and Coulomb phases of hedgehogs
supporting Majorana zero modes, the fermion dimer model.
The field theories constructed in this paper appear to
embody the physics of the fermion dimer model. However,
it is possible, in principle, that the Coulomb phase of these field
theories is different than the Coulomb phase of the fermion dimer model.
To show that they are the same, we would need to compute
the energy of the Hopfion in the fermion dimer model.
At the RK point, the Hopfion has vanishing energy gap.
We would like to know if the Hopfion gap remains zero
in the Coulomb phase. This is already an interesting question
in the ordinary dimer model (i.e. without fermions),
and a quantum Monte Carlo simulation could determine the
energy gap of the Hopfion and study its behavior near the critical point.

Can this model be used to perform topologically-protected quantum computation?  Since the Hopfion is gapless, it requires a long time to perform an operation such as twisting both quasiparticles while staying in the adiabatic regime in order to return to a state without a Hopfion at the end of the process.  That is, even though neither particle moves a large distance, the operation requires a long time; this should be no surprise given that the Hopfion must be emitted at one particle and absorbed at the other a long distance away.  This long time poses a problem since we have also argued that if defects have a quantum dynamics, then they can absorb and emit virtual Hopfions leading to a weak level splitting; if the time required to perform a twist operation is sufficiently long compared to the inverse level splitting then the desired operation will not be impliemented with high fidelity.  One way around this is to imagine that we ``pin down" the quasiparticles and stop them from twisting; that is, we imagine that we perfectly control the path along which each defect moves.

The gapless Hopfion raises the natural question of whether we can
construct a {\em gapped} Hamiltonian with
deconfined non-Abelian anyons in three dimensions -- i.e. a phase
in which non-Abelian anyons have exponentially-decaying, rather than
confining or even Coulomb interactions. At a field-theoretic level,
one possible route is the condensation of magnetic monopole pairs,
which would drive the system from our Coulomb phase to a
phase in which the U(1) gauge field is gapped.
Since the energy of a Hopfion is essentially magnetic field energy
according to the discussion in Sections \ref{sec:dimer-eft} and \ref{sec:field-theories},
monopole pair condensation should leave the Hopfion gapless
and, perhaps, even soften its dispersion relation by suppressing magnetic field
energy. Therefore, this does not lead to a fully-gapped phase
and, as a result of the softened spectrum, may even lift
the zero modes.
Within the context of the fermion dimer model, the natural approach is to consider a non-bipartite lattice.  After all, in two dimensions, the dimer model on a triangular lattice has a gapped liquid phase\cite{Moessner01}.  A possible three dimensional lattice to consider is the stacked triangular lattice.  This lattice can be chosen to have $\pi$-flux in each plaquette as required.  Although there may be problems with this approach (for example, we could find that the monomers become confined in planes, or that there is no liquid phase), it may allow the construction of a gapped three dimensional model and quantum Monte Carlo simulations of this system would be useful.  We have not been able to construct a field theoretic version of this model, however.  In this model, suppose the Hopfion is a gapped, localized object.  In this case, implementing an operation such as twisting both quasi-particles would require twisting each quasi-particle separately, emitting a Hopfion pair, and then bringing the Hopfion pair together and annihilating them.  However, it seems unnatural for the Hopfion to be gapped: suppose a quasi-particle is dragged along a long closed loop and returns to its origin with a changed sign, causing a Hopfion to be emitted.  At what point along this path does the system have to pay an energy price to emit the Hopfion?  Since the system is gapped and liquid-like, all points on the path should be the same, so why should one point be distinct?  One possible resolution is that the Hopfion is gapped in the bulk, but that there is a zero energy bound state of a Hopfion and a defect, reminiscent of what happens in the triangular lattice model in two dimensions.  Again, this is a problem for quantum Monte Carlo to study.

\acknowledgements
We would like to thank John McGreevy, T. Senthil, S. Sondhi,
Zhenghan Wang, and F. Wilczek for discussions.
We thank A. Vishwanath for comments on
a preliminary draft of this paper.
M.F. and C.N. acknowledge the hospitality of the Aspen Center for Physics
where part of this work was completed.
C.N. has been supported in part by the DARPA QuEST program.
We thank John McGreevy and Brian Swingle for sharing their
preprint\cite{McGreevy11} prior to publication.

\appendix

\section{Relation between the Hopfion and the gauge field instanton
in the (4+1)-dimensional theory}

In this appendix, we will present a more detailed discussion of the relation between the Hopfion of the order parameter and the gauge field instanton in the (4+1)-d theory discussed in Sec. \ref{sec:4d}. In the interest of generality, we consider the generic action in Eq. (\ref{actionSUN}) with $N$ copies of Weyl fermions forming the fundamental representation of SU(N), coupled to an SU(N) gauge field and a Higgs field $n^{ab}\in SU(N)/SO(N)$.

We first consider the symmetry unbroken phase $n^{ab}=0$. The axial anomaly leads to the equation
\begin{eqnarray}
\partial_\mu j^\mu=\frac1{32\pi^2}{\rm Tr}\left[\epsilon^{\mu\nu\sigma\tau}f_{\mu\nu}f_{\sigma\tau}\right]
\end{eqnarray}
with $f_{\mu\nu}$ the SU(N) gauge curvature, and
\begin{eqnarray}
j^\mu=c^\dagger\sigma^\mu c
\end{eqnarray}
is the number current of the Weyl fermion. In the $4+1$-d regularization of this theory, the Weyl fermions live on the boundary. For the slab geometry shown in Fig. (\ref{fig4d}),  an instanton configuration of $f_{\mu\nu}$ with the second Chern number $C_2=\int d^4x\frac1{32\pi^2}{\rm Tr}\left[\epsilon^{\mu\nu\sigma\tau}f_{\mu\nu}f_{\sigma\tau}\right]=N$ pumps $N$ fermions from one boundary to the other. The instanton configuration is determined by the homotopy group $\pi_3(SU(N))=\mathbb{Z}$. Consider an instanton configuration with $C_2=N$ in the (boundary) space-time $\mathbb{R}^4$ with the boundary condition $f_{\mu\nu}\rightarrow 0$ for time $t\rightarrow \pm \infty$. If we choose a gauge so that the gauge potential $a_\mu$ is continuous and $a_\mu({\bf x},t)\rightarrow 0$ for $t\rightarrow -\infty$, then $a_\mu({\bf x},t)\rightarrow -ig^{-1}({\bf x})\partial_\mu g({\bf x})$ for $t\rightarrow +\infty$, in which $g({\bf x})\in {\rm SU(N)}$ defines a mapping ${\bf x}\rightarrow g({\bf x})$ with nontrivial winding number
\begin{eqnarray}
N=\frac{1}{24\pi^2}\int d^3{\bf x}\epsilon^{ijk}{\rm Tr}\left[g^{-1}\partial_i gg^{-1}\partial_j gg^{-1}\partial_k g\right]\label{app:SUNwinding}
\end{eqnarray}

Now we consider the Higgs phase with the order parameter $n^{ab}({\bf x},t)$ defining a map from the space-time to the target space $SU(N)/SO(N)$. A Hopfion is a soliton configuration $n^{ab}({\bf x},0)$ with nontrivial winding number determined by $\pi_3\left(SU(N)/SO(N)\right)=\mathbb{Z}_2$. For unitary symmetric matrices $n^{ab}$ satisfying $n=n^T,~n^\dagger n=1$, one can always find a gauge transformation $g({\bf x})\in {\rm SU(N)}$ such that
\begin{eqnarray}
n({\bf x})=g^Tg({\bf x})
\end{eqnarray}
The choice of $g({\bf x})$ is not unique since for any $O({\bf x})\in {\rm SO(N)}$, $\tilde{g}({\bf x})=O({\bf x})g({\bf x})$ also satisfies $n({\bf x})=\tilde{g}^T\tilde{g}({\bf x})$. Thus $g({\bf x})$ is determined modulo a $SO(N)$ factor multiplied from the left, which is consistent with the fact that the order parameter space is the coset ${\rm SU(N)/SO(N)}$. The $\mathbb{Z}_2$ winding number of the configuration $n({\bf x})$ is related to the integer-valued winding number of the gauge transformation $g({\bf x})$ given in Eq. (\ref{app:SUNwinding}) by the following equation:
\begin{eqnarray}
N_{\rm Z_2}\left[n({\bf x})\right]&=&N\left[g({\bf x})\right]~{\rm mod}~2\nonumber\\
&=&\frac{1}{24\pi^2}\int d^3{\bf x}\epsilon^{ijk}{\rm Tr}\left[g^{-1}\partial_i gg^{-1}\partial_j gg^{-1}\partial_k g\right]\nonumber\\
& &{\rm mod}~2\label{windingZ2}
\end{eqnarray}
As is discussed above, $g({\bf x})$ has an SO(N) ambiguity. For two configurations $g({\bf x})$ and $\tilde{g}({\bf x})=O({\bf x})g({\bf x})$ with $O({\bf x})\in {\rm SO(N)}$, the winding number (\ref{app:SUNwinding}) is different by
\begin{eqnarray}
N\left[\tilde{g}({\bf x})\right]-N\left[g({\bf x})\right]&=&N\left[O({\bf x})\right]\nonumber\\
&\equiv &\frac{1}{24\pi^2}\int d^3{\bf x}\epsilon^{ijk}{\rm Tr}\left[O^{-1}\partial_i O\right.\nonumber\\
& &\left.\cdot O^{-1}\partial_j OO^{-1}\partial_k O\right]
\end{eqnarray}
$N\left[O({\bf x})\right]$ is the winding number of the orthogonal matrix $O({\bf x})$, which is always an even number. Thus the winding number $N_{\rm Z_2}\left[n({\bf x})\right]=N\left[g({\bf x})\right]~{\rm mod}~2$ is not affected by the choice of $g({\bf x})$ and is determined completely by $n({\bf x})=g^Tg$.

The relation between the winding numbers of ${n}({\bf x})$ and $g({\bf x})$ presented above can be verified directly, but we also would like to provide an alternative explanation in a more mathematically rigorous way. The facts we discussed here arise from the exact sequence
\begin{eqnarray}
1\rightarrow SO(N)\rightarrow SU(N)\rightarrow SU(N)/SO(N)\rightarrow 1
\end{eqnarray}
which leads to the long exact sequence
\begin{eqnarray}
...&\rightarrow& \pi_{n+1}\left(SU(N)/SO(N)\right)\rightarrow \pi_n\left(SO(N)\right)\rightarrow \pi_n\left(SU(N)\right)\nonumber\\
&\rightarrow &\pi_n\left(SU(N)/SO(N)\right)\rightarrow \pi_{n-1}\left(SO(N)\right)\rightarrow ...
\end{eqnarray}
Considering that $\pi_2(SO(N))=1$ and $\pi_4(SU(N)/SO(N))=1$ (with $1$ standing for a trivial homotopy group), we have
\begin{eqnarray}
1&\rightarrow &\pi_3\left(SO(N)\right)\rightarrow \pi_3\left(SU(N)\right)\nonumber\\
&\rightarrow &\pi_3\left(SU(N)/SO(N)\right)\rightarrow1
\end{eqnarray}
From this exact sequence we see that the homotopy group $\pi_3\left(SU(N)/SO(N)\right)=\mathbb{Z}_2$ is a quotient group of $\pi_3\left(SU(N)\right)=\mathbb{Z}$ and $\pi_3\left(SO(N)\right)=\mathbb{Z}$. In other words there is an injective projection from $\pi_3\left(SO(N)\right)$ to $\pi_3\left(SU(N)\right)$ which maps each ${\rm SO(N)}$ configuration with winding number $n\in\mathbb{Z}$ to a ${\rm SU(N)}$ configuration with winding number $2n$. The nontrivial configurations of the coset ${\rm SU(N)/SO(N)}$ are the image of the $SU(N)$ configurations with odd winding number.

With the preparation above, now we consider the gauge field coupling to the Higgs field. With the configuration $n({\bf x})$ of the Higgs field, the gauge field configuration in the ground state is a pure gauge determined by the condition
\begin{eqnarray}
D_\mu n=\partial_\mu n-ia_\mu^T n-ina_\mu=0
\end{eqnarray}
For $n=g^Tg$ one obtains
\begin{eqnarray}
a_\mu=-ig^{-1}\partial_\mu g
\end{eqnarray}
Thus we see that a Hopfion configuration with nonzero $N_{\rm Z_2}$ defined in Eq. (\ref{windingZ2}) is always associated with a pure gauge configuration with odd winding number $N\left[g({\bf x})\right]$. The Hopfion number is invariant under smooth deformation of the field $n({\bf x})$, and thus remains conserved in time evolution unless the orderparameter vanishes. If the order parameter is allowed to vanish, such as by pair creation of monopole pairs, the Hopfion number can be changed. An example of such process is illustrated in Fig. (\ref{Hopf4d}). Consider a generic Hopfion creation process, which connects the initial configuration $n_1({\bf x})$ with Hopfion number $N_{Z_2}=0$ and the final configuration $n_2({\bf x})$ with Hopfion number $N_{Z_2}=1$. Independent from the detail of the Hopfion creation process, according to the discussion above one must simultaneously create an instanton of the gauge field. Defining $n_1=g^T_1g_1({\bf x})$, $n_2=g^T_2g_2({\bf x})$, one obtains
\begin{eqnarray}
C_2&=&\frac1{32\pi^2}{\rm Tr}\left[\epsilon^{\mu\nu\sigma\tau}f_{\mu\nu}f_{\sigma\tau}\right]=N\left[g_2({\bf x})\right]-N\left[g_1({\bf x})\right]\nonumber\\
\end{eqnarray}
Consequently
\begin{eqnarray}
C_2~{\rm mod}~2&=&N_{\rm Z_2}\left[n_2({\bf x})\right]-N_{\rm Z_2}\left[n_1({\bf x})\right]=1
\end{eqnarray}
Thus we see that the gauge field configuration associated with this Hopfion creation process is an instanton with odd Chern number, which corresponds to a change of the Weyl fermion number
\begin{eqnarray}
\Delta N_F=C_2\Rightarrow \Delta N_F~{\rm mod}~2=1
\end{eqnarray}
Thus the fermion number of the boundary system changes by an odd number. Although in the Higgs phase the Fermion number conservation is explicitly broken, the Fermion number modulo $2$ is still conserved in the classical level.

In summary, the discussion above leads to the conclusion that the $\mathbb{Z}_2$ symmetry of Fermion number modular $2$ is broken anomalously for this boundary system, and the instanton configuration that produces this anomaly is a Hopfion creation process. One can also provide a complementary picture in the bulk of the $(4+1)$-d regularized theory. To understand the relation between Hopfion number and fermion number parity in the bulk, we introduce a $U(1)$ probe gauge field $A_I$ ($I=0,1,..,4$) and couple the fermions to the field $A_I+a_I$. Integrating over the fermions leads to a Chern-Simons term of the gauge field $A_I+a_I$ ({\it c.f.} Ref.\onlinecite{Qi08}):
\begin{eqnarray}
S_{\rm eff}&=&\frac1{24\pi^2}\int d^5x\epsilon^{IJKLM}{\rm Tr}\left[\left(A_I+a_I\right)\right.\nonumber\\
& &\left.\cdot\partial_J\left(A_K+a_K\right)\partial_L\left(A_M+a_M\right)+...\right]
\end{eqnarray}
with $...$ standing for the terms with higher orders of $A_I+a_I$ required by gauge invariance. The fermion number current couples to the gauge field $A_I$ and is described by the response equation
\begin{eqnarray}
j^I=\left.\frac{\delta S_{\rm eff}}{\delta A_I}\right|_{A_I=0}=\frac1{32\pi^2}\epsilon^{IJKLM}{\rm Tr}\left[f_{JK}f_{LM}\right]
\end{eqnarray}
In any $4$-d spatial region $D$, the fermion number is
\begin{eqnarray}
N_F(D)=\int_Dj^0d^4x=\int_D d^4x\frac1{32\pi^2}\epsilon^{0JKLM}{\rm Tr}\left[f_{JK}f_{LM}\right]\nonumber\\
\end{eqnarray}
In the Higgs phase, the fermion number conservation is broken but the fermion number mod $2$ is still conserved. Thus the equation above still holds modular $2$.

In the $4$-d space the Hopfion is a point-like defect. The process of creating of Hopfion on the boundary can also be thought as pulling a Hopfion defect across the surface into the bulk, as is illustrated in Fig. (\ref{Hopf4d}). Consider a spatial region $D$ including a Hopfion defect and its neighborhood. As long as the region $D$ is large enough to cover the core region of the Hopfion, the gauge field at the boundary $\Sigma=\partial D$ is a pure gauge. By definition, the configuration $n({\bf x})$ on $\Sigma$ has an odd winding number $N_{\rm Z_2}\left[n({\bf x})\right]=1$, which corresponds to a pure gauge configuration with an odd winding number $N_\Sigma$ defined in Eq. (\ref{app:SUNwinding}). Thus the fermion number in Region $D$ satisfies
\begin{eqnarray}
N_F(D)=\int_D d^4x\frac1{32\pi^2}\epsilon^{0JKLM}{\rm Tr}\left[f_{JK}f_{LM}\right]=N_\Sigma~{\rm mod}~2\nonumber\\
\end{eqnarray}
which demonstrates that each Hopfion carries an odd fermion number. When the Hopfion is created and then moved across the bulk to the other surface, one fermion (or odd number of fermions) is pumped from one boundary to the other.

\bibliographystyle{prsty}
\bibliography{projective}

\end{document}